\documentclass[%
 reprint,
 amsmath,amssymb,
 aps,
]{revtex4-2}

\usepackage{graphicx}
\usepackage{dcolumn}
\usepackage{bm}
\usepackage[dvipsnames]{xcolor}
\definecolor{linkcolor}{rgb}{0.18, 0.1875, 0.5725}
\usepackage{multirow}
\usepackage{hyperref}
\hypersetup{
	colorlinks=true,
	citecolor=linkcolor,
	linkcolor=linkcolor,
	urlcolor=linkcolor
}
\usepackage[mathlines]{lineno}
\usepackage{natbib}
\begin{document}



\preprint{APS/123-QED}

\title{Multistate Transition Dynamics by Strong Time-Dependent Perturbation in NISQ era}

\author{Yulun Wang}
\email[]{yulun.wang@stonybrook.edu} 
\author{Predrag S. Krsti{\'c}}
\email[]{krsticps@gmail.com} 
\affiliation{Institute for Advanced Computational Science, Stony Brook University, Stony Brook NY 11794-5250, USA}

\date{\today}

\begin{abstract}
We develop a quantum computing scheme utilizing McLachlan variational principle in a hybrid quantum-classical algorithm to accurately calculate the transition dynamics of a closed quantum system with many excited states subject to a strong time-dependent perturbation. A systematic approach for optimal construction of a general $N$-state ansatz with unary $N$-qubit encoding is refined. We also utilize qubit efficient encoding in McLachlan variational quantum algorithm to reduce the number of qubits to $\log_2 N$, simultaneously diminishing depths of the quantum circuits. The significant reduction of the number of time steps is achieved by use of the second order marching method. Instrumental in obtaining high accuracy are adaptations of the circuits to include time-dependent global phase correction. We illustrated, tested and optimized our quantum computing algorithm on a set of 16 bound hydrogenic eigenstates exposed to a strong laser attosecond pulse. Results for transition probabilities are obtained with accuracy better than 1$\%$, as established by comparison to the benchmark data. Use of interaction representation of the Hamiltonian reduces the effect of both NISQ noise and sampling errors accumulation while the quantum system evolves in time.
\end{abstract}
\maketitle

\section{\label{sec:intro}INTRODUCTION\protect}
All physical and chemical phenomena and reactions in the universe are undergoing constant changes, at their characteristic time scales. Simulation of dynamic systems, significant in comprehending the nature, is hence considered as one of the most important and promising applications of quantum computers. Since Feynman proposed the idea of simulating many-body dynamics using quantum computers \cite{feynman82}, various approaches have been explored over the last few decades to provide the insight of physics beyond the reach of classical computers \cite{frank20,monroe21,aspuru12,altman21}. The efforts of existing work in quantum computing are mainly focused on the static \cite{abrams97,rolando03,alan05,georgescu14,cao19} and dynamical \cite{seth96,kassal08,smith19,georgescu14,cao19} properties of many-body systems. Quantum algorithms developed for quantum computing, such as the quantum phase estimation \cite{nielsen11} and HHL algorithm \cite{harrow09} often require large circuit depth, which consequently demands a quantum device with qubits of high fidelity and long coherence time for the successful execution, mainly not available in the Noisy Intermediate-Scale Quantum (NISQ) hardware \cite{preskill18}. This has encouraged development of Variational Hybrid Quantum-Classical Algorithms (VHQCAs) \cite{peruzzo14,farhi14,li17,yuan19,higgott19,bravo19,anschuetz19,lubasch20,cerezo21} which show respectable successes. The general framework for VHQCAs is characterized by shallow quantum variational circuits followed by postprocessing with classical computational techniques. The VHQCAs have applications in a wide range of quantum computing problems during the NISQ era, such as the Variational Quantum Eigensolver (VQE) \cite{peruzzo14,kandala17}, Quantum Approximate Optimization Algorithm (QAOA) \cite{farhi14,wang18}, and Variational Quantum Linear Solver (VQLS) \cite{bravo19,patil21}

In quantum physics and chemistry, there are two classes of computationally difficult problems: (1) the eigenvalue problem of a many-body systems, including excited states, and (2) the time evolution of a quantum system, including transition dynamics between multitude of excited states. The challenge we have in mind is to develop and apply a quantum algorithm, capable for a universal quantum computer, that can evolve and atomic system with controlled accuracy ($< 1\%$) under a strong time-dependent perturbation which causing transitions between many system states. This problem is formidable even for a few-body system. We use an example of hydrogen atom in a strong, attosecond laser field pulse, to focus on and stress the important properties and difficulties of calculating accurately transitions between fully entangled many-fold of electronic states, subject to strong time dependent perturbation.

The time-dependent problems are computationally much more involved than the static ones, owing to a need to apply the calculations consecutively for many time steps as the system evolves. The number of the steps could be exceedingly large due to accuracy requirements for small time steps and a long extent of time needed for the full system evolution. Simultaneously, one must keep the propagation of the numerical error at minimum, so that the final probabilities for all states of interests reach accuracy at an acceptable level (which we set to be $<1\%$) defined by the deviation from the benchmark, which is a challenge even by classical computer and for a few-body system. 

There is a number of publications treating the quantum computing aspects of the time-dependent transition dynamics in quantum systems using the trotterization of the evolution operator $e^{-iHt}$ \cite{di21, sawaya20,fauseweh21,tranter19}. The theory of VHQCAs for a quantum computer simulation of the general real- and imaginary-time evolution has been established \cite{li17}, and developed in various aspects \cite{yuan19,endo20,mcardle19ansatz}. Error minimization by extrapolation to zero was introduced to VHQCAs in \cite{li17} for the quantum Ising model of three spins initialized in the cluster state, with significantly suppressed effect of errors. Adaptive expansion of ansatz was applied to VHQCAs for dynamical simulation of the finite-rate quantum quench in the integrable Lieb-Schultz-Mattis spin chains and sudden quench of the nonintegrable mixed-field Ising model \cite{yao21}. This resulted in a highly accurate results and much shallower circuits with two orders of magnitude smaller number of CNOT gates comparing to the first-order trotterization method. Furthermore, VHQCAs were extended to the generalized time evolution with a non-Hermitian Hamiltonian and open quantum system dynamics \cite{endo20}. This was applied to simulation of the ideal and dissipative evolution of 2D Ising model, which showed good agreement with exact solutions. However, as far as we know, none of the implementations of VHQCAs has been developed for a universal quantum computer to accurately simulate transitions in a fermionic dynamic system to a number of excited states by a strong, time-dependent perturbation. 

In this work, we apply VHQCA based on McLachlan’s variational principle (MLVP) \cite{mclachlan64} to solve transition dynamics of a closed, quantum system with potentially many excited states subject to a strong time-dependent perturbation. We apply and test quantum circuits applicable in quantum computers to calculate excitation dynamics using an example of a single atom in a strong, attosecond laser field pulse within a multitude of 16 hydrogenic states. Our contributions in this work can be summarized as following: (1) We propose a systematic approach to construct a general N-state ansatz to be used in unary encoding, using Jordan-Wigner encoding (JWE) \cite{jordan28,mcardle19ansatz,jones19} and $N$ qubits, capable to reach highly accurate transition probabilities by controlling the time-step size (discussed in detail at Sec.~\ref{sec:3.1.1}); (2) We apply a compact encoding method, so called qubit efficient encoding (QEE) \cite{shee21,sawaya20,mcardle19digital,di21} in McLachlan VHQCA to simulate the dynamic evolution of a system, reducing the number of required qubits from $N$ in unary encoding to $\log_2 N$, and significantly lessening the circuit depth (discussed in detail at Sec.~\ref{sec:2.3}); (3) The second-order parameter marching method is applied reducing the simulation time by an order of magnitude; (4) The quantum noise simulation is conducted to investigate the error accumulation and potential procedure for error mitigation; (5) All results are compared with accurate benchmark results obtained by classical computing to quantify the errors and provide verification of the successful search for the optimal computational techniques.

\begin{figure}
\includegraphics{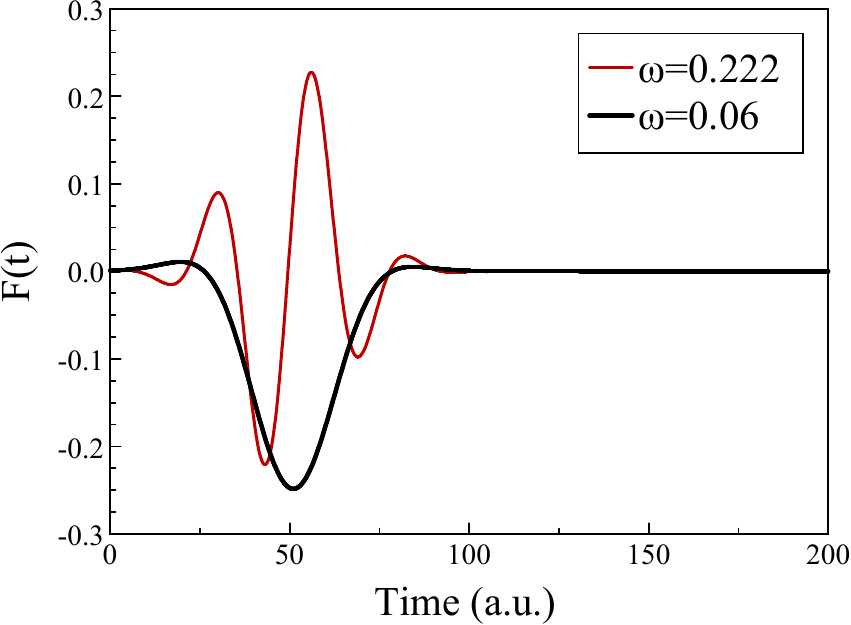}
\caption{\label{fig:1} The laser fields used in the calculation defined as $E_0 = 0.25, \tau = 20.5, t_0 = 50$ with $\omega=0.06$ (thick black line) and $\omega=0.222$ (thin red line).}
\end{figure}

In general, a dynamic atomic quantum system is described by the Time Dependent Schrodinger Equation (TDSE) which defines the time evolution of the system wavefunction $\psi\left(\bm{r},t\right)$ (atomic system of units, $\hbar=m=e=1$, is used throughout this manuscript, unless otherwise said):
\begin{equation}
\label{eq1}
\frac{\partial\psi\left(\bm{r},t\right)}{\partial t} + i H\left(\bm{r},t\right) \psi\left(\bm{r},t\right) =0
\end{equation}
Hamiltonian
\begin{equation}
\label{eq2}
H\left(\bm{r},t\right)=H_0\left(\bm{r}\right)+P\left(\bm{r},t\right)
\end{equation}
is generally a function of space and time, with unperturbed Hamiltonian $H_0(\bm{r})$ and an external time-dependent perturbation $P(\bm{r},t)$, where $\bm{r}$ is a set of electronic coordinates. There are various computational methods to solve TDSE \cite{crank47,kutta01,runge95}. For example, one can replace the Schrodinger equation by an infinite set of coupled integro-differential equations obtained upon expansion of the wave function in a complete basis set, which could be constructed by the eigenfunctions of the unperturbed Hamiltonian $H_0$. This is the approach which we apply in this work, using a single-electron, hydrogen atom, subject to a time dependent perturbation. We chose incomplete, truncated basis sets of 2,4, 8 or 16 hydrogen bound eigenfunctions to approximately model the atom.

The hydrogen atom is exposed to a short and strong laser pulse. The interaction energy of the electron with the classical, linearly polarized dipole laser field in the length gauge is defined by:
\begin{equation}
\label{eq3}
P\left(\bm{r},t\right)=\bm{F}\left(t\right)\cdot\bm{r}
\end{equation}
where
\begin{subequations}
\label{eq4}
\begin{eqnarray}
\bm{F}\left(t\right)=F\left(t\right)\hat{z} \label{eq4a}
\end{eqnarray}
\begin{equation}
F\left(t\right)=E\left(t\right)\cos\left(\omega t\right) \label{eq4b}
\end{equation}
\end{subequations}
and the time-dependent amplitude of electric field has Gaussian switching conditions:
\begin{equation}
\label{eq5}
E\left(t\right)=E_0 e^{-\left(\frac{t-t_0}{\tau}\right)^2}
\end{equation}
with Full Width at Half Maximum (FWHM) $=$ $2\left(\ln2\right)^{\frac{1}{2}}\tau$. The unperturbed $H_0$ in this case is the standard hydrogen Hamiltonian $H\left(\bm{r}\right)=-\frac{\nabla^2_{\bm{r}}}{2}+V\left(r\right)$, where $V\left(r\right)=-\frac{1}{r}$ is the potential energy of electron.

The hydrogen bound states in the truncated basis are all with $m=0$ magnetic quantum number when one starts from the initial state $1s$, and follow the optical selection rules with a linearly polarized laser $\Delta l=\pm 1,\Delta m=0$. The sets of the hydorgenic orbitals used to define the model atom are listed in Sec. SI of Supplemental Material (SM). 

By choosing two different frequencies $\omega$ in Eqs.~\ref{eq4}, with $E_0=0.25, \tau=20.5, t_0=50$, we obtain two characteristic time-dependent perturbing fields, shown in Fig.~\ref{fig:1}, with FWHM$\sim$825.7 as. Thus, for $\omega=0.06$ (wavelength $\lambda\sim700$ nm, i.e., single photon energy about 1.63 eV), we obtain approximately a half cycle pulse (HCP) within the Gaussian envelope, which is a characteristic representation of a time dependent perturbation met in non-harmonic processes. For example, a HCP-like electric field appears as the orthogonal component of electric field in electron-ion collisions. A HCP in general can be approximated by Dirac delta function if the pulse is short enough \cite{krstic94} (which is not a case here, as discussed in Sec. SIIIA of SM) and can be created experimentally by a careful convolution of many laser-field modes, as illustrated by its Fourier expansion in Sec. SIIIB of SM. The application of our VHQCA to the atom and the $\omega=0.06$ laser is studied in detail in Sec.~\ref{sec:3}, while in Sec.~\ref{sec:4} we set $\omega=0.222$ ($\lambda\sim189$ nm, i.e., single photon energy about 6.04 eV). In that case $F\left(t\right)$ in Eqs.~\ref{eq4} has a few cycles during the pulse as shown in Fig.~\ref{fig:1} (thin red line). The laser will reach its maximum amplitude at about $t\sim50$, and is considered to be practically zero at $t\sim T=200$. This enables calculation of the $S$-matrix elements for the transition of the atom from initial state defined before the laser is switched on at $t=0$ to a final state after the laser is switched off at $t=T$ \cite{krstic90}. 

In order to obtain benchmarks for testing the numeric accuracy of our quantum algorithms which we derive in in Sec.~\ref{sec:2}, \ref{sec:3} and \ref{sec:4}, the system wave function is expanded in a finite truncated set of $N$ bound hydrogenic eigenfunctions $\vert \varphi_i\rangle$, where $N$ is 2,4,8, or 16:
\begin{equation}
\label{eq6}
\psi\left(\bm{r},t\right)=\sum^N_{i=1} c_i \vert \varphi_i\rangle
\end{equation}
where $c_i$ is the amplitude of each $\vert \varphi_i\rangle$. When replaced in Eq.~\ref{eq1}, this expansion yields a finite set of coupled, time-dependent Ordinary Differential Equations (ODEs) which approximate the Schrodinger equation. 

\begin{figure}
\includegraphics{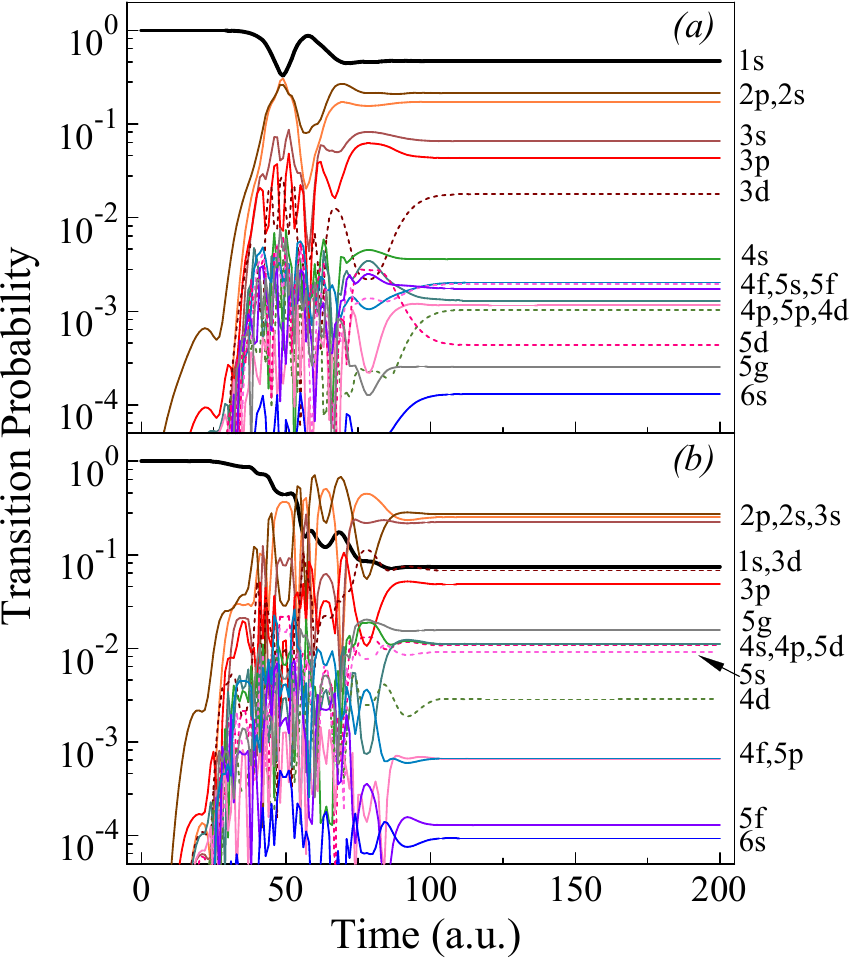}
\caption{\label{fig:2} The benchmark results for the transition dynamics of the 16-state $H$ model systems with laser pulse of (a) $\omega=0.06$ and (b) $\omega=0.222$.}
\end{figure}

The set of coupled ODEs is solved highly accurately using standard classical numerical methods with backward differentiation formula (Python \texttt{scipy.integrate.ode} function \cite{scipy}). The absolute and relative tolerance at each step are set at $10^{-12}$ and $10^{-6}$ respectively to provide sufficient accuracy. The obtained time-dependent transition probabilities from 1$s$ state for $N$=2,4,8,16 systems with HCP laser field ($\omega=0.06$) are plotted in Fig. S1 of SM, and for $N$=4,8,16 systems with laser field $\omega=0.222$ are plotted in Fig. S2 of SM. The benchmark transition probabilities at $t=T=200$ are listed in Sec. SII of SM. In Fig.~\ref{fig:2} we plot the evolution of the transition probabilities in time for $N$=16 system using laser field with (a) $\omega=0.06$ and with (b) $\omega=0.222$.

The details on the VHQCAs applied in this work and corresponding quantum circuits are shown in Sec.~\ref{sec:2}. The results, tests and discussions using different encoding techniques are presented in Sec.~\ref{sec:3}. The performance of the developed quantum algorithms in presence of NISQ noise and sampling errors is investigated and discussed in Sec.~\ref{sec:4}. Finally, our conclusions are given in Sec.~\ref{sec:5}. 

\section{\label{sec:2}Methods\protect}
\subsection{\label{sec:2.1}VHQCA for time-dependent problems}
To simulate the dynamic system using variational algorithms, the system wavefunction $\psi(\bm{r},t)$ can be approximated by a parameterized ansatz as a trial state $\phi(\bm{\theta}(t))$, where the time dependent parameter vector $\bm{\theta}(t)$ has components $\theta_i(t),i=1,\ldots,L$, and $L$ is the total number of parameters. The Eq.~\ref{eq1} then takes the form:
\begin{equation}
\label{eq7}
\sum_i \frac{\partial \Big\vert \phi\left(\bm{\theta}\left(t\right)\right)\Big\rangle }{\partial \theta_i} \dot{\theta}_i+i H \Big\vert \phi\left(\bm{\theta}\left(t\right)\right)\Big\rangle\approx 0
\end{equation}

In the limit $\phi\rightarrow\psi$, the Eq.~\ref{eq7} tends to the Schrödinger equation. The algorithms for evolving the vector $\bm{\theta}(t)$, convenient for computer processing, can be obtained using the Dirac-Frenkel \cite{dirac30,frenkel34} or MLVP \cite{mclachlan64}. These variational principles are equivalent \cite{yuan19} when the components of $\bm{\theta}(t)$ are complex numbers. However, all operators in quantum computing of a closed quantum system are unitary operators, with exception of measurement and reset operations. These can be represented by unitary matrices, preserving the inner product of two arbitrary states, and overall unitarity of the system. It requires that all variational parameters which represent real angles of the appropriate single qubit or two-qubits controlled rotation gates in quantum circuits are real numbers. The MLVP is derived assuming real variational parameters, and the solutions for derivatives in real time are always real. This makes MLVP our method of choice in implementations of the VHQCAs to the time-dependent problems in quantum computing.

The MLVP aims to minimize the squared norm of the left side of Eq.~\ref{eq7}:
\begin{equation}
\label{eq8}
\delta \bigg\|\left(\frac{d}{dt}+i H\right) \Big\vert \phi\left(\bm{\theta}\left(t\right)\right)\Big\rangle\bigg\|^2=0
\end{equation}

The variation of real $\dot{\bm{\theta}}$ yields the system of coupled algebraic equations:
\begin{equation}
\label{eq9}
\sum_j A^R_{i,j}\dot{\theta}_j=C^I_i
\end{equation}
where
\begin{subequations}
\label{eq10}
\begin{eqnarray}
A^R_{i,j}= \Re\Bigg[ \frac{\partial \Big\langle \phi\left(\bm{\theta}\left(t\right)\right)\Big\vert }{\partial \theta_i}\frac{\partial \Big\vert \phi\left(\bm{\theta}\left(t\right)\right)\Big\rangle }{\partial \theta_j} \Bigg] \label{eq10a}\\
C^I_i=\Im\Bigg[ \frac{\partial \Big\langle \phi\left(\bm{\theta}\left(t\right)\right)\Big\vert }{\partial \theta_i}H \Big\vert \phi\left(\bm{\theta}\left(t\right)\right)\Big\rangle\Bigg] \label{eq10b}
\end{eqnarray}
\end{subequations}

The full derivation of Eqs.~\ref{eq9} and ~\ref{eq10} is provided in \cite{yuan19} and in Sec. SIV of SM. For a given value of $\bm{\theta}(t)$, the matrix $\bm{A}^R$ and vector $\bm{C}^I$ are computed using quantum circuits based on a chosen form of the quantum ansatz and obtained by quantum measurement. Then $\dot{\bm{\theta}}$ can be calculated by inversion of matrix $\bm{A}^R$ using the classical numerical methods.

Since the time derivative of the ansatz, $\frac{\partial \langle \phi(\bm{\theta}(t))\vert }{\partial t}$, is included in the time-dependent variational algorithm (Eqs.~\ref{eq10}), it could suffer a substantial deviation from the exact $\frac{\partial \langle \psi(t)\vert }{\partial t}$ if global phase is not included into consideration \cite{yuan19}. Such a counter-intuitive argument does not appear in the time-independent applications of VHQCA (for example in VQE), where overall phase plays no role in the measured results. 

Assuming ansatz in the form: 
\begin{equation}
\label{eq11}
\vert \Phi\left(t\right)\rangle=e^{i\alpha\left(t\right)}\Big\vert \phi\left(\bm{\theta}\left(t\right)\right) \Big\rangle,
\end{equation}
replacing it in Eq.~\ref{eq1}, and performing the variation of $\dot{\bm{\theta}}$ and $\dot{\alpha}$, the improved equations for the time derivative of the vector $\bm{\theta}(t)$ with extra global phase correction (GPC) terms take the form \cite{yuan19}:
\begin{equation}
\label{eq12}
\sum_j M_{i,j}\dot{\theta}_j=V_i
\end{equation}
where
\begin{subequations}
\label{eq13}
\begin{eqnarray}
&&M_{i,j}= A^R_{i,j}\nonumber\\
&&\quad+\frac{\partial \big\langle \phi(\bm{\theta}(t))\big\vert }{\partial \theta_i} \big\vert \phi(\bm{\theta}(t))\big\rangle  
\frac{\partial \big\langle \phi(\bm{\theta}(t))\big\vert }{\partial \theta_j}\big\vert \phi(\bm{\theta}(t))\big\rangle \label{eq13a}\\
&&V_i=C^I_i\nonumber\\
&&\quad+i\frac{\partial \big\langle \phi(\bm{\theta}(t))\big\vert }{\partial \theta_i} \big\vert \phi(\bm{\theta}(t))\big\rangle  
\big\langle \phi(\bm{\theta}(t))\big\vert H \big\vert \phi(\bm{\theta}(t))\big\rangle \label{eq13b}
\end{eqnarray}
\end{subequations}

The full derivation of Eqs.~\ref{eq12} and ~\ref{eq13} is provided in Sec. SV of SM. It is important to stress that inclusion of the GPC does not include the phase $\alpha$ or its derivative in Eqs.~\ref{eq13}, i.e., it does not increase the number of variational parameters. Our calculations in Sec.~\ref{sec:3} show that algorithm equipped with the GPC produces significantly more accurate results than Eqs.~\ref{eq9} with the use of the same number of variational parameters and the same size of the time steps. This leads to more accurate calculations of the transition probabilities, at least in case of an atom in a laser field.   

\begin{figure}
\includegraphics{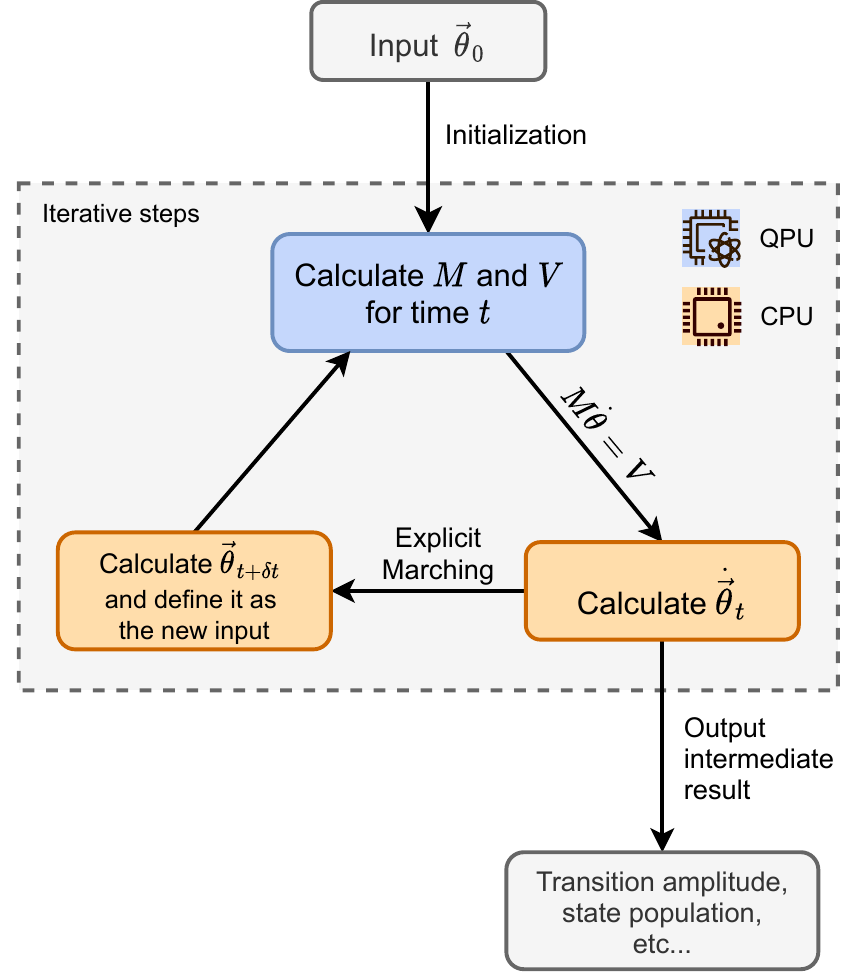}
\caption{\label{fig:3} The basic outline of our variational approach to the quantum dynamics simulations.}
\end{figure}

With $\bm{\theta}(t)$ as input at time $t$, $\bm{M}$ and $\bm{V}$ are computed from quantum circuits, as explained later in Sec.~\ref{sec:2.4}. Similarly to Eqs.~\ref{eq9}, the Eqs.~\ref{eq12} are solved for $\dot{\bm{\theta}}$ using classical computing, by inversion of matrix $\bm{M}$ with LU factorization using Python Numpy function \texttt{numpy.linalg.inv} \cite{numpy}. Vector $\bm{\theta}$ for the next time step is obtained via explicit marching methods with the knowledge of $\dot{\bm{\theta}}$, which enables the quantum computing of $\bm{M}$ and $\bm{V}$ at the new time. The whole process is sketched in Fig.~\ref{fig:3}, leading to the wave function at the targeted final time $T$. Projecting the eigenstates of the unperturbed Hamiltonian onto the $\phi(\bm{\theta}(T))$, one obtains the $S$-matrix elements for transition to all states of the used truncated basis set starting from a chosen initial state of the system, defined by $\phi(\bm{\theta}(t))$ at initial time $t=0$. We also note that we calculate the “transition amplitudes” at times $0<t<T$, while the perturbation $P$ is on. In spirit of defining initial and final states of the system when the laser is off, the intermediate amplitudes and respective transition probabilities do not have a measurable physical meaning beyond being coefficients in Eq.~\ref{eq6}, since the system does not have eigenstates while the time-dependent perturbation is on.

A common time-marching method for updating the variational parameters used in VHQCAs \cite{li17,yao21,mcardle19ansatz} is the explicit forward Euler method \cite{press92} which is denoted here as the First-Order Marching (FOM):
\begin{equation}
\label{eq14}
\bm{\theta}_{t+\delta t}=\bm{\theta}_t+\dot{\bm{\theta}}_t\cdot\delta t+\mathcal{O}\left(\delta t^2\right)
\end{equation}
where $\delta t$ is predefined fixed step size of the evolution algorithm. 

To mitigate the need for a smaller step in our time marching, and reduce the number of steps needed to reach the final time $T$, we rather use the explicit forward Adams-Bashforth second order (AB2) scheme \cite{bashforth83}, based on the second-order Taylor series expansion, denoted here as the Second-Order Marching (SOM):
\begin{equation}
\label{eq15}
\bm{\theta}_{t+\delta t}=\bm{\theta}_t+\frac{3\dot{\bm{\theta}}_t-\dot{\bm{\theta}}_{t-\delta t}}{2}\cdot\delta t+\mathcal{O}\left(\delta t^3\right)
\end{equation}

We find that the use of SOM increases the speed of marching allowing to reach the same accuracy as with FOM, using larger-size time-steps, i.e., with smaller overall number of steps. Thus the SOM saves the number of needed steps throughout the system evolution by about an order of magnitude, with proportional saving of the computing time. It is noteworthy that SOM produces the $\bm{\theta}_{t+\delta t}$ using the historical evolution data $\dot{\bm{\theta}}$ from only two previous steps.

\subsection{\label{sec:2.2}Encoding of the Hamiltonian}
The central issue in simulation of a quantum system using universal quantum computers is to encode the system into a form accessible by a quantum circuit. The occupations of atomic orbitals which form different fermionic configurations are mapped to the corresponding qubit configurations, and similarly the fermionic state operators are mapped to qubit state operators. To start the encoding, it is convenient to express the single electronic states in the Fock population basis \cite{fock32}. For a system with 1 electron and $N$ atomic orbitals, there is a total of $N$ fermionic configurations. Each fermionic configuration can be defined as a vector $\vert \bm{f} \rangle=\vert x_{N-1},\ldots,x_0 \rangle$, where $x_k\in\{0,1\}$ representing the occupation of electron (0 for vacant and 1 for occupied) on the $k$th orbital defined by the basis function $\vert \varphi_k\rangle$, where orbital indexes $k$ are sorted in a descending order of the orbital energy. Hence, we can obtain the configuration set $\{\vert\bm{f}_0 \rangle,\ldots,\vert\bm{f}_{N-1} \rangle\}$, where each $\vert\bm{f}_k \rangle$ is aligned in an ascending order of $k$, indicating that the electron is present in the $k^{th}$ atomic orbital. Then the fermionic configuration set is ready to be mapped to qubit configuration set $\{\vert \bm{q}_0 \rangle,\ldots,\vert \bm{q}_{N-1} \rangle\}$ where $\vert \bm{q}_k \rangle$ is also sorted in an ascending order.

The fermionic Hamiltonian needs to be mapped to qubit Hamiltonian. By using the secondary quantization, one can obtain the single-electron Hamiltonian \cite{whitfield11}:
\begin{equation}
\label{eq16}
H=\sum^{N-1}_{i,j}h_{ij} a^\dagger_i a_j
\end{equation}
where $h_{ij}$ are the one-electron integrals for the chosen basis set $\vert \varphi \rangle$ defined as:
\begin{equation}
\label{eq17}
h_{ij}=\langle\varphi_i\vert H\left(\bm{r},t\right)\vert \varphi_j\rangle=\langle\varphi_i\vert H_0\left(\bm{r}\right)+P\left(\bm{r},t\right)\vert \varphi_j\rangle.
\end{equation}
$a^\dagger_k$ and $a_k$ are electron creation and annihilation operators, respectively, acting on fermionic states, defined as:

\begin{subequations}
\label{eq18}
\begin{eqnarray}
a^\dagger_k\vert \ldots&&,x_k,\ldots\rangle= \nonumber\\
&& \left(1-x_k\right)\left(-1\right)^{\sum^{k-1}_{i=0}x_i}\vert\ldots,x_k+1,\ldots\rangle \label{eq18a}\\
a_k\vert \ldots&&,x_k,\ldots\rangle= \nonumber\\
&& x_k\left(-1\right)^{\sum^{k-1}_{i=0}x_i}
\vert\ldots,x_k-1,\ldots\rangle \label{eq18b}
\end{eqnarray}
\end{subequations}

\subsubsection{\label{sec:2.2.1}Unary encoding}
Encoding methods such as JWE \cite{jordan28}, parity encoding \cite{seeley12} and Bravyi-Kitaev encoding \cite{bravyi02} are commonly used in simulation of many-body systems using quantum algorithms such as VQE \cite{cao19,peruzzo14,kandala17,tranter18,bravyi17}, quantum simulation via trotterization \cite{fauseweh21,tranter19}, variational fast forwarding for quantum simulation \cite{cirstoiu20} and McLachlan VHQCA \cite{mcardle19ansatz,jones19}. The common feature of the listed methods is that the occupation of the $k^{th}$ atomic basis state is directly mapped to the state of the $k^{th}$ qubit, resulting in mapping of $N$ atomic states to $N$ qubits, known as the unary encoding method \cite{sawaya20}. In the unary encoding, the $k^{th}$ qubit configuration $\vert \bm{q}_k \rangle$ is one of the computational basis of $N$ qubits $\vert \bm{q}_k\rangle= \vert y_{N-1}=0,\ldots,y_k=1,\ldots,y_0=0\rangle$, where $y_k\in\{0,1\}$ is a basis state of an individual qubit. An example of encoding of a 4-state system using 4 qubits is shown in Table~\ref{table1}.

In this work, we apply the JWE as an example of unary encoding. According to this transformation, the operators $a^\dagger$ and $a$ are mapped to the qubit raising and lowering operators $\sigma^\dagger$ and $\sigma$:
\begin{subequations}
\label{eq19}
\begin{eqnarray}
a^\dagger_k = Z\otimes\ldots\otimes Z\otimes\sigma^\dagger_k\otimes I\otimes\ldots\otimes I \label{eq19a}\\
a_k = Z\otimes\ldots\otimes Z\otimes\sigma_k\otimes I\otimes\ldots\otimes I \label{eq19b}
\end{eqnarray}
\end{subequations}
where
\begin{subequations}
\label{eq20}
\begin{eqnarray}
\sigma^\dagger&&= \begin{bmatrix}
  0 & 0\\ 
  1 & 0
\end{bmatrix}=\frac{1}{2}\left(X-iY\right)\label{eq20a}\\
\sigma&&= \begin{bmatrix}
  0 & 1\\ 
  0 & 0
\end{bmatrix}=\frac{1}{2}\left(X+iY\right)\label{eq20b}
\end{eqnarray}
\end{subequations}

$I,X,Y,Z$ are Pauli matrices, and index $k$ denotes the operation on the $k^{th}$ atomic state, encoded by the qubits. The qubit Hamiltonian obtained from the JWE using 4 qubits for the 4 mutually coupled states ($1s, 2p, 3s, 3d$) of the hydrogen atom-laser system is shown in Eq. S15 of SM. The derivation for a general undetermined 4-state Hamiltonian encoded by JWE is provided in Eq. S16 of SM.  

Note that the unary encoding for a $N$-state system is defined in an $2^N$-dimensional Hilbert space of $N$ qubits and therefore it is not using the full power of quantum advantage. For example, in Table~\ref{table1}, the 4 qubits Hilbert space contains a total of 16 basis states forming the computational basis but only 4 basis states are utilized, while the other 12 computational basis states are left unused.

\begin{figure*}
\includegraphics{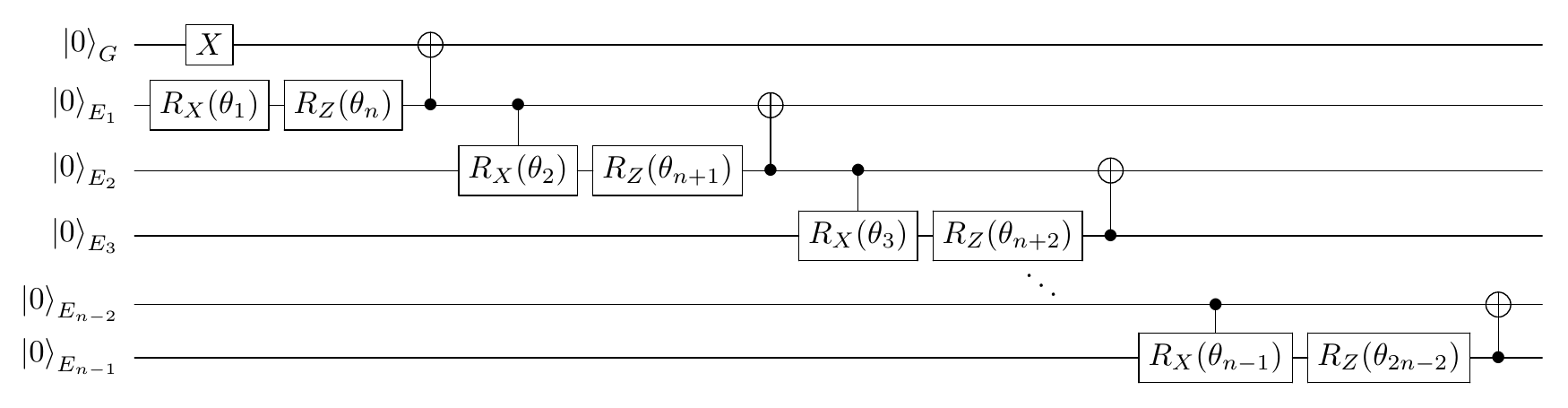}
\caption{\label{fig:4} Variational ansatz for JWE with $N$ qubits. $G$ represents the ground state and $E_i$ represents the $i^{th}$ excited state}
\end{figure*}

\subsubsection{\label{sec:2.2.2}Compact encoding}

To make advantage of all states in the computational basis and reduce the number of qubits to represent the ansatz, we utilize the compact, QEE method. With the compact mapping, a $N$-state system is encoded using $N_q=\log_2N$ qubits. For example, for a 4-state system, only 2 qubits are needed to describe the system, as shown in Table~\ref{table1}. 

\begin{table}
\caption{\label{table1}Examples of encoding 4-state system using JWE and QEE methods.}
\begin{ruledtabular}
\begin{tabular}{ccc}
Fermionic configuration & \multicolumn{2}{c}{Qubit configuration}\\
& JWE & QEE\\
$\vert\bm{f}\rangle=\vert x_3,x_2,x_1,x_0\rangle$ & $\vert\bm{q}\rangle=\vert y_3,y_2,y_1,y_0\rangle$&$\vert\bm{q}\rangle=\vert y_1,y_0\rangle$\\
\colrule
0001 & 0001 & 00\\
0010 & 0010 & 01\\
0100 & 0100 & 10\\
1000 & 1000 & 11\\
\end{tabular}
\end{ruledtabular}
\end{table}

To obtain the qubit Hamiltonian for QEE of the hydrogen laser-atom system, one can start from the secondary quantized Hamiltonian in Eq.~\ref{eq16}, and rewrite the excitation operators as \cite{shee21}: 
\begin{equation}
\label{eq21}
a^\dagger_i a_j=\vert\bm{f}_i \rangle  \langle\bm{f}_j\vert
\end{equation}
where the fermionic excitation operators $\vert\bm{f}_i \rangle  \langle\bm{f}_j \vert$ enable the electron transition from state $\vert\bm{f}_j \rangle$ to state $\vert\bm{f}_i \rangle$: $(\vert\bm{f}_i \rangle  \langle\bm{f}_j \vert)\vert\bm{f}_j \rangle=\vert\bm{f}_i \rangle$. Since the $i^{th}$ and $j^{th}$ fermionic configurations are mapped to corresponding qubit configurations \cite{sawaya20,mcardle19digital}, the qubit Hamiltonian can be written in form:
\begin{equation}
\label{eq22}
H=\sum^{N-1}_{i,j}h_{ij} \vert \bm{q}_i \rangle  \langle\bm{q}_j \vert
\end{equation}
where
\begin{equation}
\label{eq23}
\vert \bm{q}\rangle= \vert y_{N_q-1},\ldots,y_0\rangle
\end{equation}

The qubit excitation operator $\vert \bm{q}_i \rangle  \langle\bm{q}_j \vert$ can be further factorized in individual qubit as $\otimes\prod_{k=0}^{N_q-1}\vert y^i_k \rangle  \langle y^j_k \vert$ where $k$ is the qubit index, and then replaced by Pauli operations to obtain a full qubit Hamiltonian \cite{sawaya20,mcardle19digital}: 
\begin{eqnarray}
\label{eq24}
\vert 0 \rangle \langle 0\vert&&=\frac{1}{2}\left(I+Z\right),\quad \vert 0 \rangle \langle 1\vert=\frac{1}{2}\left(X+iY\right)  \nonumber\\
\vert 1 \rangle \langle 1\vert&&=\frac{1}{2}\left(I-Z\right),\quad \vert 1 \rangle \langle 0\vert=\frac{1}{2}\left(X-iY\right)
\end{eqnarray}

The two-qubit Hamiltonian encoded by QEE for the same 4 mutually coupled states ($1s$, $2p$, $3s$, $3d$) system is shown in Eq. S17 of SM. The derivation for a general 4-state Hamiltonian encoded by QEE is provided in Eq. S18 of SM.  

Comparing with the JWE, the qubit Hamiltonian for the 4-state system by QEE requires only $\log_24=2$ qubits, reducing the dimension of the Hilbert space from 16 (JWE) to 4(QEE). One can expect that for a larger size problem QEE will save quantum resources dramatically. For example, in the 1024-state system, the unary JWE requires a total of 1024 qubits, while QEE requires only $\log_21024=10$ qubits. It is noteworthy that QEE is also applicable to many-body systems to reduce the number of needed qubits. In a many-body system with $m$ electrons and $N$ spin-orbitals, there are $\begin{pmatrix} N \\ m \end{pmatrix}$ possible electronic configurations. A unary encoding scheme would require N qubits to simulate the system evolution, while with QEE one needs only  $\bigg \lceil \log_2{\begin{pmatrix} N \\ m \end{pmatrix}}\bigg \rceil$ qubits \cite{shee21}. For example, in case of $N=1,000,000$ states and two electrons, 39 qubits would be enough to describe all $\sim 5\times10^{11}$ configurations.

\subsection{\label{sec:2.3}Quantum variational ansatz}

In McLachlan VHQCA, the quality of time-dependent simulation is tied to the ability of the variational ansatz to correctly describe the many-state time-evolved wave function, which raises the challenge for constructing a sufficiently expressible and fully entangled ansatz \cite{lau21} when using either JWE or QEE. In the quantum computing, the ansatz is prepared with a set of consecutive quantum gates to approximate the wavefunction. The general single qubit parameterized gates $R_p(\theta_k)$ are defined as single qubit rotations by an angle $\theta_k$ about $p$ axis at the Bloch sphere, defined by a Pauli operator $p$ of a set $\{X,Y,Z\}$. The two-qubit parameterized gates $CR_p(\theta_k)$ are defined as controlled rotations applying a rotation $R_p(\theta_k)$ on a target qubit upon the state of the control qubit. Thus,
\begin{subequations}
\label{eq25}
\begin{eqnarray}
R_p\left(\theta_k\right)\quad&&=\quad e^{-i\theta_k p/2} \label{eq25a}\\
CR_p\left(\theta_k\right)\quad &&=\quad\vert 0\rangle \langle 0\vert\otimes I + \vert 1\rangle \langle 1\vert\otimes R_p\left(\theta_k\right) \label{eq25b}
\end{eqnarray}
\end{subequations}

The derivative of $R_p\left(\theta_k\right)$ and $CR_p\left(\theta_k\right)$ can be efficiently decomposed in the following forms:
\begin{subequations}
\label{eq26}
\begin{eqnarray}
\frac{\partial R_p\left(\theta_k\right)}{\partial \theta_k}  \quad &&=\quad -\frac{i}{2}pR_p\left(\theta_k\right) \label{eq26a}\\
\frac{\partial CR_p\left(\theta_k\right)}{\partial \theta_k} \quad &&=\quad -\frac{i}{2} \left(\vert 1\rangle \langle 1\vert\otimes pR_p\left(\theta_k\right)\right) \label{eq26b}
\end{eqnarray}
\end{subequations}

\begin{figure*}
\includegraphics{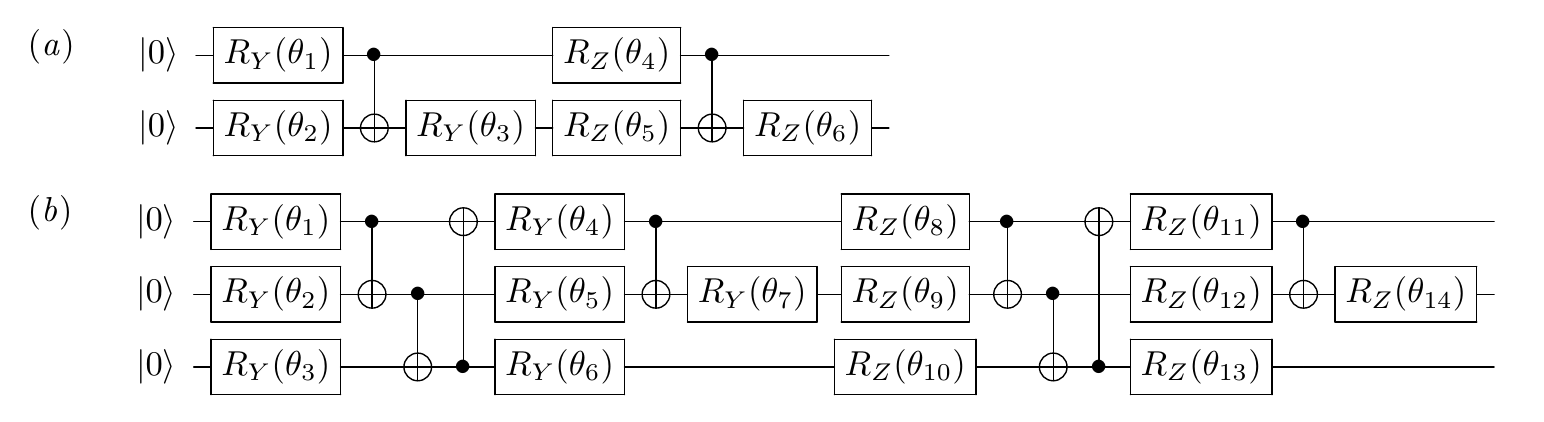}
\caption{\label{fig:5} Variational ansatz for QEE with 2 (a) and 3 (b) qubits.}
\end{figure*}

Therefore, the derivative of $\theta_k$ applied to ansatz can be obtained by replacing the corresponding parameterized gate with their derivative forms. For example, defining ansatz with three rotations applied to the initial qubit state $\vert 0\rangle$, $\tilde{\phi}(\bm{\theta})=R_Z(\theta_3)R_X(\theta_2)R_Z(\theta_1)\vert 0\rangle$, the derivative of ansatz over $\theta_2$ can be written as:
\begin{eqnarray}
\frac{\partial\tilde{\phi}\left(\bm{\theta}\right)}{\partial\theta_2}&&=R_Z\left(\theta_3\right)\frac{\partial R_X\left(\theta_2\right)}{\partial\theta_2}R_Z\left(\theta_1\right)\vert 0\rangle \nonumber\\
&&=-\frac{i}{2}R_Z\left(\theta_3\right)X R_X\left(\theta_2\right)R_Z\left(\theta_1\right)\vert 0\rangle \label{eq27}
\end{eqnarray}

It is noteworthy that the derivative of a $CR_p(\theta_k)$ over $\theta_k$ is not a unitary matrix due to the presence of $0$'s along the main diagonal. Thus, to apply the derivatives in a quantum circuit, one needs to decompose them to a linear combination of unitary operators. For example, the derivative of $CR_X$ is expressed as $\frac{\partial CR_X(\theta_k)}{\partial \theta_k}=-\frac{i}{4}(I\otimes(X\cdot CR_X(\theta_k)))+\frac{i}{4}(Z\otimes(X\cdot CR_X(\theta_k)))$. Hence, the derivative of ansatz over the parameter of a controlled rotation gate is treated as a linear combination of two circuits.

\subsubsection{\label{sec:2.3.1}Ansatz for unary encoding}
We construct the variational ansatz for unary encoding (JWE in this work), by an iterative layered structure, using the parameterized single-qubit and two-qubits controlled rotations around $X$ and $Z$-axis, as well as CNOT and $X$ gates. The circuit for a general $N$-state ansatz is shown in Fig.~\ref{fig:4}, which has symbolic form shown in Eq.~\ref{eq28}. The combination of $R_X$ and $R_Z$ gates at each qubit ensures freedom of variation in both phases and amplitudes for each qubit basis state, also essential in enabling GPC to increase accuracy of the results.

Thus the formula for the ansatz in Fig.~\ref{fig:4} has the form:
\begin{widetext}
\begin{eqnarray}
\label{eq28}
&&\Big\vert\phi\left(\theta_1, \ldots,\theta_{2N-2}\right)\Big\rangle
=c_0 e^{-\frac{i}{2}\sum\limits^{2N-2}_{x=N}\theta_x}\cos\left(\frac{\theta_{N-1}}{2}\right)\vert\bm{q}_0\rangle \nonumber\\
&&+\sum^{N-1}_{k-1}\Bigg[c_k e^{\frac{i}{2}\left(-\sum\limits^{2N-2-K}_{x=N}\theta_x+\sum\limits^{2N-2}_{x=2N-1-K}\theta_x\right)}\Bigg(\prod^{n-1}_{x=N-k}\sin\left(\frac{\theta_x}{2}\right)\Bigg)\cos\left(\frac{\theta_{N-1-k}}{2}\right)\vert\bm{q}_k\rangle\Bigg]
\end{eqnarray}
\end{widetext}
where $N>2,\theta_0=0$, $c_k=(-1)^{\frac{k}{2}}$ for even $k$, and $c_k=(-1)^{\frac{k+1}{2}}i$ for odd $k$. From Eq.~\ref{eq28}, one can see that the $R_X$ and $CR_X$ rotations in the ansatz contribute to amplitude, while the $R_Z$ rotations update phase of each basis state. The $N$-qubits problem representing $N$ atomic states requires a total of $2(N-1)$ parameters and $(3N-2)$ gates, specifically $(N-2)$ Control-$R_X$ gates, $(N-1)$ CNOT gates, $(N-1)$ $R_Z$ gates, one $R_X$ gate and one $X$ gate, as listed in Table~\ref{table2}.

\begin{table}[b]
\caption{\label{table2}Comparison of two encoding methods for a $N$-state system.}
\begin{ruledtabular}
\begin{tabular}{ccc}
Encoding method & JWE & QEE\\
\hline
Number of qubits    &  $N$   &   $\log_2N$ \\[5pt]
Total gates         & $3N-2$ & $2(2N-2-\log_2N)$\\[5pt]
\begin{tabular}{@{}c@{}} Total variational\\parameters \end{tabular} & $2(N-1)$ & $2(N-1)$ \\[10pt]
\begin{tabular}{@{}c@{}} Number of single\\parameterized gates  \end{tabular} & $N$ & $2(N-1)$ \\[10pt]
\begin{tabular}{@{}c@{}} Number of controlled\\parameterized gates  \end{tabular} & $N-2$ & $0$ \\[10pt]
\begin{tabular}{@{}c@{}} Number of\\CNOT gates \end{tabular}  & $N-1$ & $2(2N-1-\log_2N)$\\
\end{tabular}
\end{ruledtabular}
\end{table}

\subsubsection{\label{sec:2.3.2}Ansatz for compact encoding}
The QEE, as a compact encoding method, uses the computational basis in full, utilizing a streamlined hardware-efficient variational ansatz to ensure the coverage of the full Hilbert space \cite{di21}. For $N$ atomic states represented by $\log_2N$ qubits, the same number of parameters as in corresponding unary cases are used to ensure the same “freedom” of variations for each state. For a general $N$-state case, there are $2(N-1)$ single parameterized rotations and $2(N-1-\log_2N)$ CNOT gates \cite{di21}. Similarly, the combination of $R_Y$ and $R_Z$ gates provides an approximation for each basis state phase and amplitude, respectively. The circuits for 2 and 3 qubits are shown in Fig.~\ref{fig:5}. The symbolic form of the 2-qubit ansatz takes the form:
\begin{eqnarray}
\label{eq29}
&&\Big\vert\phi\left(\bm{\theta}\right)\Big\rangle \nonumber\\
=&&e^{\frac{i}{2}\left(-\theta_4-\theta_5-\theta_6\right)}\cos\left(\frac{\theta_1}{2}\right)\cos\left(\frac{\theta_2+\theta_3}{2}\right)\vert 00\rangle \nonumber\\
+&&e^{\frac{i}{2}\left(-\theta_4+\theta_5+\theta_6\right)}\cos\left(\frac{\theta_1}{2}\right)\sin\left(\frac{\theta_2+\theta_3}{2}\right)\vert 01\rangle \nonumber\\
+&&e^{\frac{i}{2}\left(\theta_4+\theta_5-\theta_6\right)}\sin\left(\frac{\theta_1}{2}\right)\cos\left(\frac{\theta_2-\theta_3}{2}\right)\vert 10\rangle\nonumber\\ +&&e^{\frac{i}{2}\left(\theta_4-\theta_5+\theta_6\right)}\sin\left(\frac{\theta_1}{2}\right)\sin\left(\frac{\theta_2-\theta_3}{2}\right)\vert 11\rangle 
\end{eqnarray}
The symbolic forms of ansatz for more than 2 qubits become too messy to be written here.
\begin{table*}
\centering
\caption{\label{table3}The definition of unitary operations $U_0$, $U_A$, $U_h$ and $U_B$ in the Hadamard test.}
\begin{ruledtabular}
\begin{tabular}{ccccc}
& $U_0\vert 0\rangle$\footnotemark[1]
& $U_A\vert 0\rangle$\footnotemark[1]
& $U_B^\dagger\vert 0\rangle$\footnotemark[1]
& $U_h$ \\
\hline
$A^R_{i,j}$ & $\vert 0\rangle$ & $\frac{\partial \vert \phi(\bm{\theta}(t))\rangle }{\partial \theta_j}$ & $\frac{\partial \vert \phi(\bm{\theta}(t))\rangle }{\partial \theta_i}$ & $I$ \\
$\frac{\partial \langle \phi(\bm{\theta}(t))\vert }{\partial \theta_i}\vert\phi(\bm{\theta}(t))\rangle$ & $\vert 0\rangle$ & $ \vert \phi(\bm{\theta}(t))\rangle$ & $\frac{\partial \vert \phi(\bm{\theta}(t))\rangle }{\partial \theta_i}$ & $I$ \\
$C^I_i$ & $\vert 0\rangle$ & $ \vert \phi(\bm{\theta}(t))\rangle$ & $\frac{\partial \vert \phi(\bm{\theta}(t))\rangle }{\partial \theta_i}$ & Pauli terms \\
$\langle \phi(\bm{\theta}(t))\vert H \vert \phi(\bm{\theta}(t))\rangle$ & $ \vert \phi(\bm{\theta}(t))\rangle$ & $\vert 0\rangle$ & $\vert 0\rangle$ & Pauli terms \\
\end{tabular}
\end{ruledtabular}
\footnotetext[1]{$U\vert 0\rangle=\vert 0\rangle$ means the unitary operation $U$ is identity operation $I$.}
\end{table*}
The JWE and QEE for the $N$-state problem have similar total number of parameterized gates in the ansatz computation. However, nearly a half of parameterized gates in JWE ansatz are two-qubit $CR_X$ gates, while in the QEE ansatz all parameterized rotation gates are the single-qubit operations. It is noteworthy that the controlled parameterized rotation $CR_X$ in the unary encoding cannot directly be executed by the quantum computers. Instead, it has to be decomposed to a set of one and two-qubit gates which are considered as elementary basis gate and can be operated in the quantum computers. The decomposition of $CR_X$ gates generates a much deeper circuits than single-qubit $R_Y$ gates in QEE ansatz. An example of the gate decomposition for the real IBM device backend “\texttt{ibmq\_jakarta}” is shown in Fig. S4 of the SM. After decomposition, the depth of a $CR_X$ circuit is more than twice as big as that of a decomposed $R_Y$ circuit. Hence the total number of 2-qubit gates and single qubits gates actually executed on quantum computers is significantly smaller with the QEE ansatz than with JWE ansatz. 

Besides, using QEE the quantum devices are capable for mapping exponentially larger size of systems than with JWE. The reduction in number of qubits by QEE is also important for NISQ implementations since it reduces the demand for larger scale quantum computers which might be exposed to undesirable external disturbances and unwanted couplings between signal paths (“crosstalk”, \cite{sarovar20,mundada19}). Saving both width (number of qubits) and depth of quantum circuits leads to the smaller noise effects. However, the downside of the QEE is its complexity to generalize ansatz in a symbolic form for large N, which downgrades the simulation speed.

\begin{figure}
\includegraphics{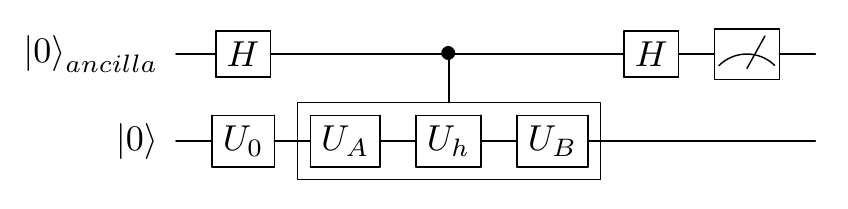}
\caption{\label{fig:6} General quantum circuits for the evaluating $A^R_{i,j}$, $\frac{\partial \langle \phi(\bm{\theta}(t))\vert }{\partial \theta_i}\vert\phi(\bm{\theta}(t))\rangle$, $C^I_i$ and $\langle \phi(\bm{\theta}(t))\vert H \vert \phi(\bm{\theta}(t))\rangle$}
\end{figure} 

\subsection{\label{sec:2.4}Hadamard test circuits}
To obtain expectation values of all terms in Eqs.~\ref{eq13}, we use the Hadamard test with the circuits shown in Fig.~\ref{fig:6}, where only an ancilla qubit is measured. The Hadamard test shown here describes a general circuit for measurement of the real part of expectation value of an operator $U$, $\Re\langle\phi(\bm{\theta}(t))\vert U\vert\phi(\bm{\theta}(t))\rangle$. The expectation value is calculated as the difference between probabilities measuring state $\vert0\rangle$ and $\vert1\rangle$. Unitary operations $U_0$, $U_A$, $U_h$ and $U_B$ in Fig.~\ref{fig:6} are defined by various measured quantities, listed in Table~\ref{table3}. $U_0$ represents the unitary operations for preparing an ansatz. $U_A$ and $U_B$ are the circuits for calculations of various terms in $\bm{M}$ and $\bm{V}$. $U_h$ can be replaced by Pauli gate in various qubit Hamiltonian terms, while the matrix elements $h_{ij}$ of $H$ are multiplied with measured results for the final expectation values. It is noteworthy that a derivative of ansatz gives an imaginary coefficient $-\frac{i}{2}$, as shown in Eqs.~\ref{eq26}, which is extracted to be postprocessed with measurement results. Hence the evaluation of $\frac{\partial \langle \phi(\bm{\theta}(t))\vert }{\partial \theta_i}\vert\phi(\bm{\theta}(t))\rangle$ and $C^I_i$ by the circuits in Fig.~\ref{fig:6} is equivalent to the measurement of the real parts of $-i\frac{\partial \langle \phi(\bm{\theta}(t))\vert }{\partial \theta_i} \vert\phi(\bm{\theta}(t))\rangle$ and $-i\frac{\partial \langle \phi(\bm{\theta}(t))\vert }{\partial \theta_i} H\vert\phi(\bm{\theta}(t))\rangle$.

\section{\label{sec:3}Results\protect}
\subsection{\label{sec:3.1}HCP laser field with $\omega$=0.06}
The evolution from $t=0$ to $t=200$ a.u. of a hydrogen atom, modeled by a finite number of bound states and irradiated by a pulse of a linearly polarized HCP laser field involves 2000 to 200,000 marching steps, depending on the choice of a step size ($\Delta t=10^{-1}$-$10^{-3}$). To obtain the results quickly and efficiently, the symbolic simulation is applied to all tests in this section. Ansatz is a function of $\bm{\theta}$ (e.g., Eqs.~\ref{eq28} and \ref{eq29}), from which the formulas of derivative of ansatz are obtained. The qubit Hamiltonian is a matrix function of $t$. In each step, $\bm{\theta}$ and $t$ are loaded to update the numerical Hamiltonian matrix and ansatz (or the derivative of ansatz) vector, which are used to calculate the expectation values of the terms of $\bm{M}$ and $\bm{V}$ in Eqs.~\ref{eq13}, enabling calculation of $\dot{\bm{\theta}}$ and then marching by the next step. The symbolic simulation is carried out using Python Sympy and Numpy packages. The symbolic simulation with JWE can be conveniently done for arbitrary $N$ having the general expression for ansatz (Eq.~\ref{eq28}). However, in the QEE approach the symbolic calculation becomes difficult when $N_q>3$ since the expressions for ansatz become formidable. 

In the experiments in this section, only ground state is fully populated at the beginning. The GPC (Eq.~\ref{eq12}) and SOM (Eq.~\ref{eq15}) are applied in all computations to achieve the best accuracy. For comparison, two sets of tests are conducted to investigate the performance of algorithms with different techniques: a) without GPC (Eq.~\ref{eq9}) and with SOM and b) with GPC and FOM (Eq.~\ref{eq14}). The results of these two tests are listed in Sec. SVIII of SM. Note that the common quantum gate-based simulator is not used here due to a need for large computation power and time needed for simulation of the system evolution. However, we partially provided the simulation results with QEE from gate-based simulator, Pennylane \cite{pennylane}, in Sec. SIX of SM to prove that the obtained transition probabilities are at the same level of accuracy as the symbolic simulation results. 

\subsubsection{\label{sec:3.1.1}Results using JWE}

\begin{table}[b]
\caption{\label{table4}The relative deviations (in \%) of final transition probabilities $P(T)$ from symbolic simulation comparing to the benchmark for the 2,4,8,16-state systems using JWE with SOM and GPC. The probabilities for staying in the ground state and for transitions to the excited states are listed bottom-up.}
\begin{ruledtabular}
\begin{tabular}{ccccc}
$N$ & Orbitals & $\Delta t=10^{-1}$& $\Delta t=10^{-2}$ & $\Delta t=10^{-3}$\\
\hline
\multirow{2}{*}{2 states} & $2p$& \textrm{ $ $ 6.99e$-$1}&\textrm{ $ $ 6.71e$-$2}&\textrm{$-$3.48e$-$2}\\
                          & $1s$& \textrm{$-$3.10e$-$4}&\textrm{$-$3.00e$-$5}&\textrm{ $ $ 2.00e$-$5}\\[5pt]
\multirow{4}{*}{4 states} & $3d$& \textrm{$-$3.10e$-$1}&\textrm{$-$7.07e$-$2}&\textrm{$-$7.80e$-$2}\\
                          & $3s$& \textrm{$-$1.44e$-$1}&\textrm{$-$4.43e$-$2}&\textrm{$-$5.57e$-$2}\\
                          & $2p$& \textrm{ $ $ 1.64e$+$0}&\textrm{$-$2.68e$-$2}&\textrm{$-$6.73e$-$2}\\
                          & $1s$& \textrm{$-$2.15e$-$2}&\textrm{ $ $ 1.59e$-$3}&\textrm{ $ $ 2.37e$-$3}\\[5pt]
\multirow{8}{*}{8 states} & $4p$& \textrm{ $ $ 3.64e$+$3}&\textrm{ $ $ 8.96e$-$1}&\textrm{$-$1.29e$-$1}\\
                          & $4s$& \textrm{ $ $ 9.58e$+$4}&\textrm{$-$4.05e$+$0}&\textrm{ $ $ 2.19e$-$1}\\
                          & $3d$& \textrm{$-$9.43e$+$1}&\textrm{ $ $ 1.98e$-$1}&\textrm{ $ $ 4.68e$-$3}\\
                          & $3p$& \textrm{ $ $ 2.30e$+$2}&\textrm{ $ $ 6.37e$-$2}&\textrm{ $ $ 2.17e$-$2}\\
                          & $3s$& \textrm{ $ $ 3.09e$+$2}&\textrm{$-$1.82e$-$1}&\textrm{$-$3.26e$-$2}\\
                          & $2p$& \textrm{$-$3.28e$+$1}&\textrm{$-$5.54e$-$3}&\textrm{$-$1.56e$-$2}\\
                          & $2s$& \textrm{$-$5.38e$+$1}&\textrm{$-$4.77e$-$3}&\textrm{$-$5.74e$-$3}\\
                          & $1s$& \textrm{$-$4.62e$+$1}&\textrm{ $ $ 1.68e$-$2}&\textrm{ $ $ 1.37e$-$2}\\[5pt]
\multirow{16}{*}{16 states} & $6s$& \textrm{ $ $ 3.76e$+$4}&\textrm{ $ $ 1.48e$+$1}&\textrm{ $ $ 1.93e$-$1}\\
                            & $5g$& \textrm{ $ $ 1.08e$+$4}&\textrm{ $ $ 4.90e$+$1}&\textrm{ $ $ 8.90e$-$2}\\
                            & $5f$& \textrm{ $ $ 8.83e$+$3}&\textrm{$-$1.38e$+$1}&\textrm{$-$2.07e$-$1}\\
                            & $5d$& \textrm{ $ $ 8.11e$+$4}&\textrm{ $ $ 5.59e$+$0}&\textrm{ $ $ 3.31e$-$1}\\
                            & $5p$& \textrm{ $ $ 2.55e$+$3}&\textrm{ $ $ 2.25e$+$1}&\textrm{ $ $ 5.01e$-$1}\\
                            & $5s$& \textrm{ $ $ 4.94e$+$3}&\textrm{ $ $ 1.13e$+$1}&\textrm{$-$2.88e$-$1}\\
                            & $4f$& \textrm{ $ $ 1.20e$+$3}&\textrm{ $ $ 3.90e$+$0}&\textrm{ $ $ 6.34e$-$3}\\
                            & $4d$& \textrm{$-$3.44e$+$0}&\textrm{ $ $ 2.13e$+$0}&\textrm{$-$2.04e$-$1}\\
                            & $4p$& \textrm{ $ $ 1.23e$+$3}&\textrm{$-$7.27e$+$0}&\textrm{$-$6.83e$-$2}\\
                            & $4s$& \textrm{ $ $ 3.13e$+$2}&\textrm{$-$1.33e$+$0}&\textrm{$-$5.94e$-$2}\\
                            & $3d$& \textrm{ $ $ 5.43e$+$2}&\textrm{$-$2.54e$+$0}&\textrm{$-$8.42e$-$2}\\
                            & $3p$& \textrm{$-$7.76e$+$1}&\textrm{ $ $ 1.27e$+$0}&\textrm{$-$3.91e$-$2}\\
                            & $3s$& \textrm{$-$9.91e$+$1}&\textrm{$-$6.95e$-$2}&\textrm{$-$3.42e$-$2}\\
                            & $2p$& \textrm{$-$9.96e$+$1}&\textrm{$-$6.71e$-$2}&\textrm{$-$3.44e$-$2}\\
                            & $2s$& \textrm{$-$4.56e$+$1}&\textrm{$-$3.09e$-$1}&\textrm{$-$5.37e$-$2}\\
                            & $1s$& \textrm{$-$9.95e$+$1}&\textrm{ $ $ 5.51e$-$2}&\textrm{ $ $ 4.83e$-$2}\\
\end{tabular}
\end{ruledtabular}
\end{table}

\begin{figure}[b]
\includegraphics{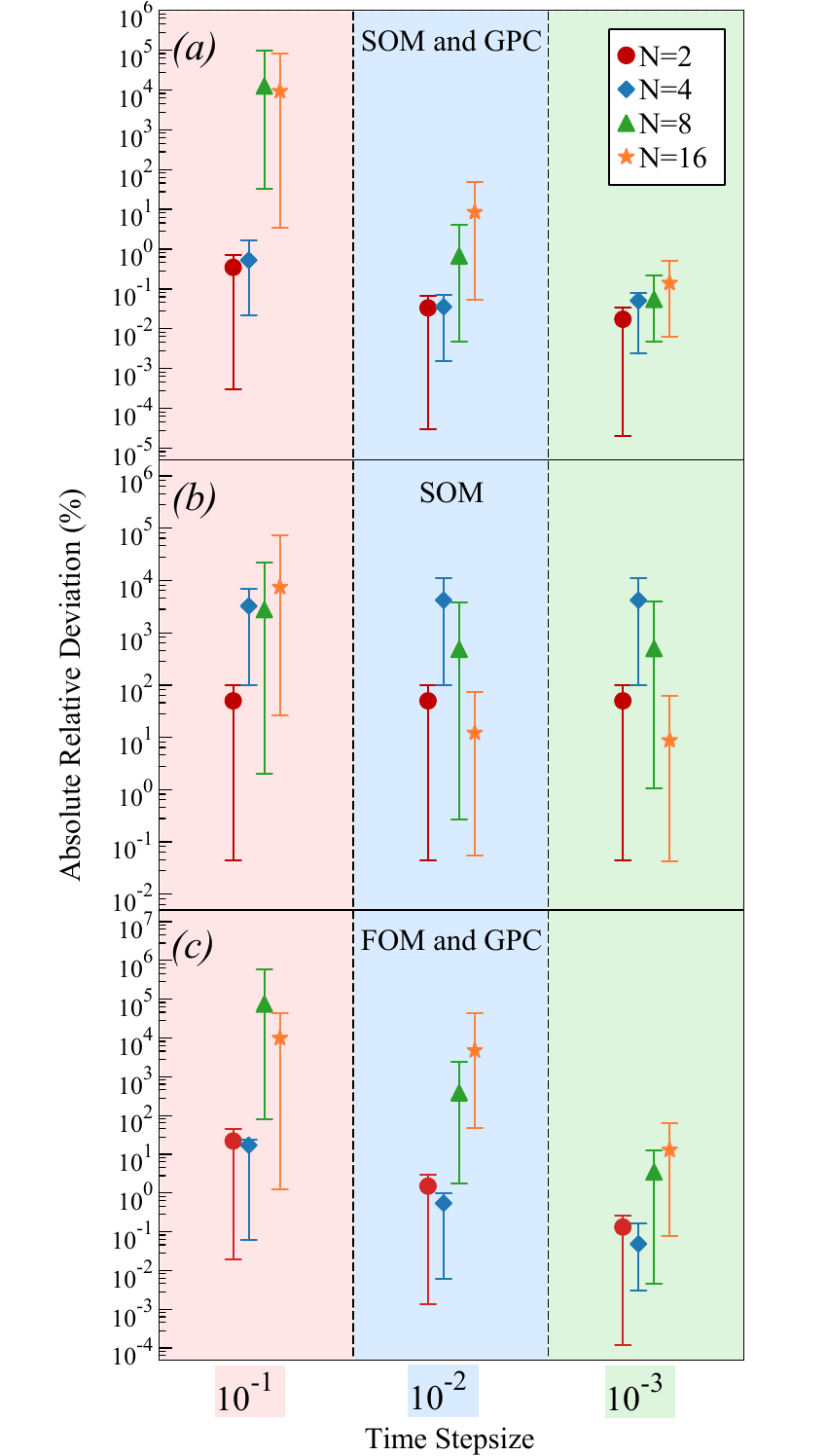}
\caption{\label{fig:7} The absolute relative deviations of transition probabilities versus different step sizes with JWE from tests (a) with SOM and GPC, (b) with SOM and no-GPC and (c) with FOM and GPC. The symbols are average absolute relative deviations of all states. Error bar provides the maximum and minimum absolute relative deviation among all states.}
\end{figure}

JWE method, explained in Secs.~\ref{sec:2.2.1} and \ref{sec:2.3.1}, is applied for the hydrogen model with $N$ = 2, 4, 8, 16 hydrogen eigenstates. The final transition probabilities $P(T)$ are recorded and compared with the benchmark results $P_B$ (listed in Table S2 of SM) for the models with various $N$. The relative deviations of $P(T)$ from the benchmark values $P_B$, defined as
\begin{equation}
\label{eq30}
\frac{P(T)-P_B}{P_B}\times 100\%
\end{equation}
are also calculated for various $N$ and different time step sizes $\Delta t$, and presented in Table~\ref{table4}. We define the absolute value of relative deviation at 1\% as the accuracy threshold, below which the results are considered accurate enough. 

With the step size of $10^{-3}$ (~0.024 a.s.), the accepted accuracy is obtained for all $N$ cases. For 2 and 4-state cases, the accuracy is achieved with step size $\Delta t=10^{-1}$. When lowering the step size to $10^{-2}$, the absolute relative deviation decreases by 5-10 times. The results stop improving with further lowering the step size, indicating that the time step size is not the only constraint for the accuracy. The results with $\Delta t=10^{-1}$ show large relative deviations from the benchmark in 8 and 16-state cases, though these are somewhat improving with $\Delta t=10^{-2}$, reaching the acceptable accuracy with $\Delta t=10^{-3}$. 

The results obtained by varying the choice of SOM or FOM, and with or without GPC are presented in Fig.~\ref{fig:7}, to quantify the improvements GPC and SOM brought to the results. The data are listed in Sec. SVIIIA of SM. From Fig. ~\ref{fig:7}b, when GPC is disabled, huge relative deviations are observed in all cases, indicating existence of a significant global phase mismatch. The relative deviations are not improving for $N$=2 and 4 when applying smaller step size. For $N$=8 and 16, improvements are obtained by reducing the step size to $10^{-2}$ but relative deviations without GPC remain at an unacceptable level even with further reduction of the step size to $10^{-3}$. The substantial relative deviations in Fig.~\ref{fig:7}b imply that the GPC is necessary for accurate results in implementation of the McLachlan in VHQCA. 

Comparing Fig.~\ref{fig:7}c with Fig.~\ref{fig:7}a, one concludes that a significant reduction of absolute relative deviation (about 5-10 times) can be found in all cases when using SOM rather than FOM. With SOM, the results for all $N$ at $\Delta t=10^{-2}$ reach the same level of accuracy as with FOM at $\Delta t=10^{-3}$. In conclusion, using SOM allows one magnitude larger step size to reach accuracy threshold than using FOM, which saves the computation time by one order of magnitude. 

\subsubsection{\label{sec:3.1.2}Results using QEE}

\begin{table}[b]
\caption{\label{table5}The relative deviations (in \%) of final transition probabilities from symbolic simulation comparing to the benchmark for the 2,4,8-state systems using QEE with SOM and GPC. The probabilities for staying in the ground state and for transitions to the excited states are listed bottom-up.}

\begin{ruledtabular}
\begin{tabular}{ccccc}
$N$ & Orbitals &  $\Delta t=10^{-1}$& $\Delta t=10^{-2}$ &$ \Delta t=10^{-3}$ \\
\colrule
\multirow{2}{*}{2 states} & $2p$& \textrm{ $ $ 1.50e$-$1}&\textrm{ $ $ 2.40e$-$2}&\textrm{ $ $ 2.61e$-$2}\\
                          & $1s$& \textrm{$-$7.00e$-$5}&\textrm{$-$1.00e$-$5}&\textrm{$-$1.00e$-$5}\\[5pt]
\multirow{4}{*}{4 states} & $3d$& \textrm{$-$2.40e$-$1}&\textrm{$-$3.20e$-$2}&\textrm{ $ $ 2.49e$-$2}\\
                          & $3s$& \textrm{$-$1.62e$-$1}&\textrm{$-$3.52e$-$2}&\textrm{$-$4.40e$-$4}\\
                          & $2p$& \textrm{ $ $ 1.64e$+$0}&\textrm{$-$5.03e$-$2}&\textrm{$-$8.33e$-$3}\\
                          & $1s$& \textrm{$-$2.27e$-$2}&\textrm{ $ $ 1.35e$-$3}&\textrm{$-$2.60e$-$4}\\[5pt]
\multirow{8}{*}{8 states} & $4p$& \textrm{ $ $ 8.69e$+$3}&\textrm{ $ $ 5.45e$-$1}&\textrm{ $ $ 5.27e$-$2}\\
                          & $4s$& \textrm{ $ $ 4.36e$+$5}&\textrm{$-$3.16e$+$0}&\textrm{ $ $ 4.47e$-$1}\\
                          & $3d$& \textrm{ $ $ 2.42e$+$2}&\textrm{ $ $ 3.14e$-$1}&\textrm{ $ $ 7.56e$-$3}\\
                          & $3p$& \textrm{$-$6.97e$+$1}&\textrm{$-$1.70e$-$2}&\textrm{$-$2.31e$-$2}\\
                          & $3s$& \textrm{ $ $ 3.46e$+$1}&\textrm{$-$3.50e$-$2}&\textrm{ $ $ 4.06e$-$3}\\
                          & $2p$& \textrm{$-$4.53e$+$1}&\textrm{$-$1.66e$-$2}&\textrm{$-$9.43e$-$3}\\
                          & $2s$& \textrm{$-$3.17e$+$1}&\textrm{$-$4.81e$-$2}&\textrm{$-$2.32e$-$2}\\
                          & $1s$& \textrm{$-$8.39e$+$1}&\textrm{ $ $ 2.28e$-$2}&\textrm{ $ $ 1.46e$-$2}\\
\end{tabular}
\end{ruledtabular}
\end{table}

\begin{figure}[b]
\includegraphics{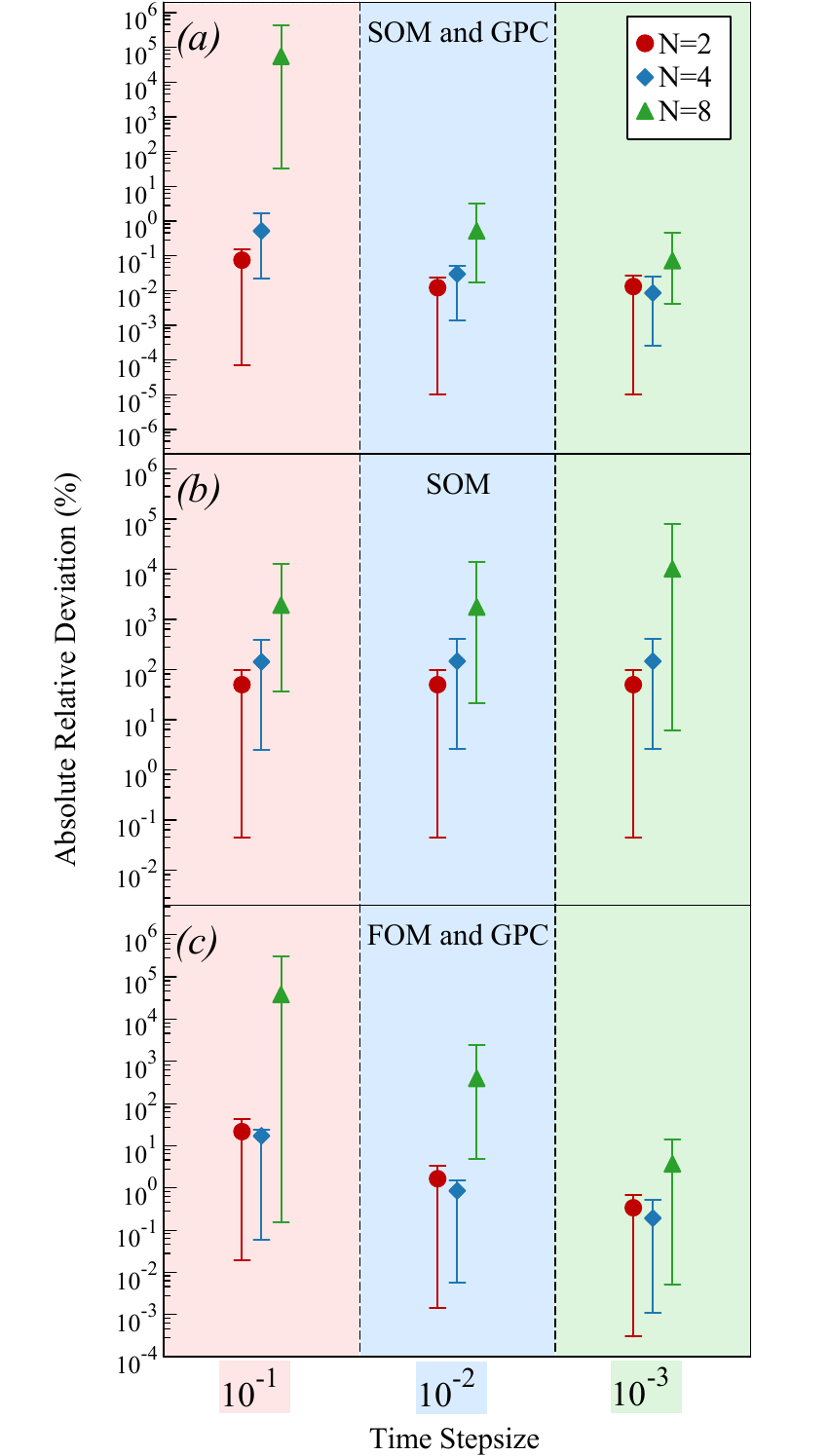}
\caption{\label{fig:8} The absolute relative deviations from the benchmarks of the transition probabilities versus different step sizes with QEE from tests (a) with SOM and GPC, (b) with SOM and no-GPC and (c) with FOM and GPC. The symbols are average absolute relative deviations of all states. Error bars provide the ranges of the absolute relative deviations about the average.}
\end{figure}

The QEE is applied to 2, 4 and 8-state systems but with a logarithmic reduction in the number of qubits, i.e., with 1, 2 and 3 qubits, respectively. The tested results using SOM with GPC are shown in Table~\ref{table5}. If the number of states is not a power of 2, one can still do the QEE by adding null states to get $N=2^{N_q}$. The null states have no interaction with any other state and are initialized to the zero population. 

Relative deviations of the transition probabilities from the benchmark values with step sizes $\Delta t=10^{-1}$ and $10^{-2}$ for all QEE cases have the similar values as the unary results in Table~\ref{table4}. Interestingly, with $\Delta t=10^{-3}$ in the 4 and 8-state systems, the relative deviations reduce up to one order of magnitudes in comparison to the ones in Table~\ref{table4}. 

The comparisons of the results when varying the choices of SOM, FOM, and GPC are shown in Fig.~\ref{fig:8} while the data are listed in Sec. SVIIIB of SM. The results are very similar to the ones with JWE (Fig.~\ref{fig:7}), indicating advantage of using SOM and GPC with QEE, too.

\begin{table*}[htb!]
\caption{\label{table6}The relative deviations (in \%) of the final transition probabilities from the benchmark for the 16-state with JWE and 4, 8-state with QEE, using SOM and GPC. The probabilities for staying in the ground state and transitions to the excited states are listed bottom-up.}
\begin{ruledtabular}
\begin{tabular}{cccccccc}
& & \multicolumn{2}{c}{JW} & & &\multicolumn{2}{c}{QEE}\\
$N$ & Orbitals & $\Delta t=10^{-2}$&$\Delta t=10^{-3}$ & $N$ & Orbitals & $\Delta t=10^{-2}$& $\Delta t=10^{-3}$ \\
\hline
\multirow{16}{*}{16 states} & $6s$& \textrm{ $ $ 5.95e$-$1}  &\textrm{ $ $ 4.98e$-$1}&                          & \\
                            & $5g$& \textrm{ $ $ 3.81e$-$1}  &\textrm{$-$2.36e$-$2}&\multirow{4}{*}{4 states} & $3d$&\textrm{ $ $ 3.61e$-$2}&\textrm{ $ $ 3.51e$-$2}\\
                            & $5f$& \textrm{ $ $ 5.55e$+$1}  &\textrm{$-$1.88e$-$1}&                          & $3s$&\textrm{$-$1.13e$-$1}&\textrm{ $ $ 3.19e$-$1}\\
                            & $5d$& \textrm{ $ $ 3.10e$+$0}  &\textrm{$-$3.99e$-$3}&                          & $2p$&\textrm{$-$1.20e$-$1}&\textrm{$-$1.21e$-$1}\\
                            & $5p$& \textrm{ $ $ 1.39e$$+$$1}&\textrm{ $ $ 3.94e$-$1}&                          & $1s$&\textrm{$-$1.43e$-$2}&\textrm{$-$2.60e$-$2}\\
                            & $5s$& \textrm{$-$1.78e$-$1}  &\textrm{$-$2.77e$-$3}&                          & \\
                            & $4f$& \textrm{$-$8.76e$+$0}  &\textrm{ $ $ 3.87e$-$2}&                          & \\
                            & $4d$& \textrm{ $ $ 2.78e$$+$$0}&\textrm{$-$2.57e$-$2}&\multirow{8}{*}{8 states} & $4p$&\textrm{$-$1.38e$$+$$1}&\textrm{$-$1.83e$-$2}\\
                            & $4p$& \textrm{ $ $ 8.36e$+$0}  &\textrm{ $ $ 7.87e$-$3}&                          & $4s$&\textrm{ $ $ 3.74e$+$0}&\textrm{$-$2.64e$-$2}\\
                            & $4s$& \textrm{ $ $ 2.84e$+$0}  &\textrm{ $ $ 2.15e$-$2}&                          & $3d$&\textrm{$-$2.09e$+$0}&\textrm{$-$1.80e$-$2}\\
                            & $3d$& \textrm{ $ $ 3.40e$-$1}  &\textrm{$-$3.22e$-$2}&                          & $3p$&\textrm{ $ $ 3.04e$+$0}&\textrm{$-$1.83e$-$2}\\
                            & $3p$& \textrm{ $ $ 3.58e$-$1}  &\textrm{ $ $ 1.35e$-$3}&                          & $3s$&\textrm{$-$7.83e$-$1}&\textrm{$-$1.08e$-$2}\\
                            & $3s$& \textrm{$-$7.25e$-$1}    &\textrm{ $ $ 4.30e$-$3}&                          & $2p$&\textrm{$-$1.27e$+$0}&\textrm{$-$4.90e$-$4}\\
                            & $2p$& \textrm{ $ $ 6.32e$-$1}  &\textrm{$-$2.02e$-$2}&                          & $2s$&\textrm{ $ $ 1.68e$+$0}&\textrm{ $ $ 2.90e$-$4}\\
                            & $2s$& \textrm{$-$1.03e$+$0}    &\textrm{ $ $ 4.00e$-$5}&                          & $1s$&\textrm{ $ $ 4.08e$+$0}&\textrm{ $ $ 8.86e$-$2}\\
                            & $1s$& \textrm{ $ $ 3.72e$-$1}  &\textrm{ $ $ 8.74e$-$2}&                          & \\
\end{tabular}
\end{ruledtabular}
\end{table*}

\subsection{\label{sec:3.2}Laser field with $\omega$=0.222}

Here we test the system interacting with short laser pulse with high frequency of $\omega$=0.222. The laser electric field in this case has a few cycles, causing different transition probabilities than HCP, as shown in Fig. S2 of SM. Three cases are tested here using symbolic simulations: 16-state system using JWE and 4, 8-state systems using QEE. The relative deviations of the results from the benchmark are presented in Table~\ref{table6}.

All results with time step size $\Delta t=10^{-3}$ show relative deviations from the benchmarks well below the threshold of 1\%. This is also a case for the 4-states system with $\Delta t=10^{-2}$. However, for 8- and 16-states, the required relative deviation threshold of 1\% is not reached for all probabilities. We note that the conclusions on the accuracy of the calculated results with respect to the time step size are similar for the two types of the strong laser field in Secs.~\ref{sec:3.1} and \ref{sec:3.2}.   

\section{\label{sec:4}Results in presence of quantum noise \protect}
\begin{figure}[b]
\includegraphics{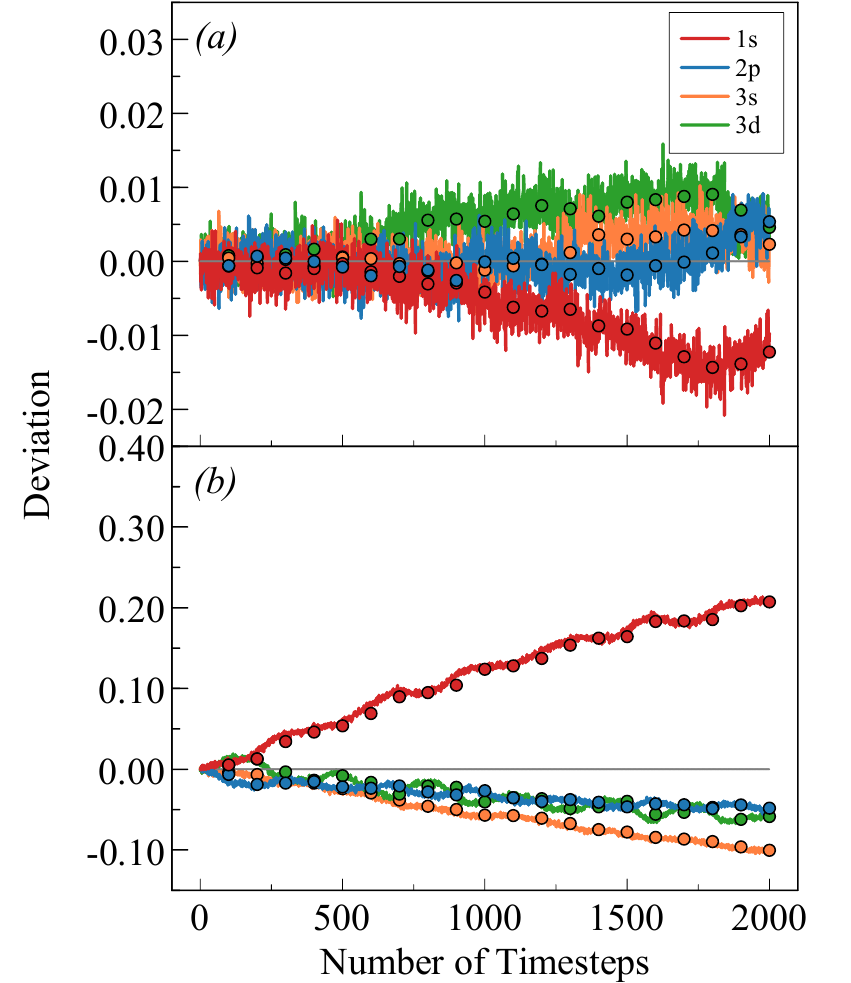}
\caption{\label{fig:9} The deviations of the state probabilities from the benchmark values with SR Hamiltonian. Only sampling is applied in (a). Both sampling and the device noise model are applied in (b). The plots show the evolution of deviations for the ground state (1s) and the three excited states (2p, 3s, 3d). The circle symbols show the averaged values over 100 time steps intervals.}
\end{figure}

\begin{figure}[b]
\includegraphics{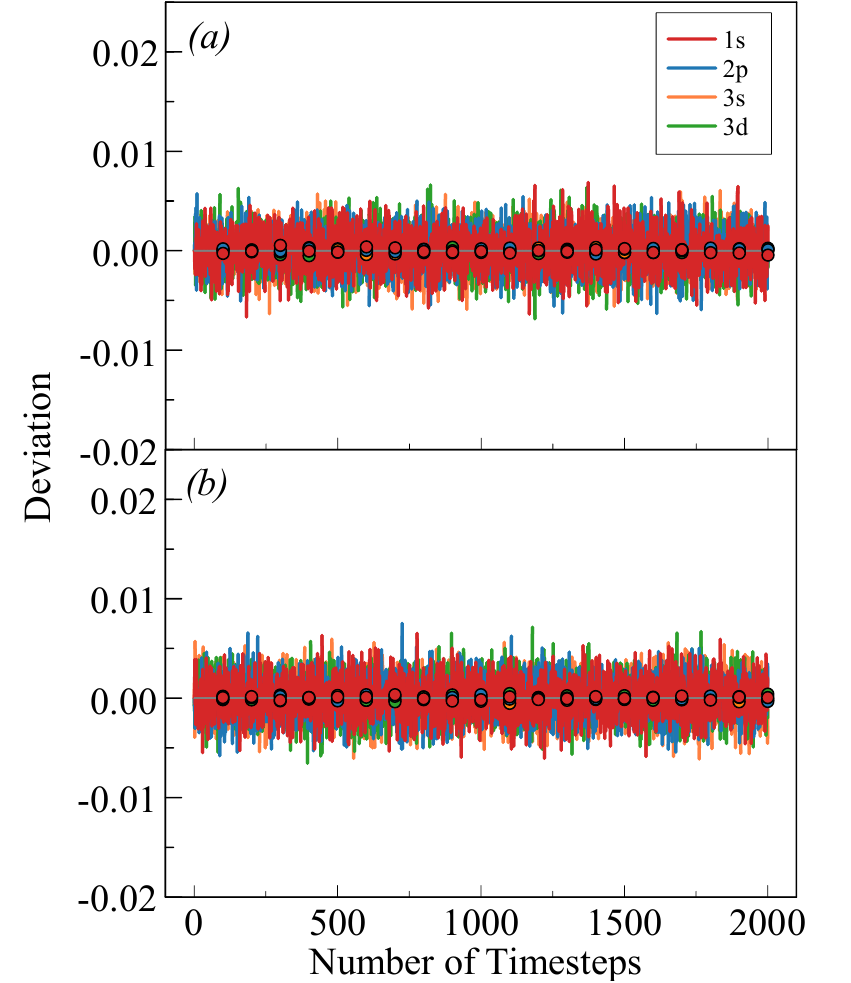}
\caption{\label{fig:10} The deviations of the state probabilities from the benchmark values with IR Hamiltonian. Only sampling is applied in (a). Both sampling and the device noise model are applied in (b). The plots show the evolution of deviations for the ground state (1s) and the three excited states (2p, 3s, 3d). The circle symbols show the averaged values over 100 time steps intervals.}
\end{figure}
Since quantum computing version of McLachlan variational algorithm is initially proposed as a solution for NISQ era \cite{li17}, it is also very important to investigate the algorithm performance in presence of quantum noise. For example, in \cite{li17} Li et al. explored the noise performance and potential error mitigation techniques of McLachlan VHQCA with an example of simulating a quantum Ising model of 3 spins. But the noise effects and accumulation when simulating a strongly time-perturbed systems with many states is missing. The evaluation of the algorithm performance due to the sampling errors and with inclusion of a hardware noise model are carried out and discussed in this section. In the tests of the noise effects, quantum circuits are constructed with the IBM Quantum Information Science Kit (QISKit, version 0.30.0) \cite{qiskit} with the quantum noise modelled from the properties of the IBM Q quantum device backends “\texttt{ibmq\_jakarta}”. This device noise model is constructed from the calibration data of quantum device containing depolarizing errors at both single and two-qubit gates, thermal relaxation errors for simulating decoherence at all gates, and single-qubit readout errors on all individual measurements. The details of calibration data are provided in the Sec. SX of SM. While the quantum noise will be reduced with the technology advancements, the sampling errors are inevitable and impossible to mitigate even with a better hardware. The collapse of the wavefunction due to quantum measurements requires a statistical approach by preparing and measuring circuits in repetitions (known as shots), which cause the sampling errors. The sampling errors can be reduced by increasing the number of shots until a desired statistical accuracy is achieved, but at the cost of using more quantum resources and longer time. The system dynamics with only sampling errors involved are also tested here as the reference. All quantum circuits in this section are measured with 50,000 shots. 

For demonstration, we chose 4-state system utilizing QEE (2 qubits), with SOM and GPC included in all tests of this section. To investigate the noise resistance of the algorithm under various representations, we applied both Hamiltonian in Schrodinger Representation (SR) (Eq.~\ref{eq22}) and in Interaction Representation (IR) \cite{sakurai20}. In IR we assume a new $\Psi(\bm{r},t)=e^{-iH_0(\bm{r})t}\psi(\bm{r},t)$, which unitarily transforms matrix elements $h_{ij}$ of the Hamiltonian (Eq.~\ref{eq22}) into:
\begin{eqnarray}
\label{eq31}
\tilde{h}_{ij}&&=\langle\Psi_i\vert P\left(\bm{r},t\right)\vert\Psi_j\rangle\nonumber\\
&&=e^{i(E_i-E_j)t} \langle\psi_i\vert P\left(\bm{r},t\right)\vert\psi_j\rangle
\end{eqnarray}
In the noise-free simulations, Schrodinger and interaction representations yield the same results.

\subsection{\label{sec:4.1}Noise and sampling error with perturbation-free Hamiltonian}
The algorithm is first tested without laser field, where all 4-state initial ($t$=0) amplitudes are set to equal values (=0.5). Since the perturbation $P(\bm{r},t)=0$, the variational parameters and the transition probabilities are expected to stay constant with time. To evaluate the noise accumulation in time, two sets of tests are conducted: 1) The system with only sampling errors due to 50,000 shots for all circuit measurements, and 2) evolution with both sampling errors and the device noise model present. A total of 2000 time steps (step size 0.1) are used in simulation and the intermediate data are recorded at each iteration.

The deviations of the state probabilities from the initial values in time are shown in Figs.~\ref{fig:9} and \ref{fig:10}. In absence of noise, small accumulated deviations are observed when using Hamiltonian in SR due to sampling (Fig.~\ref{fig:9}a). However, when using Hamiltonian in IR, small fluctuating deviations throughout the time indicates that almost no sampling errors are accumulated in this case (Fig.~\ref{fig:10}a). 

\begin{figure}[b]
\includegraphics{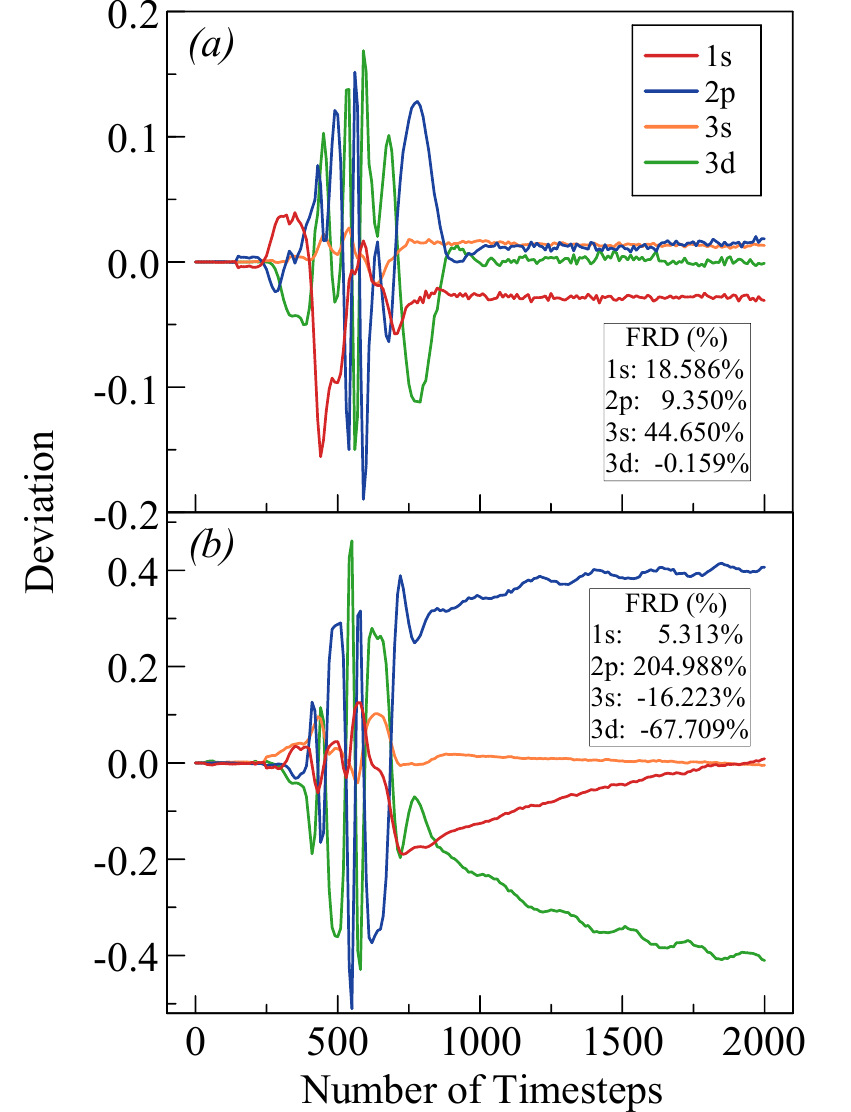}
\caption{\label{fig:11} The deviations of the state probabilities in time using the SR Hamiltonian, as explained in the text. Only sampling is applied in (a). Both sampling and the device noise model are applied in (b). The final relative deviation (FRD) at T for all states is give in the box.}
\end{figure}

\begin{figure}[b]
\includegraphics{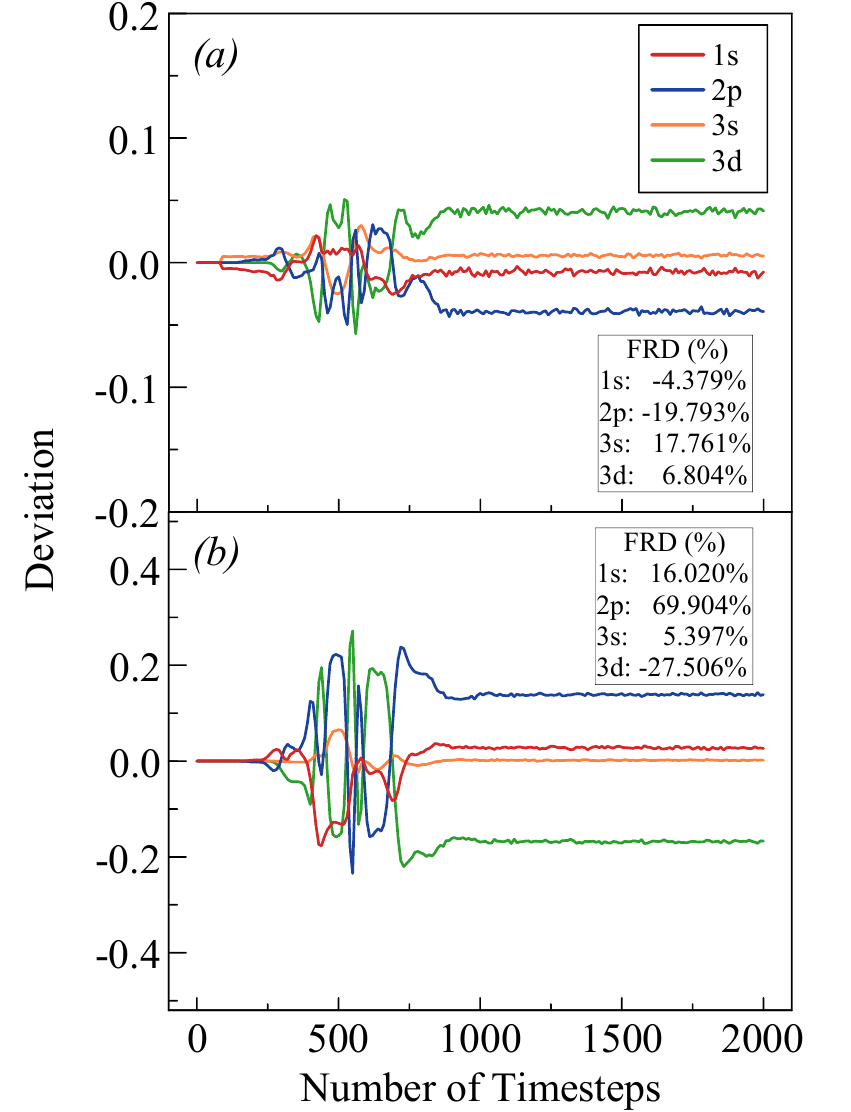}
\caption{\label{fig:12} The deviations of the state probabilities in time using the IR Hamiltonian, as explained in the text. Only sampling is applied in (a). Both sampling and the device noise model are applied in (b). The final relative deviation (FRD) at T for all states is give in the box.}
\end{figure}

When the device noise is added, a significant error accumulation is observed when using the SR of the Hamiltonian (Fig.~\ref{fig:9}b). At $T$, the maximum deviations from the initial values are more than 10 times larger than those with only sampling error in Fig.~\ref{fig:9}a. Thus, by using SR Hamiltonian, serious errors accumulate throughout the time even in absence of any external time-dependent perturbation. These are caused by both sampling and the hardware noise.

However, the results with the IR Hamiltonian show no error accumulation even with the device noise model applied (Fig.~\ref{fig:10}b). Since the $P(\bm{r},t)=0$, there is no interaction between the basis states in the system, i.e., the Hamiltonian matrix in IR is a zero matrix. Hence, the $V$ vector (Eq.~\ref{eq13b}) is always zero and $\bm{\theta}$ has no change during the whole evolution. The observed small fluctuations in the deviations are the result of sampling errors. Obviously, for quantum computing applications with the algorithms presented in this work, use of Hamiltonian in IR is a must in the NISQ era. 

\subsection{\label{sec:4.2}Noise accumulation in time in presence of the laser-atom interaction}
The impact of the quantum noise and sampling to the transition dynamics in a hydrogen atom modeled with only 4 states under the short laser pulse perturbation studied in Sec.~\ref{sec:3.2} ($\omega$=0.222) is evaluated. The noise and sampling-free transition probabilities for this case, obtained by symbolic simulation in Sec.~\ref{sec:3.2}, are used as the reference to compare with the noisy results. 

The deviations of the transition probabilities in presence of sampling only and with both device noise and sampling from the noise and sampling-free simulations are shown in Figs.~\ref{fig:11} and \ref{fig:12}, for times between $t_0$ and $T$, with time step $\Delta t=10^{-1}$. The accumulation of the noise effects appears mainly while the laser field is strong (for $t$ between 20 and 80). With SR Hamiltonian, sampling errors are in acceptable range (Fig.~\ref{fig:11}a), however the device noise errors accumulate even when the laser field is off, increasing almost linearly with time (Fig.~\ref{fig:11}b). When IR Hamiltonian is used (Fig.~\ref{fig:12}), the state probabilities, and therefore deviations stay constant as expected after $t>100$. In this case the sampling error does not deteriorate the accuracy of the final results more than a few percent (Fig.~\ref{fig:12}a). The device noise contributes a few times more to the errors (Fig.~\ref{fig:12}b). These results are in accord with the conclusions of Sec.~\ref{sec:4.1} that the IR Hamiltonian yields a noise resistance when the time-dependent perturbation is off. This brings the presented algorithms a step closer to the practical implementations for the time dependent quantum systems in the NISQ devices with improving noise characteristics.

\section{\label{sec:5}CONCLUSIONS\protect}
We propose hybrid, quantum-classical algorithm for simulation of evolution and multi-state transition dynamics of an atom, subject to a strong time-dependent perturbation at the NISQ era. This problem is still formidable to apply to a multi-electron system at a universal quantum computer. We focus to the important properties, advantages and difficulties when computing dynamics of a fully entangled, hydrogen atom model of 16 states in a strong, attosecond laser field, utilizing McLachlan variational principle. 

One challenge for the algorithm is to construct a sufficiently expressible variational ansatz, capable to describe the fast time evolution and transition dynamics fully and accurately for a system wave function, with a need for repeated calculations at many time steps during the strong, time dependent perturbation of the Hamiltonian (up to 200,000 in the studied case to reach accuracy $<1\%$ for all 16 states). Since McLachlan VHQCA is dependent on the time derivatives of the ansatz, the global phase correction has to be taken into account for an accurate application of the algorithm. The second-order time marching for updating the variational parameters reduces needed number of steps by an order of magnitude, when compared with Euler first- order marching. We developed ansatzs with both unary (Jordan-Wigner) and quantum efficient encoding (QEE) to compare their advantages and shortcomings. Both types of ansatzes were efficient in reaching highly accurate wave function under the laser pulse perturbation. With unary encoding we were able to provide simple quantum circuits and analytical formula for construction a general $N$-state ansatz, using $N$ qubits. QEE simulate the dynamic evolution with significantly lessen number of qubits, $\log_2 N$, and with reduced quantum depth. The latter is achieved by replacing $N-2$ two-qubit control-rotation gates with $N-2$ single-qubit rotation gates, of important for NISQ applications. However, when number of states increases, the QEE ansatz becomes difficult for simulation, requiring the quantum circuit approach at a universal quantum computer. 

We study the algorithm response to a quantum noise, modelled by a real hardware noise of “\texttt{ibmq jakarta}”, as well as to the sampling errors. Accumulation of the errors is reduced several times during the laser-field pulse when using interaction representation rather than Schrodinger representation of the Hamiltonian (in hydrogenic basis). No error accumulation is present with IR when the laser interaction was off. 
We systematically do quantification of the accuracy of our results by comparison with benchmarks, obtained by solving time-dependent Schrodinger equation with classical ODE methods. The error bound of the resulting transition probabilities of all excited states is strictly set to 1$\%$ at the end of the evolution (after the time-dependent perturbation is off) as a condition for acceptance of results.

We hope that this work is an important stepstone toward the quantum advantage in simulating transition dynamics with large number of excited states of a few-electron perturbed system.

\begin{acknowledgments}
YW acknowledge financial support from the Institute for Advanced Computational Science at Stony Brook University. We are grateful to Stony Brook University for the access to the SeaWulf HPC, to XSEDE for the access to SDSC expanse HPC through grant TG-DMR110037 and allocation UTK101, and to IBM Quantum HUB at ORNL (project mat127) for the access to the quantum hardware.
\end{acknowledgments}

\bibliography{ms}

\end{document}


\title{Supplemental Material\\
\small Multistate Transition Dynamics by Strong Time-Dependent Perturbation in NISQ era}

\author{Yulun Wang}
\email[]{yulun.wang@stonybrook.edu} 
\author{Predrag S. Krsti{\'c}}
\email[]{krsticps@gmail.com} 
\affiliation{Institute for Advanced Computational Science, Stony Brook University, Stony Brook NY 11794-5250, USA}

\maketitle
\tableofcontents
\newpage
\renewcommand{\thefigure}{S\arabic{figure}}
\renewcommand{\thetable}{S\Roman{table}}
\renewcommand{\thesection}{S\Roman{section}}
\renewcommand{\theequation}{S\arabic{equation}}

\makeatletter
\renewcommand{\p@subsection}{}
\renewcommand{\p@subsubsection}{}
\makeatother
\captionsetup{font={normalsize,stretch=0.9}}

\section{\label{sec:s1}Hydrogen Atomic Orbitals}
The orbitals are designated by principal ($n$), angular ($l$) and magnetic ($m$) quantum numbers.

\begin{table}[H]
\centering
\renewcommand{\arraystretch}{0.85}
\caption{\label{table:s1}The hydrogen atomic orbitals used in this work for the N-state system. Ground state to higher excited states are listed bottom-up.}
\setlength\tabcolsep{8pt}
\begin{tabular}{|c|c|c|c|c|}
\hline
\rule{0pt}{12pt} \emph{\textbf{N}} & 2 & 4 & 8 & 16\\[2pt]
\hline
\multirow{16}{*}{\textbf{Orbitals}} &    &   &   &6$s$\\
                                    &    &   &   &5$g$\\
                                    &    &   &   &5$f$\\
                                    &    &   &   &5$d$\\
                                    &    &   &4$p$&5$p$\\
                                    &    &   &4$s$&5$s$\\
                                    &    &3$d$&3$d$&4$f$\\
                                    &2$p$&3$s$&3$p$&4$d$\\
                                    &1$s$&2$p$&3$s$&4$p$\\
                                    &    &1$s$&2$p$&4$s$\\
                                    &    &   &2$s$&3$d$\\
                                    &    &   &1$s$&3$p$\\
                                    &    &   &   &3$s$\\
                                    &    &   &   &2$p$\\
                                    &    &   &   &2$s$\\
                                    &    &   &   &1$s$\\
\hline
\end{tabular}
\end{table}

\section{\label{sec:s2}Benchmark Results}
\vspace{-10pt}
\begin{figure}[H]
\centering
\includegraphics[width=5.5in]{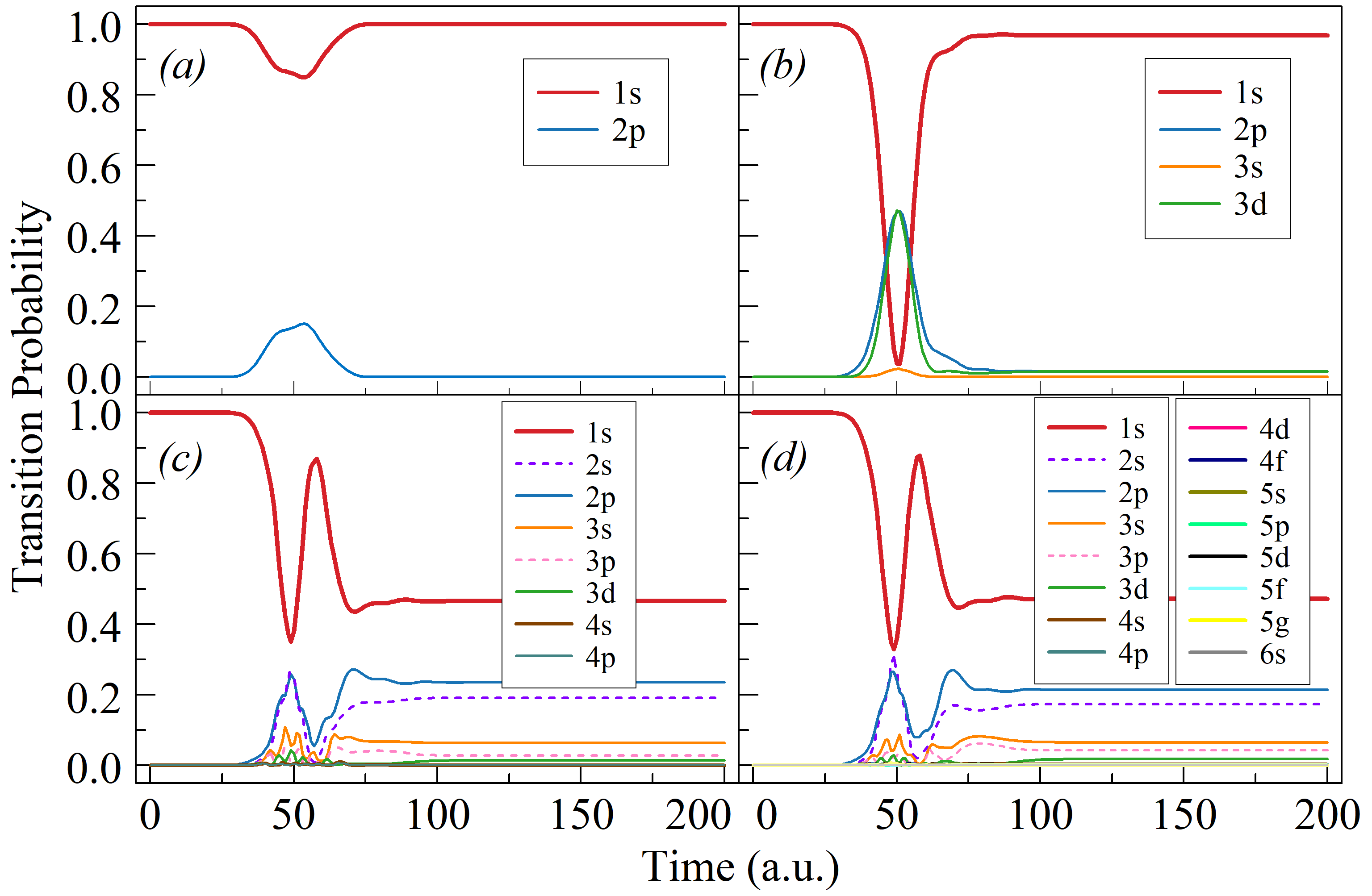}
\caption{\label{fig:s1} The results for the transition dynamics from the ground, 1$s$ state, for the (a) 2, (b) 4, (c) 8 and (d) 16-state  $H$ model systems with HCP laser field ($\omega$=0.06 in Eq. 4b of the manuscript). These results are obtained by accurate numerical solution of ODE’s and are used as benchmarks for solutions by quantum algorithms in Sec. III.}
\end{figure}

\begin{table}[H]
\renewcommand{\arraystretch}{0.9}
\centering
\small
\caption{\label{table:s2}The results for the final transition probabilities at $t$=200, for the $N$-state system with HCP laser field ($\omega$=0.06 in Eq. 4b of the manuscript). These results are obtained by accurate numerical solution of ODE’s and are used as benchmarks for solutions by quantum algorithms in Sec. III.}
\setlength\tabcolsep{8pt}
\begin{tabular}{|c|c|c|c|c|}
\hline
\rule{0pt}{12pt} \emph{\textbf{N}} & 2 & 4 & 8 & 16\\[2pt]
\hline
                            &               &               &               &6$s$ 0.00013056\\
                            &               &               &               &5$g$ 0.00025562\\
                            &               &               &               &5$f$ 0.00174622\\
                            &               &               &               &5$d$ 0.00043734\\
                            &               &               &4$p$ 0.00191683&5$p$ 0.00115603\\
                            &               &               &4$s$ 0.00008123&5$s$ 0.00195571\\
                            &               &3$d$ 0.01534255&3$d$ 0.01412010&4$f$ 0.00203183\\
\textbf{Final transition}   &2$p$ 0.00044517&3$s$ 0.00074915&3$p$ 0.02816392&4$d$ 0.00103995\\
\textbf{probability}        &1$s$ 0.99955483&2$p$ 0.01571505&3$s$ 0.06336444&4$p$ 0.00130138\\
                            &               &1$s$ 0.96819325&2$p$ 0.23518301&4$s$ 0.00359375\\
                            &               &               &2$s$ 0.19188919&3$d$ 0.01784057\\
                            &               &               &1$s$ 0.46528127&3$p$ 0.04349100\\
                            &               &               &               &3$s$ 0.06559239\\
                            &               &               &               &2$p$ 0.21398703\\
                            &               &               &               &2$s$ 0.17376639\\
                            &               &               &               &1$s$ 0.47167423\\
\hline
\end{tabular}
\end{table}

\vspace{-18pt}
\begin{figure}[H]
\centering
\includegraphics[width=2.8in]{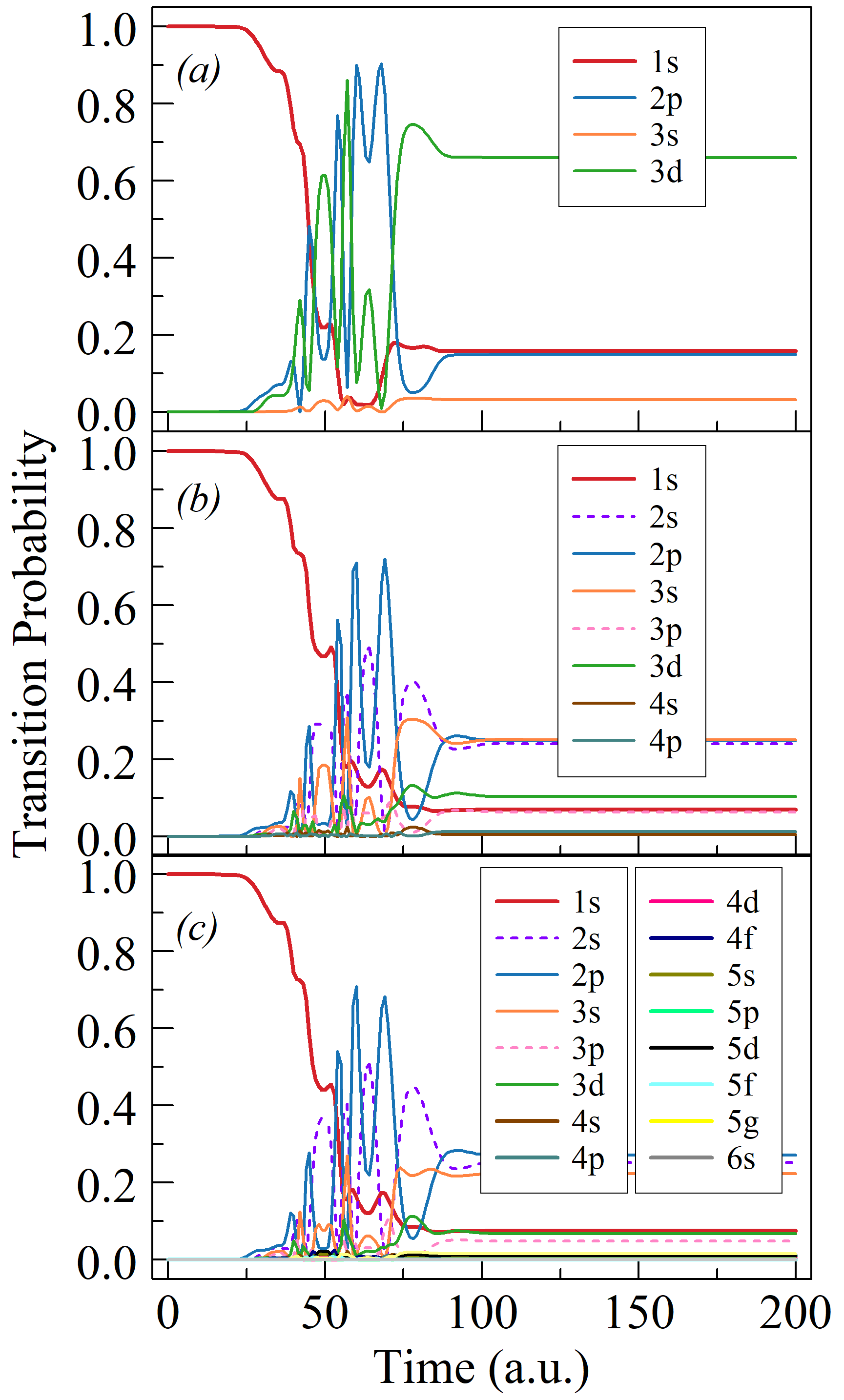}
\caption{\label{fig:s2} The results for the transition dynamics from the ground, 1$s$ state, for the (a) 4, (b) 8 and (c) 16-state $H$ model systems with laser field of $\omega$=0.222 (in Eq. 4b of the manuscript). These results are obtained by accurate numerical solution of ODE’s and are used as benchmarks for solutions by quantum algorithms in Secs. III and IV.}
\end{figure}

\begin{table}[H]
\centering
\small
\renewcommand{\arraystretch}{0.9}
\caption{\label{table:s3}The results for the final transition probabilities at $t$=200, for the $N$-state system with laser field of $\omega$=0.222. These results are obtained by accurate numerical solution of ODE’s and are used as benchmarks for solutions by quantum algorithms in Secs. III and IV.}
\setlength\tabcolsep{8pt}
\begin{tabular}{|c|c|c|c|}
\hline
\rule{0pt}{12pt} \emph{\textbf{N}} & 4 & 8 & 16\\[2pt]
\hline
                            &               &               &6$s$ 0.00009350\\
                            &               &               &5$g$ 0.01563733\\
                            &               &               &5$f$ 0.00012968\\
                            &               &               &5$d$ 0.01077185\\
                            &               &4$p$ 0.01284282&5$p$ 0.00066253\\
                            &               &4$s$ 0.00575194&5$s$ 0.00917148\\
                            &3$d$ 0.65956825&3$d$ 0.10446710&4$f$ 0.00066586\\
\textbf{Final transition}   &3$s$ 0.03220548&3$p$ 0.06442210&4$d$ 0.00285904\\
\textbf{probability}        &2$p$ 0.14928599&3$s$ 0.25150509&4$p$ 0.01117909\\
                            &1$s$ 0.15894029&2$p$ 0.24968068&4$s$ 0.01118176\\
                            &               &2$s$ 0.24131801&3$d$ 0.06802595\\
                            &               &1$s$ 0.07001227&3$p$ 0.04831312\\
                            &               &               &3$s$ 0.22341135\\
                            &               &               &2$p$ 0.27168428\\
                            &               &               &2$s$ 0.25160038\\
                            &               &               &1$s$ 0.07461280\\
\hline
\end{tabular}
\end{table}

\section{\label{sec:s3}Discussion on the HCP pulses}
\subsection{\label{sec:s3a}Transition probabilities for the delta-function electric field pulse}
When modelling  a short half cycle pulse (HCP) by a Dirac delta function, $F=F_0\delta$, the exact transition amplitudes from the ground state 1$s$ are given by the following integral [44]:
\begin{equation}
\label{eqs1}
\alpha_i(t)=\int dV \phi^*_i(r)e^{-ig(t)r}\phi_{1s}(r)
\end{equation}
with
\begin{equation}
\label{eqs2}
\gamma(t)=\int^t_{-T}F(t^\prime)dt^\prime=F_0h(t)
\end{equation}
where $h(t)$ is the Heaviside step function. When $F(t)$ is a short pulse rather than a delta function, Eq.~\ref{eqs1} still can be used with acceptable accuracy providing $\gamma(T)$, calculated from Eq.~\ref{eqs2}, satisfies $\gamma(t)\tau^2<1$ [44], where $\tau$ is the pulse duration. Calculations of $\gamma(T)$, using the laser fields in Sec. III yield values $\gamma(T)=3.112$ when $\omega=0.06$ and $\gamma(T)=0.0256$ when $\omega=0.222$. Taking for $\tau$ value of 20.5 implies $\gamma(t)\tau^2\gg1$ in both cases and Eq.~\ref{eqs1} is not applicable for the laser fields used in this work. 

\begin{figure}[h!]
\includegraphics[width=4in]{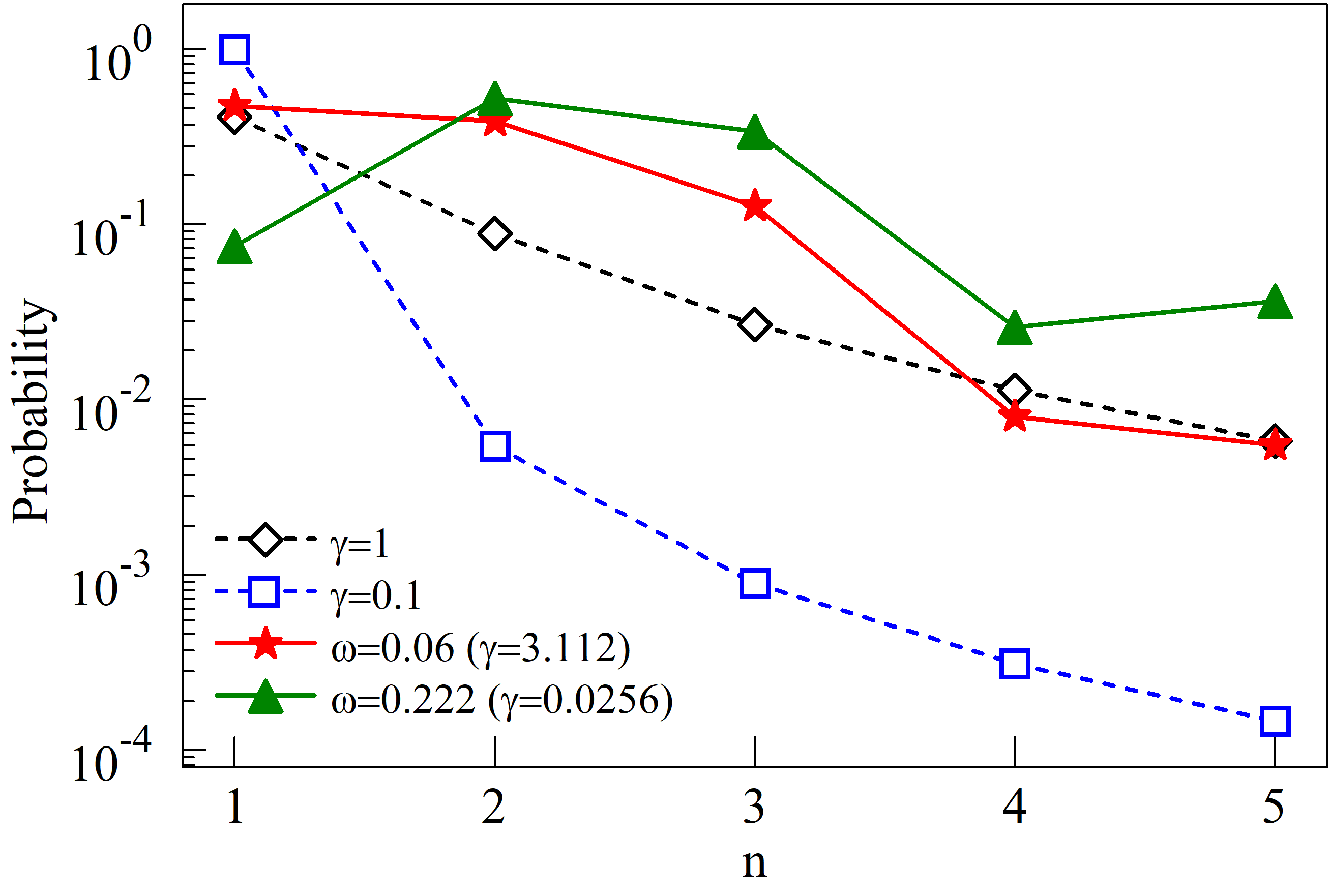}
\caption{\label{fig:s3} The $n$-mixing transition probabilities for various $\gamma$ values, in case of delta function pulse (dashed lines) and 16-state model in Tables~\ref{table:s2} and \ref{table:s3}.}
\end{figure}
Still, it is instructive to show the trends of an exact solution to the problem with an ultra-short, delta-function pulse and comparing with trends of the calculated data in Sec.~\ref{sec:s2}. In case of the delta-function pulse we take two cases: $\gamma_1=1$ and $\gamma_2=0.1$, calculate the transition probabilities for the hydrogen atom starting from 1$s$ to all states $n$=1 to $n$=5, sum them over $l$ for each $n$, and present in Fig.\ref{fig:s3} as function of $n$ (dashed lines). We also sum transition probabilities for various $l$ of each $n$ for the data in Tables~\ref{table:s2} and \ref{table:s3} ($N=16$), and present them at the same figure (solid lines). The transition probabilities for the model of atom with 16 states, shown in Tables~\ref{table:s2} and \ref{table:s3} appear larger than the data calculated from Eq.~\ref{eqs1}. One reason is certainly the normalization. For the cases with delta-function pulse in Fig.~\ref{fig:s3}, only a limited subset of the full set of infinitely many data that obtained from Eq.~\ref{eqs1}, is normalized, while the transition probabilities of a hydrogen model with 16 states, obtained in Tables~\ref{table:s2} and \ref{table:s3}, present here the full normalized set of data.

\subsection{\label{sec:s3b}Fourier analysis of the HCP in Sec.~\ref{sec:s3a}}
A common half cycle pulse can be designed from the linear combination of a carefully chosen set of the laser field modes. If a pulse of electric field is given, one can obtain its Fourier series expansion to see of what components it is built, following:
\begin{equation}
\label{eqs3}
F(t)=A_0+\sum^\infty_{n=1}A_n\cos\left(\frac{n\pi t}{L}\right)+\sum^\infty_{n=1}B_n\sin\left(\frac{n\pi t}{L}\right)
\end{equation}
where
\begin{subequations}
\label{eqs4}
\begin{equation}
A_0=\frac{1}{2L}\int^L_{-L} F(t)dt
\end{equation}
\begin{equation}
A_n=\frac{1}{L}\int^L_{-L} F(t)\cos\left(\frac{n\pi t}{L}\right)dt
\end{equation}
\begin{equation}
B_n=\frac{1}{L}\int^L_{-L} F(t)\sin\left(\frac{n\pi t}{L}\right)dt
\end{equation}
\end{subequations}

As an example, here we use the half cycle laser pulse studied in Sec. III A , with peak shifted to $t$=0:
\begin{equation}
\label{eqs5}
F(t)=0.25e^{-\left(\frac{t}{20.5}\right)^2}\cos(0.06t)
\end{equation}

The coefficients of the corresponding Fourier expansion up to $n$=8 are listed in Table~\ref{table:s4}, calculated with $L$=100. If these first harmonics are used in Eq.~\ref{eqs3}, the resulting function is almost identical to the HCP function in Fig. 1.

\begin{table}[H]
\centering
\small
\caption{\label{table:s4}The coefficients of Fourier expansion for laser pulse in Eq.~\ref{eqs5}.}
\setlength\tabcolsep{8pt}
\begin{tabular}{|c|c|c|}
\hline
& \multicolumn{2}{c|}{$F(t)$} \\
\cline{2-3}
\emph{\textbf{n}} & $A_n$ & $B_n$\\
\hline
0&0.0311156&0\\\hline
1&0.0605599&0\\\hline
2&0.0546881&0\\\hline
3&0.0438827&0\\\hline
4&0.0300882&0\\\hline
5&0.0171955&0\\\hline
6&0.0080835&0\\\hline
7&0.0031056&0\\\hline
8&0.0009722&0\\
\hline
\end{tabular}
\end{table}

\section{\label{sec:s4}The Derivation of McLachlan Variational Principle}
From Eq. 7 in the manuscript:
\begin{eqnarray}
\label{eqs6}
&&\| (\frac{d}{dt}+iH)\Big\vert\phi(\boldsymbol{\theta}(t))\Big\rangle \|^2\nonumber\\
=&&\left((\frac{d}{dt}+iH)\Big\vert\phi(\boldsymbol{\theta}(t))\Big\rangle\right)^\dagger\left((\frac{d}{dt}+iH)\Big\vert\phi(\boldsymbol{\theta}(t))\Big\rangle\right)\nonumber\\
=&&\frac{\partial \Big\langle\phi(\boldsymbol{\theta}(t))\Big\vert }{\partial t} \frac{\partial \Big\vert\phi(\boldsymbol{\theta}(t))\Big\rangle }{\partial t}
+i\frac{\partial \Big\langle\phi(\boldsymbol{\theta}(t))\Big\vert }{\partial t} H\Big\vert\phi(\boldsymbol{\theta}(t))\Big\rangle\nonumber\\
-&&i\Big\langle\phi(\boldsymbol{\theta}(t))\Big\vert H\frac{\partial \Big\vert\phi(\boldsymbol{\theta}(t))\Big\rangle }{\partial t}
+\Big\langle\phi(\boldsymbol{\theta}(t))\Big\vert H^2\Big\vert\phi(\boldsymbol{\theta}(t))\Big\rangle \nonumber\\
=&&\sum_{i,j}\frac{\partial \Big\langle\phi(\boldsymbol{\theta}(t))\Big\vert }{\partial \theta_i} \frac{\partial \Big\vert\phi(\boldsymbol{\theta}(t))\Big\rangle }{\partial \theta_j}\dot{\theta}_i^\ast\dot{\theta}_j
+i\sum_i\frac{\partial \Big\langle\phi(\boldsymbol{\theta}(t))\Big\vert }{\partial \theta_i} H\Big\vert\phi(\boldsymbol{\theta}(t))\Big\rangle\dot{\theta}_i^\ast\nonumber\\
-&&i\sum_i \Big\langle\phi(\boldsymbol{\theta}(t))\Big\vert H\frac{\partial \Big\vert\phi(\boldsymbol{\theta}(t))\Big\rangle }{\partial \theta_i}\dot{\theta}_i
+\Big\langle\phi(\boldsymbol{\theta}(t))\Big\vert H^2\Big\vert\phi(\boldsymbol{\theta}(t))\Big\rangle
\end{eqnarray}

Suppose $\dot{\theta}$ is real, then Eq. 7 in the manuscript, upon variation, can be expanded as:

\begin{eqnarray}
&&\quad\delta\| (\frac{d}{dt}+iH)\Big\vert\phi(\boldsymbol{\theta}(t))\Big\rangle \|^2 = 0\label{eqs7}\\
\longrightarrow&&\quad\left(
\sum_{j}\frac{\partial \Big\langle\phi(\boldsymbol{\theta}(t))\Big\vert }{\partial \theta_i} \frac{\partial \Big\vert\phi(\boldsymbol{\theta}(t))\Big\rangle }{\partial \theta_j}\dot{\theta}_j
+i\frac{\partial \Big\langle\phi(\boldsymbol{\theta}(t))\Big\vert }{\partial \theta_i} H\Big\vert\phi(\boldsymbol{\theta}(t))\Big\rangle
\right)\delta\theta_i \nonumber\\
&&+\left(
\sum_{j}\frac{\partial \Big\langle\phi(\boldsymbol{\theta}(t))\Big\vert }{\partial \theta_j} \frac{\partial \Big\vert\phi(\boldsymbol{\theta}(t))\Big\rangle }{\partial \theta_i}\dot{\theta}_j
-i\Big\langle\phi(\boldsymbol{\theta}(t))\Big\vert H\frac{\partial \Big\vert\phi(\boldsymbol{\theta}(t))\Big\rangle }{\partial \theta_i}
\right)\delta\theta_i=0 \label{eqs8}
\end{eqnarray}
\begin{eqnarray}
\longrightarrow\quad&&\sum_{j}\left(
\frac{\partial \Big\langle\phi(\boldsymbol{\theta}(t))\Big\vert }{\partial \theta_i} \frac{\partial \Big\vert\phi(\boldsymbol{\theta}(t))\Big\rangle }{\partial \theta_j}
+\frac{\partial \Big\langle\phi(\boldsymbol{\theta}(t))\Big\vert }{\partial \theta_j} \frac{\partial \Big\vert\phi(\boldsymbol{\theta}(t))\Big\rangle }{\partial \theta_i}
\right)\dot{\theta}_j \nonumber\\
&&=-i\left(
\frac{\partial \Big\langle\phi(\boldsymbol{\theta}(t))\Big\vert }{\partial \theta_i} H\Big\vert\phi(\boldsymbol{\theta}(t))\Big\rangle
-\Big\langle\phi(\boldsymbol{\theta}(t))\Big\vert H\frac{\partial \Big\vert\phi(\boldsymbol{\theta}(t))\Big\rangle }{\partial \theta_i}
\right)\label{eqs9}
\end{eqnarray}
Defining the matrix $A$ and vector $C$ with elements:

\begin{subnumcases}{\qquad\qquad\qquad\qquad\qquad\qquad\qquad}
\label{eqs11}
A_{i,j}=\frac{\partial \Big\langle\phi(\boldsymbol{\theta}(t))\Big\vert }{\partial \theta_i} \frac{\partial \Big\vert\phi(\boldsymbol{\theta}(t))\Big\rangle }{\partial \theta_j}\\
C_i=\frac{\partial \Big\langle\phi(\boldsymbol{\theta}(t))\Big\vert }{\partial \theta_i} H\Big\vert\phi(\boldsymbol{\theta}(t))\Big\rangle
\end{subnumcases}

and the system of algebraic equations for real $\dot{\boldsymbol{\theta}}$ can be written in form:
\begin{equation}
\label{eqs9}
\sum_j A^R_{i,j}\dot{\theta}_j=C^I_i
\end{equation}

\vspace{20pt}
\section{\label{sec:s5}The Derivation of McLachlan Variational Principle with Global Phase Correction}
Replacing the wavefunction $\psi(\boldsymbol{r},t)$ in the Eq. 1 with $\vert\Phi(t)\rangle=e^{i\alpha(t)}\vert\phi(\boldsymbol{\theta}(t))\rangle$, we get:
\begin{eqnarray}
&&e^{i\alpha}\frac{\partial \Big\vert\phi(\boldsymbol{\theta}(t))\Big\rangle }{\partial \theta}\dot{\theta}+i\dot{\alpha}e^{i\alpha} \Big\vert\phi(\boldsymbol{\theta}(t))\Big\rangle \approx-iHe^{i\alpha} \Big\vert\phi(\boldsymbol{\theta}(t))\Big\rangle \label{eqs12}\\
\longrightarrow&&\frac{\partial \Big\vert\phi(\boldsymbol{\theta}(t))\Big\rangle }{\partial \theta}\dot{\theta}+i\dot{\alpha}\Big\vert\phi(\boldsymbol{\theta}(t))\Big\rangle \approx-iH \Big\vert\phi(\boldsymbol{\theta}(t))\Big\rangle \label{eqs13}
\end{eqnarray}

The squared norm of the deviation between both sides of Eq.~\ref{eqs13} can be expanded, with the assumption that $\dot{\boldsymbol{\theta}}$ is a real vector:
\allowdisplaybreaks
\begin{eqnarray}
\label{eqs14}
&&\qquad\| (\frac{d}{dt}+i\dot{\alpha}+iH)\Big\vert\phi(\boldsymbol{\theta}(t))\Big\rangle \|^2\nonumber\\
&&=\left((\frac{d}{dt}+i\dot{\alpha}+iH)\Big\vert\phi(\boldsymbol{\theta}(t))\Big\rangle\right)^\dagger\left((\frac{d}{dt}+i\dot{\alpha}+iH)\Big\vert\phi(\boldsymbol{\theta}(t))\Big\rangle\right)\nonumber\\
&&=\frac{\partial \Big\langle\phi(\boldsymbol{\theta}(t))\Big\vert }{\partial t} \frac{\partial \Big\vert\phi(\boldsymbol{\theta}(t))\Big\rangle }{\partial t}
+\Big\langle\phi(\boldsymbol{\theta}(t))\Big\vert H^2\Big\vert\phi(\boldsymbol{\theta}(t))\Big\rangle\nonumber\\
&&+i\frac{\partial \Big\langle\phi(\boldsymbol{\theta}(t))\Big\vert }{\partial t} H\Big\vert\phi(\boldsymbol{\theta}(t))\Big\rangle 
-i\Big\langle\phi(\boldsymbol{\theta}(t))\Big\vert H\frac{\partial \Big\vert\phi(\boldsymbol{\theta}(t))\Big\rangle }{\partial t}\nonumber\\
&& 
+i\dot{\alpha}\frac{\partial \Big\langle\phi(\boldsymbol{\theta}(t))\Big\vert }{\partial t} \Big\vert\phi(\boldsymbol{\theta}(t))\Big\rangle
-i\dot{\alpha}\Big\langle\phi(\boldsymbol{\theta}(t))\Big\vert \frac{\partial \Big\vert\phi(\boldsymbol{\theta}(t))\Big\rangle }{\partial t}
+\dot{\alpha}^2
+2\dot{\alpha}\Big\langle\phi(\boldsymbol{\theta}(t))\Big\vert H\Big\vert\phi(\boldsymbol{\theta}(t))\Big\rangle \nonumber\\
&&=\sum_{i,j}\frac{\partial \Big\langle\phi(\boldsymbol{\theta}(t))\Big\vert }{\partial \theta_i} \frac{\partial \Big\vert\phi(\boldsymbol{\theta}(t))\Big\rangle }{\partial \theta_j}\dot{\theta}_i\dot{\theta}_j
+i\sum_i\frac{\partial \Big\langle\phi(\boldsymbol{\theta}(t))\Big\vert }{\partial \theta_i} H\Big\vert\phi(\boldsymbol{\theta}(t))\Big\rangle\dot{\theta}_i\nonumber\\
&&-i\sum_i \Big\langle\phi(\boldsymbol{\theta}(t))\Big\vert H\frac{\partial \Big\vert\phi(\boldsymbol{\theta}(t))\Big\rangle }{\partial \theta_i}\dot{\theta}_i
+i\dot{\alpha}\sum_i\frac{\partial \Big\langle\phi(\boldsymbol{\theta}(t))\Big\vert }{\partial \theta_i} \Big\vert\phi(\boldsymbol{\theta}(t))\Big\rangle\dot{\theta}_i\nonumber\\
&&-i\dot{\alpha}\sum_i \Big\langle\phi(\boldsymbol{\theta}(t))\Big\vert \frac{\partial \Big\vert\phi(\boldsymbol{\theta}(t))\Big\rangle }{\partial \theta_i}\dot{\theta}_i
+\dot{\alpha}^2
+\Big\langle\phi(\boldsymbol{\theta}(t))\Big\vert H^2\Big\vert\phi(\boldsymbol{\theta}(t))\Big\rangle
+2\dot{\alpha}\Big\langle\phi(\boldsymbol{\theta}(t))\Big\vert H\Big\vert\phi(\boldsymbol{\theta}(t))\Big\rangle
\end{eqnarray}
Then, upon variation, we get:
\begin{eqnarray}
&&\quad\delta\| (\frac{d}{dt}+i\dot{\alpha}+iH)\Big\vert\phi(\boldsymbol{\theta}(t))\Big\rangle \|^2 = 0 \label{eqs15}\\
\longrightarrow&&\quad\Bigg( \Bigg.
\sum_{j}\frac{\partial \big\langle\phi(\boldsymbol{\theta})\big\vert }{\partial \theta_i} \frac{\partial \big\vert\phi(\boldsymbol{\theta})\big\rangle }{\partial \theta_j}\dot{\theta}_j
+\sum_{j}\frac{\partial \big\langle\phi(\boldsymbol{\theta})\big\vert }{\partial \theta_j} \frac{\partial \big\vert\phi(\boldsymbol{\theta})\big\rangle }{\partial \theta_i}\dot{\theta}_j
+i\frac{\partial \big\langle\phi(\boldsymbol{\theta})\big\vert }{\partial \theta_i} H\big\vert\phi(\boldsymbol{\theta})\big\rangle\nonumber\\
&&-i\big\langle\phi(\boldsymbol{\theta})\big\vert H\frac{\partial \big\vert\phi(\boldsymbol{\theta})\big\rangle }{\partial \theta_i}
+i\dot{\alpha}\frac{\partial \big\langle\phi(\boldsymbol{\theta})\big\vert }{\partial \theta_i} \big\vert\phi(\boldsymbol{\theta})\big\rangle
-i\dot{\alpha}\big\langle\phi(\boldsymbol{\theta})\big\vert \frac{\partial \big\vert\phi(\boldsymbol{\theta})\big\rangle }{\partial \theta_i}
\Bigg. \Bigg)\delta\theta_i \nonumber\\
&&+\Bigg( \Bigg.
i\sum_i\Bigg(\frac{\partial \big\langle\phi(\boldsymbol{\theta})\big\vert }{\partial \theta_i} \big\vert\phi(\boldsymbol{\theta})\big\rangle
-\big\langle\phi(\boldsymbol{\theta})\big\vert \frac{\partial \big\vert\phi(\boldsymbol{\theta})\big\rangle }{\partial \theta_i}\Bigg)\dot{\theta}_i
+2\dot{\alpha}+2\big\langle\phi(\boldsymbol{\theta})\big\vert H\big\vert\phi(\boldsymbol{\theta})\big\rangle
\Bigg. \Bigg)\delta{\alpha}=0 \label{eqs16}
\end{eqnarray}
\begin{subnumcases}{\longrightarrow\quad}
\label{eqs17}
\sum_j\Big(\frac{\partial \langle\phi(\boldsymbol{\theta})\vert }{\partial \theta_i} \frac{\partial \vert\phi(\boldsymbol{\theta})\rangle }{\partial \theta_j}
+\frac{\partial \langle\phi(\boldsymbol{\theta})\vert }{\partial \theta_j} \frac{\partial \vert\phi(\boldsymbol{\theta})\rangle }{\partial \theta_i}\Big)\dot{\theta}_j \nonumber\\
=-i\Big(
\frac{\partial \langle\phi(\boldsymbol{\theta})\vert }{\partial \theta_i} H\vert\phi(\boldsymbol{\theta})\rangle
-\langle\phi(\boldsymbol{\theta})\vert H\frac{\partial \vert\phi(\boldsymbol{\theta})\rangle }{\partial \theta_i}
\Big)
-i\dot{\alpha}\Big(
\frac{\partial \langle\phi(\boldsymbol{\theta})\vert }{\partial \theta_i} \vert\phi(\boldsymbol{\theta})\rangle
-\langle\phi(\boldsymbol{\theta})\vert \frac{\partial \vert\phi(\boldsymbol{\theta})\rangle }{\partial \theta_i}
\Big)\\[8pt]
i\sum_i\Big(\frac{\partial \langle\phi(\boldsymbol{\theta})\vert }{\partial \theta_i} \vert\phi(\boldsymbol{\theta})\rangle
-\langle\phi(\boldsymbol{\theta})\vert \frac{\partial\vert\phi(\boldsymbol{\theta})\rangle }{\partial \theta_i}\Big)\dot{\theta}_i
+2\dot{\alpha}+2\langle\phi(\boldsymbol{\theta})\vert H\vert\phi(\boldsymbol{\theta}(t))\rangle=0
\end{subnumcases}
\begin{subnumcases}{\quad\quad\quad\longrightarrow\quad}
\label{eqs18}
\sum_j A^R_{i,j}\dot{\theta}_j=C^I_i+\dot{\alpha}\Im\left(\frac{\partial \langle\phi(\boldsymbol{\theta})\vert }{\partial \theta_i} \vert\phi(\boldsymbol{\theta})\rangle\right)\\[8pt]
\dot{\alpha}=\sum_i\Im\left(\frac{\partial \langle\phi(\boldsymbol{\theta})\vert }{\partial \theta_i} \vert\phi(\boldsymbol{\theta})\rangle\right)\dot{\theta}_i-\big\langle\phi(\boldsymbol{\theta})\big\vert H\big\vert\phi(\boldsymbol{\theta})\big\rangle
\end{subnumcases}
However, $\frac{\partial \langle\phi(\boldsymbol{\theta})\vert }{\partial \theta_i} \vert\phi(\boldsymbol{\theta})\rangle$ is fully imaginary. Then combining the Eqs. S18, we get:
\begin{eqnarray}
\label{eqs19}
\sum_j\left(
A^R_{i,j}
+\frac{\partial \big\langle\phi(\boldsymbol{\theta})\big\vert }{\partial \theta_i} \big\vert\phi(\boldsymbol{\theta})\big\rangle
\frac{\partial \big\langle\phi(\boldsymbol{\theta})\big\vert }{\partial \theta_j} \big\vert\phi(\boldsymbol{\theta})\big\rangle
\right)\dot{\theta}_j
=C^I_i
+i\frac{\partial \big\langle\phi(\boldsymbol{\theta})\big\vert }{\partial \theta_i} \big\vert\phi(\boldsymbol{\theta})\big\rangle
\big\langle\phi(\boldsymbol{\theta})\big\vert H\big\vert\phi(\boldsymbol{\theta})\big\rangle
\end{eqnarray}
\begin{eqnarray}
\longrightarrow\sum_j M_{i,j}\dot{\theta}_j=V_i \label{eqs20}
\end{eqnarray}
where
\begin{subnumcases}{\qquad\qquad\qquad\qquad\qquad\qquad}
\label{eqs21}
M=A^R_{i,j}
+\Big(\frac{\partial \langle\phi(\boldsymbol{\theta})\vert }{\partial \theta_i} \vert\phi(\boldsymbol{\theta})\rangle\Big)
\Big(\frac{\partial \langle\phi(\boldsymbol{\theta})\vert }{\partial \theta_j} \vert\phi(\boldsymbol{\theta})\rangle\Big)\label{eqs21}\\[8pt]
V=C^I_i
+i\Big(\frac{\partial \langle\phi(\boldsymbol{\theta})\vert }{\partial \theta_i} \vert\phi(\boldsymbol{\theta})\rangle\Big)
\langle\phi(\boldsymbol{\theta})\vert H\vert\phi(\boldsymbol{\theta})\rangle\label{eqs22}
\end{subnumcases}

\vspace*{20pt}
\section{\label{sec:s6}Examples of Encoding of the Hamiltonian}
\subsection{\label{sec:s6a}Jordan-Wigner encoding}
The  qubit  Hamiltonian,  directly  obtained from the Jordan-Wigner encoding (JWE), using 4 qubits for the 4 mutually coupled states (1s, 2p, 3s, 3d) of the hydrogen atom-laser system has the form:
\begin{eqnarray}
\label{eqs22}
H=&&-\frac{53}{144}I+\frac{1}{4}Z_0+\frac{1}{16}Z_1+\frac{1}{36}Z_2+\frac{1}{36}Z_4\nonumber\\
&&+\frac{64\sqrt{2}}{243}F\left(t\right)X_0X_1+\frac{64\sqrt{2}}{243}F\left(t\right)Y_0Y_1+\frac{1728\sqrt{6}}{15625}F\left(t\right)X_1X_2\nonumber\\
&&+\frac{1728\sqrt{6}}{15625}F\left(t\right)Y_1Y_2+\frac{55296\sqrt{3}}{78125}F\left(t\right)X_1Z_2X_3+\frac{55296\sqrt{3}}{78125}F\left(t\right)Y_1Z_2Y_3
\end{eqnarray}

Taking a general 4-state system, the qubit Hamiltonian by JWE with 4 qubits is:
\allowdisplaybreaks
\begin{eqnarray}
\label{eqs23}
H&&=h_{00}a^\dagger_0 a_0
+h_{11}a^\dagger_1 a_1
+h_{22}a^\dagger_2 a_2
+h_{33}a^\dagger_3 a_3
+h_{01}a^\dagger_0 a_1
+h_{10}a^\dagger_1 a_0
+h_{02}a^\dagger_0 a_2
+h_{20}a^\dagger_2 a_0\nonumber\\
&&+h_{03}a^\dagger_0 a_3
+h_{30}a^\dagger_3 a_0
+h_{12}a^\dagger_1 a_2
+h_{21}a^\dagger_2 a_1
+h_{13}a^\dagger_1 a_3
+h_{31}a^\dagger_3 a_1
+h_{23}a^\dagger_2 a_3
+h_{32}a^\dagger_3 a_2\nonumber\\
&&=\frac{1}{2}h_{00}\left(I-Z_0\right)
+\frac{1}{2}h_{11}\left(I-Z_1\right)
+\frac{1}{2}h_{22}\left(I-Z_2\right)
+\frac{1}{2}h_{33}\left(I-Z_3\right)\nonumber\\
&&+\frac{1}{4}h_{01}\left(X_0X_1+iX_0Y_1-iY_0X_1+Y_0Y_1\right)
+\frac{1}{4}h_{10}\left(X_0X_1-iX_0Y_1+iY_0X_1+Y_0Y_1\right)\nonumber\\
&&+\frac{1}{4}h_{02}\left(X_0Z_1X_2+iX_0Z_1Y_2-iY_0Z_1X_2+Y_0Z_1Y_2\right)\nonumber\\
&&+\frac{1}{4}h_{20}\left(X_0Z_1X_2-iX_0Z_1Y_2+iY_0Z_1X_2+Y_0Z_1Y_2\right)\nonumber\\
&&+\frac{1}{4}h_{03}\left(X_0Z_1Z_2X_3+iX_0Z_1Z_2Y_3-iY_0Z_1Z_2X_3+Y_0Z_1Z_2Y_3\right)\nonumber\\
&&+\frac{1}{4}h_{30}\left(X_0Z_1Z_2X_3-iX_0Z_1Z_2Y_3+iY_0Z_1Z_2X_3+Y_0Z_1Z_2Y_3\right)\nonumber\\
&&+\frac{1}{4}h_{12}\left(X_1X_2+iX_1Y_2-iY_1X_2+Y_1Y_2\right)
+\frac{1}{4}h_{21}\left(X_1X_2-iX_1Y_2+iY_1X_2+Y_1Y_2\right)\nonumber\\
&&+\frac{1}{4}h_{13}\left(X_1Z_2X_3+iX_1Z_2Y_3-iY_1Z_2X_3+Y_1Z_2Y_3\right)\nonumber\\
&&+\frac{1}{4}h_{31}\left(X_1Z_2X_3-iX_1Z_2Y_3+iY_1Z_2X_3+Y_1Z_2Y_3\right)\nonumber\\
&&+\frac{1}{4}h_{23}\left(X_2X_3+iX_2Y_3-iY_2X_3+Y_2Y_3\right)
+\frac{1}{4}h_{32}\left(X_2X_3-iX_2Y_3+iY_2X_3+Y_2Y_3\right)\nonumber\\
&&=\frac{1}{2}\left(h_{00}+h_{11}+h_{22}+h_{33}\right)I
-\frac{1}{2}h_{00}Z_0-\frac{1}{2}h_{11}Z_1-\frac{1}{2}h_{22}Z_2-\frac{1}{2}h_{33}Z_3\nonumber\\
&&+\frac{1}{4}\left(h_{01}+h_{10}\right)\left(X_0X_1+Y_0Y_1\right)
+\frac{i}{4}\left(h_{01}-h_{10}\right)\left(X_0Y_1-Y_0X_1\right)\nonumber\\
&&+\frac{1}{4}\left(h_{02}+h_{20}\right)\left(X_0Z_1X_2+Y_0Z_1Y_2\right)
+\frac{i}{4}\left(h_{02}-h_{20}\right)\left(X_0Z_1Y_2-Y_0Z_1X_2\right)\nonumber\\
&&+\frac{1}{4}\left(h_{03}+h_{30}\right)\left(X_0Z_1Z_2X_3+Y_0Z_1Z_2Y_3\right)
+\frac{i}{4}\left(h_{03}-h_{30}\right)\left(X_0Z_1Z_2Y_3-Y_0Z_1Z_2X_3\right)\nonumber\\
&&+\frac{1}{4}\left(h_{12}+h_{21}\right)\left(X_1X_2+Y_1Y_2\right)
+\frac{i}{4}\left(h_{12}-h_{21}\right)\left(X_1Y_2-Y_1X_2\right)\nonumber\\
&&+\frac{1}{4}\left(h_{13}+h_{31}\right)\left(X_1Z_2X_3+Y_1Z_2Y_3\right)
+\frac{i}{4}\left(h_{13}-h_{31}\right)\left(X_1Z_2Y_3-Y_1Z_2X_3\right)\nonumber\\
&&+\frac{1}{4}\left(h_{23}+h_{32}\right)\left(X_2X_3+Y_2Y_3\right)
+\frac{i}{4}\left(h_{23}-h_{32}\right)\left(X_2Y_3-Y_2X_3\right)
\end{eqnarray}
where index $i$ of a Pauli matrix $P_i$ from the Pauli set {X, Y, Z} indicates the operation on the $i^{th}$ qubit.
There are totally 28 Pauli terms in the qubit Hamiltonian.

\vspace{20pt}
\subsection{\label{sec:s6b}Qubit efficient encoding}
The qubit Hamiltonian encoded by QEE for the same 4-state system as in Eq.~\ref{eqs22} with the two qubit takes the form:
\begin{eqnarray}
\label{eqs24}
H=&&-\frac{53}{288}I-\frac{37}{288}Z_0-\frac{3}{32}Z_1-\frac{3}{32}Z_0Z_1\nonumber\\
&&+\frac{55296\sqrt{3}}{78125}F\left(t\right)X_0+\frac{64\sqrt{2}}{243}F\left(t\right)X_1-\frac{55296\sqrt{3}}{78125}F\left(t\right)X_0Z_1\nonumber\\
&&+\frac{64\sqrt{2}}{243}F\left(t\right)Z_0X_1+\frac{1728\sqrt{6}}{15625}F\left(t\right)X_0X_1+\frac{1728\sqrt{6}}{15625}F\left(t\right)Y_0Y_1
\end{eqnarray}
Taking a general 4-state system, the qubit efficient encoding (QEE) for 2 qubits and 4 states  gives:
\begin{eqnarray}
\label{eqs25}
H&&=\sum^{N-1}_{i,j}h_{ij}\vert\boldsymbol{q}\rangle_i\langle\boldsymbol{q}\vert_j\nonumber\\
&&=h_{00}\vert00\rangle\langle00\vert
+h_{11}\vert01\rangle\langle01\vert
+h_{22}\vert10\rangle\langle10\vert
+h_{33}\vert11\rangle\langle11\vert\nonumber\\
&&+h_{01}\vert00\rangle\langle01\vert
+h_{10}\vert01\rangle\langle00\vert
+h_{12}\vert01\rangle\langle10\vert
+h_{21}\vert10\rangle\langle01\vert\nonumber\\
&&+h_{13}\vert01\rangle\langle11\vert
+h_{31}\vert11\rangle\langle01\vert
+h_{02}\vert00\rangle\langle10\vert
+h_{20}\vert10\rangle\langle00\vert\nonumber\\
&&+h_{03}\vert00\rangle\langle11\vert
+h_{30}\vert11\rangle\langle00\vert
+h_{23}\vert10\rangle\langle11\vert
+h_{32}\vert11\rangle\langle10\vert\nonumber\\
&&=h_{00}\left(\vert0\rangle\langle0\vert\otimes\vert0\rangle\langle0\vert\right)
+h_{11}\left(\vert0\rangle\langle0\vert\otimes\vert1\rangle\langle1\vert\right)\nonumber\\
&&+h_{22}\left(\vert1\rangle\langle1\vert\otimes\vert0\rangle\langle0\vert\right)
+h_{33}\left(\vert1\rangle\langle1\vert\otimes\vert1\rangle\langle1\vert\right)\nonumber\\
&&+h_{01}\left(\vert0\rangle\langle0\vert\otimes\vert0\rangle\langle1\vert\right)
+h_{10}\left(\vert0\rangle\langle0\vert\otimes\vert1\rangle\langle0\vert\right)\nonumber\\
&&+h_{12}\left(\vert0\rangle\langle1\vert\otimes\vert1\rangle\langle0\vert\right)
+h_{21}\left(\vert1\rangle\langle0\vert\otimes\vert0\rangle\langle1\vert\right)\nonumber\\
&&+h_{13}\left(\vert0\rangle\langle1\vert\otimes\vert1\rangle\langle1\vert\right)
+h_{31}\left(\vert1\rangle\langle0\vert\otimes\vert1\rangle\langle1\vert\right)\nonumber\\
&&+h_{02}\left(\vert0\rangle\langle1\vert\otimes\vert0\rangle\langle0\vert\right)
+h_{20}\left(\vert1\rangle\langle0\vert\otimes\vert0\rangle\langle0\vert\right)\nonumber\\
&&+h_{03}\left(\vert0\rangle\langle1\vert\otimes\vert0\rangle\langle1\vert\right)
+h_{30}\left(\vert1\rangle\langle0\vert\otimes\vert1\rangle\langle0\vert\right)\nonumber\\
&&+h_{23}\left(\vert1\rangle\langle1\vert\otimes\vert0\rangle\langle1\vert\right)
+h_{32}\left(\vert1\rangle\langle1\vert\otimes\vert1\rangle\langle0\vert\right)\nonumber\\
&&=\frac{1}{4}\left(h_{00}+h_{11}+h_{22}+h_{33}\right)I
+\frac{1}{4}\left(h_{00}+h_{11}-h_{22}-h_{33}\right)Z_0\nonumber\\
&&+\frac{1}{4}\left(h_{00}-h_{11}+h_{22}-h_{33}\right)Z_1
+\frac{1}{4}\left(h_{00}-h_{11}-h_{22}+h_{33}\right)Z_0Z_1\nonumber\\
&&+\frac{1}{4}\left(h_{13}+h_{31}+h_{02}+h_{20}\right)X_0
+\frac{1}{4}\left(h_{01}+h_{10}+h_{23}+h_{32}\right)X_1\nonumber\\
&&+\frac{i}{4}\left(h_{13}-h_{31}+h_{02}-h_{20}\right)Y_0
+\frac{i}{4}\left(h_{01}-h_{10}+h_{23}-h_{32}\right)Y_1\nonumber\\
&&-\frac{1}{4}\left(h_{13}+h_{31}+h_{02}+h_{20}\right)X_0Z_1
-\frac{i}{4}\left(h_{13}-h_{31}-h_{02}+h_{20}\right)Y_0Z_1\nonumber\\
&&-\frac{1}{4}\left(h_{01}+h_{10}+h_{23}+h_{32}\right)Z_0X_1
+\frac{i}{4}\left(h_{01}-h_{10}-h_{23}+h_{32}\right)Z_0Y_1\nonumber\\
&&+\frac{1}{4}\left(h_{12}+h_{21}+h_{03}+h_{30}\right)X_0X_1
+\frac{1}{4}\left(h_{12}+h_{21}-h_{03}-h_{30}\right)Y_0Y_1\nonumber\\
&&-\frac{i}{4}\left(h_{12}-h_{21}-h_{03}+h_{30}\right)X_0Y_1
+\frac{i}{4}\left(h_{12}-h_{21}+h_{03}-h_{30}\right)Y_0X_1
\end{eqnarray}
There are totally 15 Pauli terms in the qubit Hamiltonian.

\vspace{30pt}
\section{\label{sec:s7}Example of the gate decomposition at IBM Q quantum computer}
The $R_Y$ and $CR_X$ gates are decomposed to backend basis operations: ['\texttt{id}', '\texttt{rz}', '\texttt{sx}', '\texttt{x}', '\texttt{cx}', '\texttt{reset}'] of IBM device backend “\texttt{ibmq\_jakarta}” as:
\begin{figure}[h!]
\includegraphics[width=6in]{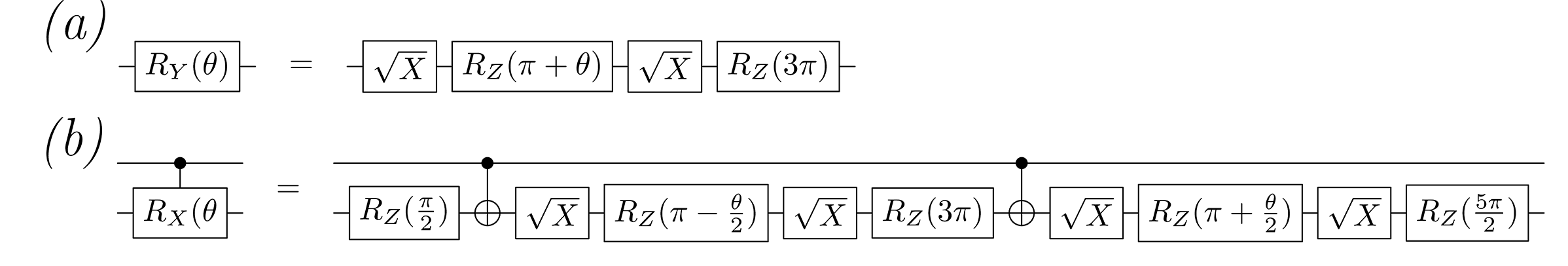}
\caption{\label{fig:s4} The decomposition of $R_Y$ and $CR_X$ gates into basis gates on IBM Q backend.}
\end{figure}

\section{\label{sec:s8}Symbolic Simulation Results}
\subsection{\label{sec:s8a}Results Using JWE}

\begin{table}[H]
\centering
\small
\caption{\label{table:s5}The results of symbolic simulations for the 2,4,8,16-state systems using JWE with SOM and without GPC. The order of the states in each cell is the same as in Table~\ref{table:s2}. Relative deviations of the final transition probabilities are calculated with respect to the benchmark data in Table~\ref{table:s2}, using Eq. 32.}
\setlength\tabcolsep{8pt}
\begin{tabular}{|c|c|c|c|c|c|c|}
\hline
\rule{0pt}{10pt} $\Delta t$& \multicolumn{2}{c|}{$10^{-1}$} & \multicolumn{2}{c|}{$10^{-2}$}& \multicolumn{2}{c|}{$10^{-3}$}\\
\cline{2-7}
& \begin{tabular}{@{}c@{}}Relative\\Deviation(\%)\end{tabular}& \begin{tabular}{@{}c@{}}Final\\Probability\end{tabular}& \begin{tabular}{@{}c@{}}Relative\\Deviation(\%)\end{tabular}& \begin{tabular}{@{}c@{}}Final\\Probability\end{tabular}& \begin{tabular}{@{}c@{}}Relative\\Deviation(\%)\end{tabular}& \begin{tabular}{@{}c@{}}Final\\Probability\end{tabular}\\
\hline
2       &\textrm{$-$9.99e$+$01}&\textrm{ $\ \,$5.30e$-$07}&\textrm{$-$9.99e$+$01}&\textrm{ $\ \,$5.10e$-$07}&\textrm{$-$9.99e$+$01}&\textrm{ $\ \,$4.80e$-$07}\\
states  &\textrm{ $\ \,$4.40e$-$02}&\textrm{ $\ \,$1.00e$+$00}&\textrm{ $\ \,$4.45e$-$02}&\textrm{ $\ \,$1.00e$+$00}&\textrm{ $\ \,$4.45e$-$02}&\textrm{ $\ \,$1.00e$+$00}\\ \hline
        &\textrm{ $\ \,$5.44e$+$03}&\textrm{ $\ \,$8.50e$-$01}&\textrm{ $\ \,$4.80e$+$03}&\textrm{ $\ \,$7.52e$-$01}&\textrm{ $\ \,$4.79e$+$03}&\textrm{ $\ \,$7.51e$-$01}\\ 
4       &\textrm{ $\ \,$7.00e$+$03}&\textrm{ $\ \,$5.32e$-$02}&\textrm{ $\ \,$1.10e$+$04}&\textrm{ $\ \,$8.35e$-$02}&\textrm{ $\ \,$1.10e$+$04}&\textrm{ $\ \,$8.34e$-$02}\\ 
states  &\textrm{ $\ \,$4.77e$+$02}&\textrm{ $\ \,$9.07e$-$02}&\textrm{ $\ \,$8.70e$+$02}&\textrm{ $\ \,$1.52e$-$01}&\textrm{ $\ \,$8.75e$+$02}&\textrm{ $\ \,$1.53e$-$01}\\
        &\textrm{$-$9.93e$+$01}&\textrm{ $\ \,$6.38e$-$03}&\textrm{$-$9.87e$+$01}&\textrm{ $\ \,$1.25e$-$02}&\textrm{$-$9.87e$+$01}&\textrm{ $\ \,$1.25e$-$02}\\ \hline
        &\textrm{$-$8.40e$+$01}&\textrm{ $\ \,$3.07e$-$04}&\textrm{ $\ \,$1.18e$+$01}&\textrm{ $\ \,$2.14e$-$03}&\textrm{ $\ \,$1.74e$+$01}&\textrm{ $\ \,$2.25e$-$03}\\ 
        &\textrm{ $\ \,$2.18e$+$04}&\textrm{ $\ \,$1.78e$-$02}&\textrm{ $\ \,$3.83e$+$03}&\textrm{ $\ \,$3.20e$-$03}&\textrm{ $\ \,$3.96e$+$03}&\textrm{ $\ \,$3.30e$-$03}\\
        &\textrm{$-$4.14e$+$01}&\textrm{ $\ \,$8.27e$-$03}&\textrm{$-$4.33e$-$01}&\textrm{ $\ \,$1.41e$-$02}&\textrm{ $\ \,$1.08e$+$00}&\textrm{ $\ \,$1.43e$-$02}\\ 
8       &\textrm{$-$8.49e$+$01}&\textrm{ $\ \,$4.25e$-$03}&\textrm{ $\ \,$2.69e$-$01}&\textrm{ $\ \,$2.82e$-$02}&\textrm{ $\ \,$1.96e$+$00}&\textrm{ $\ \,$2.87e$-$02}\\ 
states  &\textrm{$-$2.78e$+$01}&\textrm{ $\ \,$4.57e$-$02}&\textrm{$-$3.77e$+$00}&\textrm{ $\ \,$6.10e$-$02}&\textrm{$-$3.71e$+$00}&\textrm{ $\ \,$6.10e$-$02}\\ 
        &\textrm{ $\ \,$2.46e$+$01}&\textrm{ $\ \,$2.93e$-$01}&\textrm{ $\ \,$1.11e$+$01}&\textrm{ $\ \,$2.61e$-$01}&\textrm{ $\ \,$1.13e$+$01}&\textrm{ $\ \,$2.62e$-$01}\\
        &\textrm{$-$8.98e$+$00}&\textrm{ $\ \,$1.75e$-$01}&\textrm{$-$3.02e$+$00}&\textrm{ $\ \,$1.86e$-$01}&\textrm{$-$3.27e$+$00}&\textrm{ $\ \,$1.86e$-$01}\\ 
        &\textrm{$-$2.01e$+$00}&\textrm{ $\ \,$4.56e$-$01}&\textrm{$-$4.55e$+$00}&\textrm{ $\ \,$4.44e$-$01}&\textrm{$-$4.77e$+$00}&\textrm{ $\ \,$4.43e$-$01}\\ \hline
        &\textrm{ $\ \,$2.72e$+$03}&\textrm{ $\ \,$3.68e$-$03}&\textrm{$-$1.95e$-$01}&\textrm{ $\ \,$1.30e$-$04}&\textrm{ $\ \,$1.46e$+$01}&\textrm{ $\ \,$1.50e$-$04}\\
        &\textrm{ $\ \,$1.11e$+$04}&\textrm{ $\ \,$2.86e$-$02}&\textrm{$-$2.32e$+$01}&\textrm{ $\ \,$1.96e$-$04}&\textrm{$-$5.51e$+$00}&\textrm{ $\ \,$2.41e$-$04}\\
        &\textrm{ $\ \,$4.55e$+$03}&\textrm{ $\ \,$8.13e$-$02}&\textrm{$-$1.22e$+$01}&\textrm{ $\ \,$1.53e$-$03}&\textrm{ $\ \,$5.92e$+$00}&\textrm{ $\ \,$1.85e$-$03}\\
        &\textrm{ $\ \,$7.39e$+$04}&\textrm{ $\ \,$3.24e$-$01}&\textrm{ $\ \,$7.25e$+$01}&\textrm{ $\ \,$7.55e$-$04}&\textrm{ $\ \,$5.78e$+$00}&\textrm{ $\ \,$4.63e$-$04}\\
        &\textrm{ $\ \,$9.55e$+$03}&\textrm{ $\ \,$1.12e$-$01}&\textrm{$-$3.41e$+$01}&\textrm{ $\ \,$7.61e$-$04}&\textrm{$-$6.20e$+$01}&\textrm{ $\ \,$4.40e$-$04}\\
        &\textrm{$-$3.84e$+$01}&\textrm{ $\ \,$1.20e$-$03}&\textrm{$-$6.71e$+$00}&\textrm{ $\ \,$1.82e$-$03}&\textrm{ $\ \,$1.18e$+$01}&\textrm{ $\ \,$2.19e$-$03}\\
        &\textrm{ $\ \,$6.42e$+$03}&\textrm{ $\ \,$1.32e$-$01}&\textrm{$-$1.25e$+$01}&\textrm{ $\ \,$1.78e$-$03}&\textrm{$-$1.63e$+$01}&\textrm{ $\ \,$1.70e$-$03}\\
16      &\textrm{ $\ \,$3.81e$+$03}&\textrm{ $\ \,$4.06e$-$02}&\textrm{$-$4.85e$+$00}&\textrm{ $\ \,$9.89e$-$04}&\textrm{ $\ \,$1.80e$+$00}&\textrm{ $\ \,$1.06e$-$03}\\
states  &\textrm{ $\ \,$2.89e$+$03}&\textrm{ $\ \,$3.89e$-$02}&\textrm{$-$1.33e$+$01}&\textrm{ $\ \,$1.13e$-$03}&\textrm{ $\ \,$8.33e$+$00}&\textrm{ $\ \,$1.41e$-$03}\\
        &\textrm{ $\ \,$3.99e$+$03}&\textrm{ $\ \,$1.47e$-$01}&\textrm{ $\ \,$1.00e$+$01}&\textrm{ $\ \,$3.95e$-$03}&\textrm{ $\ \,$2.71e$+$00}&\textrm{ $\ \,$3.69e$-$03}\\
        &\textrm{$-$6.69e$+$01}&\textrm{ $\ \,$5.90e$-$03}&\textrm{ $\ \,$2.92e$+$00}&\textrm{ $\ \,$1.84e$-$02}&\textrm{ $\ \,$2.91e$+$00}&\textrm{ $\ \,$1.84e$-$02}\\
        &\textrm{$-$9.83e$+$01}&\textrm{ $\ \,$7.49e$-$04}&\textrm{$-$1.27e$+$00}&\textrm{ $\ \,$4.29e$-$02}&\textrm{$-$1.89e$+$00}&\textrm{ $\ \,$4.27e$-$02}\\
        &\textrm{ $\ \,$2.69e$+$01}&\textrm{ $\ \,$8.33e$-$02}&\textrm{$-$5.93e$-$01}&\textrm{ $\ \,$6.52e$-$02}&\textrm{$-$2.95e$-$01}&\textrm{ $\ \,$6.54e$-$02}\\
        &\textrm{$-$9.97e$+$01}&\textrm{ $\ \,$5.80e$-$04}&\textrm{ $\ \,$2.76e$-$01}&\textrm{ $\ \,$2.15e$-$01}&\textrm{ $\ \,$9.82e$-$02}&\textrm{ $\ \,$2.14e$-$01}\\
        &\textrm{$-$9.96e$+$01}&\textrm{ $\ \,$7.58e$-$04}&\textrm{ $\ \,$5.61e$-$02}&\textrm{ $\ \,$1.74e$-$01}&\textrm{ $\ \,$3.13e$-$01}&\textrm{ $\ \,$1.74e$-$01}\\
        &\textrm{$-$9.99e$+$01}&\textrm{ $\ \,$2.47e$-$04}&\textrm{ $\ \,$6.99e$-$02}&\textrm{ $\ \,$4.72e$-$01}&\textrm{ $\ \,$4.22e$-$02}&\textrm{ $\ \,$4.72e$-$01}\\\hline
\end{tabular}
\end{table}

\begin{table}[H]
\centering
\small
\caption{\label{table:s6}The results of symbolic simulations for the 2,4,8,16-state systems using JWE with FOM and GPC. The order of the states in each cell is the same as in Table~\ref{table:s2}. Relative deviations of the final transition probabilities are calculated with respect to the benchmark data in Table~\ref{table:s2}, using Eq. 32.}
\setlength\tabcolsep{8pt}
\begin{tabular}{|c|c|c|c|c|c|c|}
\hline
\rule{0pt}{10pt} $\Delta t$& \multicolumn{2}{c|}{$10^{-1}$} & \multicolumn{2}{c|}{$10^{-2}$}& \multicolumn{2}{c|}{$10^{-3}$}\\
\cline{2-7}
& \begin{tabular}{@{}c@{}}Relative\\Deviation(\%)\end{tabular}& \begin{tabular}{@{}c@{}}Final\\Probability\end{tabular}& \begin{tabular}{@{}c@{}}Relative\\Deviation(\%)\end{tabular}& \begin{tabular}{@{}c@{}}Final\\Probability\end{tabular}& \begin{tabular}{@{}c@{}}Relative\\Deviation(\%)\end{tabular}& \begin{tabular}{@{}c@{}}Final\\Probability\end{tabular}\\
\hline
2       &\textrm{ $\ \,$4.39e$+$01}&\textrm{ $\ \,$6.41e$-$04}&\textrm{ $\ \,$3.00e$+$00}&\textrm{ $\ \,$4.58e$-$04}&\textrm{ $\ \,$2.63e$-$01}&\textrm{ $\ \,$4.46e$-$04}\\
states  &\textrm{$-$1.95e$-$02}&\textrm{ $\ \,$9.99e$-$01}&\textrm{$-$1.34e$-$03}&\textrm{ $\ \,$1.00e$+$00}&\textrm{$-$1.20e$-$04}&\textrm{ $\ \,$1.00e$+$00}\\ \hline
        &\textrm{ $\ \,$2.38e$+$01}&\textrm{ $\ \,$1.90e$-$02}&\textrm{ $\ \,$5.65e$-$01}&\textrm{ $\ \,$1.54e$-$02}&\textrm{$-$2.21e$-$02}&\textrm{ $\ \,$1.53e$-$02}\\
4       &\textrm{ $\ \,$2.38e$+$01}&\textrm{ $\ \,$9.27e$-$04}&\textrm{ $\ \,$6.39e$-$01}&\textrm{ $\ \,$7.54e$-$04}&\textrm{$-$5.73e$-$03}&\textrm{ $\ \,$7.49e$-$04}\\
states  &\textrm{$-$2.06e$+$01}&\textrm{ $\ \,$1.25e$-$02}&\textrm{$-$9.63e$-$01}&\textrm{ $\ \,$1.56e$-$02}&\textrm{$-$1.64e$-$01}&\textrm{ $\ \,$1.57e$-$02}\\
        &\textrm{$-$6.00e$-$02}&\textrm{ $\ \,$9.68e$-$01}&\textrm{ $\ \,$6.19e$-$03}&\textrm{ $\ \,$9.68e$-$01}&\textrm{ $\ \,$3.02e$-$03}&\textrm{ $\ \,$9.68e$-$01}\\\hline
        &\textrm{ $\ \,$1.29e$+$04}&\textrm{ $\ \,$2.50e$-$01}&\textrm{ $\ \,$5.00e$+$02}&\textrm{ $\ \,$1.15e$-$02}&\textrm{ $\ \,$1.24e$+$01}&\textrm{ $\ \,$2.15e$-$03}\\
        &\textrm{ $\ \,$5.88e$+$05}&\textrm{ $\ \,$4.78e$-$01}&\textrm{ $\ \,$2.44e$+$03}&\textrm{ $\ \,$2.06e$-$03}&\textrm{$-$1.21e$+$01}&\textrm{ $\ \,$7.14e$-$05}\\
        &\textrm{ $\ \,$4.77e$+$02}&\textrm{ $\ \,$8.14e$-$02}&\textrm{ $\ \,$4.30e$+$01}&\textrm{ $\ \,$2.02e$-$02}&\textrm{ $\ \,$1.10e$+$00}&\textrm{ $\ \,$1.43e$-$02}\\
8       &\textrm{$-$7.88e$+$01}&\textrm{ $\ \,$5.97e$-$03}&\textrm{ $\ \,$6.13e$+$01}&\textrm{ $\ \,$4.54e$-$02}&\textrm{ $\ \,$1.13e$+$00}&\textrm{ $\ \,$2.85e$-$02}\\
states  &\textrm{ $\ \,$1.55e$+$02}&\textrm{ $\ \,$1.62e$-$01}&\textrm{$-$2.96e$+$01}&\textrm{ $\ \,$4.46e$-$02}&\textrm{$-$1.09e$+$00}&\textrm{ $\ \,$6.27e$-$02}\\
        &\textrm{$-$9.40e$+$01}&\textrm{ $\ \,$1.40e$-$02}&\textrm{$-$7.74e$+$00}&\textrm{ $\ \,$2.17e$-$01}&\textrm{$-$1.35e$-$01}&\textrm{ $\ \,$2.35e$-$01}\\
        &\textrm{$-$9.56e$+$01}&\textrm{ $\ \,$8.39e$-$03}&\textrm{ $\ \,$5.34e$+$00}&\textrm{ $\ \,$2.02e$-$01}&\textrm{ $\ \,$1.51e$-$01}&\textrm{ $\ \,$1.92e$-$01}\\
        &\textrm{$-$9.99e$+$01}&\textrm{ $\ \,$4.43e$-$04}&\textrm{$-$1.76e$+$00}&\textrm{ $\ \,$4.57e$-$01}&\textrm{ $\ \,$4.50e$-$03}&\textrm{ $\ \,$4.65e$-$01}\\   \hline     
        &\textrm{ $\ \,$4.40e$+$04}&\textrm{ $\ \,$5.75e$-$02}&\textrm{ $\ \,$3.02e$+$03}&\textrm{ $\ \,$4.08e$-$03}&\textrm{ $\ \,$1.48e$+$01}&\textrm{ $\ \,$1.50e$-$04}\\
        &\textrm{ $\ \,$3.47e$+$04}&\textrm{ $\ \,$8.90e$-$02}&\textrm{ $\ \,$5.24e$+$02}&\textrm{ $\ \,$1.60e$-$03}&\textrm{$-$2.07e$+$01}&\textrm{ $\ \,$2.03e$-$04}\\
        &\textrm{ $\ \,$4.02e$+$02}&\textrm{ $\ \,$8.77e$-$03}&\textrm{ $\ \,$1.46e$+$04}&\textrm{ $\ \,$2.57e$-$01}&\textrm{ $\ \,$3.50e$+$00}&\textrm{ $\ \,$1.81e$-$03}\\
        &\textrm{ $\ \,$2.32e$+$04}&\textrm{ $\ \,$1.02e$-$01}&\textrm{ $\ \,$1.33e$+$03}&\textrm{ $\ \,$6.23e$-$03}&\textrm{ $\ \,$3.14e$+$01}&\textrm{ $\ \,$5.75e$-$04}\\
        &\textrm{ $\ \,$1.96e$+$04}&\textrm{ $\ \,$2.28e$-$01}&\textrm{ $\ \,$4.27e$+$04}&\textrm{ $\ \,$4.94e$-$01}&\textrm{ $\ \,$6.30e$+$01}&\textrm{ $\ \,$1.88e$-$03}\\
        &\textrm{ $\ \,$3.22e$+$02}&\textrm{ $\ \,$8.25e$-$03}&\textrm{ $\ \,$3.45e$+$02}&\textrm{ $\ \,$8.69e$-$03}&\textrm{ $\ \,$1.58e$+$01}&\textrm{ $\ \,$2.26e$-$03}\\
        &\textrm{ $\ \,$2.45e$+$03}&\textrm{ $\ \,$5.19e$-$02}&\textrm{ $\ \,$2.37e$+$03}&\textrm{ $\ \,$5.01e$-$02}&\textrm{$-$5.85e$+$00}&\textrm{ $\ \,$1.91e$-$03}\\
16      &\textrm{ $\ \,$3.07e$+$04}&\textrm{ $\ \,$3.21e$-$01}&\textrm{ $\ \,$4.78e$+$01}&\textrm{ $\ \,$1.54e$-$03}&\textrm{$-$7.60e$+$00}&\textrm{ $\ \,$9.61e$-$04}\\
states  &\textrm{ $\ \,$1.81e$+$03}&\textrm{ $\ \,$2.48e$-$02}&\textrm{ $\ \,$1.08e$+$04}&\textrm{ $\ \,$1.42e$-$01}&\textrm{$-$2.91e$+$01}&\textrm{ $\ \,$9.23e$-$04}\\
        &\textrm{$-$5.46e$+$01}&\textrm{ $\ \,$1.63e$-$03}&\textrm{ $\ \,$7.47e$+$01}&\textrm{ $\ \,$6.28e$-$03}&\textrm{ $\ \,$8.18e$+$00}&\textrm{ $\ \,$3.89e$-$03}\\
        &\textrm{ $\ \,$1.86e$+$02}&\textrm{ $\ \,$5.10e$-$02}&\textrm{$-$8.85e$+$01}&\textrm{ $\ \,$2.06e$-$03}&\textrm{ $\ \,$3.18e$-$01}&\textrm{ $\ \,$1.79e$-$02}\\
        &\textrm{$-$1.23e$+$00}&\textrm{ $\ \,$4.30e$-$02}&\textrm{$-$5.08e$+$01}&\textrm{ $\ \,$2.14e$-$02}&\textrm{ $\ \,$2.59e$+$00}&\textrm{ $\ \,$4.46e$-$02}\\
        &\textrm{$-$8.89e$+$01}&\textrm{ $\ \,$7.28e$-$03}&\textrm{$-$9.70e$+$01}&\textrm{ $\ \,$1.95e$-$03}&\textrm{$-$2.00e$+$00}&\textrm{ $\ \,$6.43e$-$02}\\
        &\textrm{$-$9.81e$+$01}&\textrm{ $\ \,$4.15e$-$03}&\textrm{$-$9.95e$+$01}&\textrm{ $\ \,$1.04e$-$03}&\textrm{$-$1.21e$-$01}&\textrm{ $\ \,$2.14e$-$01}\\
        &\textrm{$-$9.89e$+$01}&\textrm{ $\ \,$1.97e$-$03}&\textrm{$-$9.90e$+$01}&\textrm{ $\ \,$1.76e$-$03}&\textrm{$-$9.96e$-$02}&\textrm{ $\ \,$1.74e$-$01}\\
        &\textrm{$-$1.00e$+$02}&\textrm{ $\ \,$2.34e$-$05}&\textrm{$-$1.00e$+$02}&\textrm{ $\ \,$7.19e$-$05}&\textrm{$-$7.70e$-$02}&\textrm{ $\ \,$4.71e$-$01}\\\hline  
\end{tabular}
\end{table}

\subsection{\label{sec:s8b}Results Using QEE}
\begin{table}[H]
\centering
\small
\caption{\label{table:s7}The results of symbolic simulations for the 2,4,8-state systems using QEE with SOM and without GPC. The order of the states in each cell is the same as in Table~\ref{table:s2}. Relative deviations of the final transition probabilities are calculated with respect to the benchmark data in Table~\ref{table:s2}, using Eq. 32.}
\setlength\tabcolsep{8pt}
\begin{tabular}{|c|c|c|c|c|c|c|}
\hline
\rule{0pt}{10pt} $\Delta t$& \multicolumn{2}{c|}{$10^{-1}$} & \multicolumn{2}{c|}{$10^{-2}$}& \multicolumn{2}{c|}{$10^{-3}$}\\
\cline{2-7}
& \begin{tabular}{@{}c@{}}Relative\\Deviation(\%)\end{tabular}& \begin{tabular}{@{}c@{}}Final\\Probability\end{tabular}& \begin{tabular}{@{}c@{}}Relative\\Deviation(\%)\end{tabular}& \begin{tabular}{@{}c@{}}Final\\Probability\end{tabular}& \begin{tabular}{@{}c@{}}Relative\\Deviation(\%)\end{tabular}& \begin{tabular}{@{}c@{}}Final\\Probability\end{tabular}\\
\hline
2       &\textrm{$-$9.98e$+$01}&\textrm{ $\ \,$8.20e$-$07}&\textrm{$-$9.98e$+$01}&\textrm{ $\ \,$8.20e$-$07}&\textrm{$-$9.98e$+$01}&\textrm{ $\ \,$8.20e$-$07}\\
states  &\textrm{ $\ \,$4.45e$-$02}&\textrm{ $\ \,$1.00e$+$00}&\textrm{ $\ \,$4.45e$-$02}&\textrm{ $\ \,$1.00e$+$00}&\textrm{ $\ \,$4.45e$-$02}&\textrm{ $\ \,$1.00e$+$00}\\ \hline
        &\textrm{$-$8.79e$+$01}&\textrm{ $\ \,$1.86e$-$03}&\textrm{$-$8.91e$+$01}&\textrm{ $\ \,$1.68e$-$03}&\textrm{$-$8.93e$+$01}&\textrm{ $\ \,$1.64e$-$03}\\
4       &\textrm{ $\ \,$3.94e$+$02}&\textrm{ $\ \,$3.70e$-$03}&\textrm{ $\ \,$4.04e$+$02}&\textrm{ $\ \,$3.78e$-$03}&\textrm{ $\ \,$4.07e$+$02}&\textrm{ $\ \,$3.80e$-$03}\\
states  &\textrm{$-$8.87e$+$01}&\textrm{ $\ \,$1.77e$-$03}&\textrm{$-$9.04e$+$01}&\textrm{ $\ \,$1.50e$-$03}&\textrm{$-$9.09e$+$01}&\textrm{ $\ \,$1.44e$-$03}\\
        &\textrm{ $\ \,$2.53e$+$00}&\textrm{ $\ \,$9.93e$-$01}&\textrm{ $\ \,$2.57e$+$00}&\textrm{ $\ \,$9.93e$-$01}&\textrm{ $\ \,$2.57e$+$00}&\textrm{ $\ \,$9.93e$-$01}\\\hline
        &\textrm{ $\ \,$1.28e$+$04}&\textrm{ $\ \,$2.46e$-$01}&\textrm{$-$9.92e$+$01}&\textrm{ $\ \,$1.55e$-$05}&\textrm{ $\ \,$1.52e$+$03}&\textrm{ $\ \,$3.11e$-$02}\\
        &\textrm{ $\ \,$1.27e$+$03}&\textrm{ $\ \,$1.11e$-$03}&\textrm{ $\ \,$1.38e$+$04}&\textrm{ $\ \,$1.13e$-$02}&\textrm{ $\ \,$7.88e$+$04}&\textrm{ $\ \,$6.41e$-$02}\\
        &\textrm{ $\ \,$9.52e$+$02}&\textrm{ $\ \,$1.49e$-$01}&\textrm{ $\ \,$2.17e$+$01}&\textrm{ $\ \,$1.72e$-$02}&\textrm{ $\ \,$2.16e$+$02}&\textrm{ $\ \,$4.47e$-$02}\\
8       &\textrm{ $\ \,$2.52e$+$02}&\textrm{ $\ \,$9.93e$-$02}&\textrm{ $\ \,$4.63e$+$01}&\textrm{ $\ \,$4.12e$-$02}&\textrm{ $\ \,$2.58e$+$02}&\textrm{ $\ \,$1.01e$-$01}\\
states  &\textrm{$-$3.68e$+$01}&\textrm{ $\ \,$4.01e$-$02}&\textrm{$-$8.95e$+$01}&\textrm{ $\ \,$6.65e$-$03}&\textrm{$-$6.27e$+$00}&\textrm{ $\ \,$5.94e$-$02}\\
        &\textrm{$-$4.77e$+$01}&\textrm{ $\ \,$1.23e$-$01}&\textrm{$-$7.33e$+$01}&\textrm{ $\ \,$6.29e$-$02}&\textrm{$-$7.85e$+$01}&\textrm{ $\ \,$5.05e$-$02}\\
        &\textrm{$-$4.17e$+$01}&\textrm{ $\ \,$1.12e$-$01}&\textrm{ $\ \,$2.32e$+$01}&\textrm{ $\ \,$2.36e$-$01}&\textrm{$-$9.77e$+$01}&\textrm{ $\ \,$4.34e$-$03}\\
        &\textrm{$-$5.06e$+$01}&\textrm{ $\ \,$2.30e$-$01}&\textrm{ $\ \,$3.42e$+$01}&\textrm{ $\ \,$6.24e$-$01}&\textrm{ $\ \,$3.86e$+$01}&\textrm{ $\ \,$6.45e$-$01}\\\hline
\end{tabular}
\end{table}
\vspace*{30pt}
\begin{table}[H]
\centering
\small
\caption{\label{table:s8}The results of symbolic simulations for the 2,4,8-state systems using QEE with FOM and GPC. The order of the states in each cell is the same as in Table~\ref{table:s2}. Relative deviations of the final transition probabilities are calculated with respect to the benchmark data in Table~\ref{table:s2}, using Eq. 32.}
\setlength\tabcolsep{8pt}
\begin{tabular}{|c|c|c|c|c|c|c|}
\hline
\rule{0pt}{10pt} $\Delta t$& \multicolumn{2}{c|}{$10^{-1}$} & \multicolumn{2}{c|}{$10^{-2}$}& \multicolumn{2}{c|}{$10^{-3}$}\\
\cline{2-7}
& \begin{tabular}{@{}c@{}}Relative\\Deviation(\%)\end{tabular}& \begin{tabular}{@{}c@{}}Final\\Probability\end{tabular}& \begin{tabular}{@{}c@{}}Relative\\Deviation(\%)\end{tabular}& \begin{tabular}{@{}c@{}}Final\\Probability\end{tabular}& \begin{tabular}{@{}c@{}}Relative\\Deviation(\%)\end{tabular}& \begin{tabular}{@{}c@{}}Final\\Probability\end{tabular}\\
\hline
2       &\textrm{ $\ \,$4.32e$+$01}&\textrm{ $\ \,$6.38e$-$04}&\textrm{ $\ \,$3.30e$+$00}&\textrm{ $\ \,$4.60e$-$04}&\textrm{ $\ \,$6.79e$-$01}&\textrm{ $\ \,$4.48e$-$04}\\
states  &\textrm{$-$1.92e$-$02}&\textrm{ $\ \,$9.99e$-$01}&\textrm{$-$1.47e$-$03}&\textrm{ $\ \,$1.00e$+$00}&\textrm{$-$3.00e$-$04}&\textrm{ $\ \,$1.00e$+$00}\\\hline
        &\textrm{ $\ \,$2.39e$+$01}&\textrm{ $\ \,$1.90e$-$02}&\textrm{ $\ \,$1.11e$+$00}&\textrm{ $\ \,$1.55e$-$02}&\textrm{ $\ \,$1.46e$-$01}&\textrm{ $\ \,$1.54e$-$02}\\
4       &\textrm{ $\ \,$2.39e$+$01}&\textrm{ $\ \,$9.28e$-$04}&\textrm{ $\ \,$1.51e$+$00}&\textrm{ $\ \,$7.60e$-$04}&\textrm{ $\ \,$5.25e$-$01}&\textrm{ $\ \,$7.53e$-$04}\\
states  &\textrm{$-$2.07e$+$01}&\textrm{ $\ \,$1.25e$-$02}&\textrm{$-$8.13e$-$01}&\textrm{ $\ \,$1.56e$-$02}&\textrm{$-$9.90e$-$02}&\textrm{ $\ \,$1.57e$-$02}\\
        &\textrm{$-$6.06e$-$02}&\textrm{ $\ \,$9.68e$-$01}&\textrm{$-$5.62e$-$03}&\textrm{ $\ \,$9.68e$-$01}&\textrm{$-$1.12e$-$03}&\textrm{ $\ \,$9.68e$-$01}\\\hline
        &\textrm{$-$3.30e$+$01}&\textrm{ $\ \,$1.28e$-$03}&\textrm{ $\ \,$5.16e$+$02}&\textrm{ $\ \,$1.18e$-$02}&\textrm{ $\ \,$1.19e$+$01}&\textrm{ $\ \,$2.15e$-$03}\\
        &\textrm{ $\ \,$3.05e$+$05}&\textrm{ $\ \,$2.48e$-$01}&\textrm{ $\ \,$2.48e$+$03}&\textrm{ $\ \,$2.10e$-$03}&\textrm{$-$1.41e$+$01}&\textrm{ $\ \,$6.98e$-$05}\\
        &\textrm{$-$9.72e$+$01}&\textrm{ $\ \,$3.97e$-$04}&\textrm{ $\ \,$7.40e$+$01}&\textrm{ $\ \,$2.46e$-$02}&\textrm{ $\ \,$1.20e$+$00}&\textrm{ $\ \,$1.43e$-$02}\\
8       &\textrm{ $\ \,$1.48e$+$03}&\textrm{ $\ \,$4.44e$-$01}&\textrm{ $\ \,$5.04e$+$01}&\textrm{ $\ \,$4.24e$-$02}&\textrm{ $\ \,$1.23e$+$00}&\textrm{ $\ \,$2.85e$-$02}\\
states  &\textrm{$-$9.18e$+$01}&\textrm{ $\ \,$5.22e$-$03}&\textrm{$-$2.45e$+$01}&\textrm{ $\ \,$4.78e$-$02}&\textrm{$-$1.07e$+$00}&\textrm{ $\ \,$6.27e$-$02}\\
        &\textrm{$-$9.14e$+$01}&\textrm{ $\ \,$2.03e$-$02}&\textrm{$-$6.22e$+$00}&\textrm{ $\ \,$2.21e$-$01}&\textrm{$-$1.44e$-$01}&\textrm{ $\ \,$2.35e$-$01}\\
        &\textrm{ $\ \,$1.51e$-$01}&\textrm{ $\ \,$1.92e$-$01}&\textrm{ $\ \,$8.58e$+$00}&\textrm{ $\ \,$2.08e$-$01}&\textrm{ $\ \,$1.35e$-$01}&\textrm{ $\ \,$1.92e$-$01}\\
        &\textrm{$-$8.10e$+$01}&\textrm{ $\ \,$8.86e$-$02}&\textrm{$-$4.91e$+$00}&\textrm{ $\ \,$4.43e$-$01}&\textrm{ $\ \,$5.01e$-$03}&\textrm{ $\ \,$4.65e$-$01}\\\hline
\end{tabular}
\end{table}

\section{\label{sec:s9}Gate-Based Circuit Simulation Results}

The results of the gate-based circuit simulations are done by the Pennylane [62] "\texttt{default\_qubit}" simulator without sampling (no shots), which means that the device returns analytical results. The results of 8-state system with step size $10^{-3}$ are not available due to excessive amount of computation time.

\begin{table}[H]
\centering
\small
\caption{\label{table:s9}The results of Pennylane simulation for the 2,4,8-state systems using QEE with SOM and GPC. The order of the states in each cell is the same as in Table~\ref{table:s2}. Relative deviations of the final transition probabilities are calculated with respect to the benchmark data in Table~\ref{table:s2}, using Eq. 32.}
\setlength\tabcolsep{8pt}
\begin{tabular}{|c|c|c|c|c|c|c|}
\hline
\rule{0pt}{10pt} $\Delta t$& \multicolumn{2}{c|}{$10^{-1}$} & \multicolumn{2}{c|}{$10^{-2}$}& \multicolumn{2}{c|}{$10^{-3}$}\\
\cline{2-7}
& \begin{tabular}{@{}c@{}}Relative\\Deviation(\%)\end{tabular}& \begin{tabular}{@{}c@{}}Final\\Probability\end{tabular}& \begin{tabular}{@{}c@{}}Relative\\Deviation(\%)\end{tabular}& \begin{tabular}{@{}c@{}}Final\\Probability\end{tabular}& \begin{tabular}{@{}c@{}}Relative\\Deviation(\%)\end{tabular}& \begin{tabular}{@{}c@{}}Final\\Probability\end{tabular}\\
\hline
2       &\textrm{ $\ \,$1.55e$-$01}&\textrm{ $\ \,$4.46e$-$04}&\textrm{$-$9.54e$-$02}&\textrm{ $\ \,$4.45e$-$04}&\textrm{$-$1.60e$-$01}&\textrm{ $\ \,$4.44e$-$04}\\
states  &\textrm{$-$7.00e$-$05}&\textrm{ $\ \,$1.00e$+$00}&\textrm{ $\ \,$4.00e$-$05}&\textrm{ $\ \,$1.00e$+$00}&\textrm{ $\ \,$7.00e$-$05}&\textrm{ $\ \,$1.00e$+$00}\\\hline
        &\textrm{$-$3.78e$-$01}&\textrm{ $\ \,$1.53e$-$02}&\textrm{$-$1.69e$-$01}&\textrm{ $\ \,$1.53e$-$02}&\textrm{$-$1.02e$+$00}&\textrm{ $\ \,$1.52e$-$02}\\
4       &\textrm{$-$2.99e$-$01}&\textrm{ $\ \,$7.47e$-$04}&\textrm{$-$1.35e$-$01}&\textrm{ $\ \,$7.48e$-$04}&\textrm{$-$4.97e$-$01}&\textrm{ $\ \,$7.45e$-$04}\\
states  &\textrm{ $\ \,$1.62e$+$00}&\textrm{ $\ \,$1.60e$-$02}&\textrm{$-$9.69e$-$02}&\textrm{ $\ \,$1.57e$-$02}&\textrm{$-$4.23e$-$01}&\textrm{ $\ \,$1.57e$-$02}\\
        &\textrm{$-$2.01e$-$02}&\textrm{ $\ \,$9.68e$-$01}&\textrm{ $\ \,$4.35e$-$03}&\textrm{ $\ \,$9.68e$-$01}&\textrm{ $\ \,$2.34e$-$02}&\textrm{ $\ \,$9.68e$-$01}\\\hline
        &\textrm{ $\ \,$2.26e$+$04}&\textrm{ $\ \,$4.35e$-$01}&\textrm{ $\ \,$2.17e$+$04}&\textrm{ $\ \,$4.17e$-$01}&\multirow{8}{*}{n/a}&\multirow{8}{*}{n/a}\\
        &\textrm{ $\ \,$1.14e$+$05}&\textrm{ $\ \,$9.28e$-$02}&\textrm{ $\ \,$3.28e$+$02}&\textrm{ $\ \,$3.47e$-$04}&&\\
        &\textrm{$-$9.43e$+$01}&\textrm{ $\ \,$8.09e$-$04}&\textrm{ $\ \,$1.55e$+$03}&\textrm{ $\ \,$2.34e$-$01}&&\\
8       &\textrm{ $\ \,$1.07e$+$02}&\textrm{ $\ \,$5.83e$-$02}&\textrm{ $\ \,$3.53e$+$01}&\textrm{ $\ \,$3.81e$-$02}&&\\
states  &\textrm{$-$3.75e$+$01}&\textrm{ $\ \,$3.96e$-$02}&\textrm{ $\ \,$1.58e$+$02}&\textrm{ $\ \,$1.64e$-$01}&&\\
        &\textrm{$-$7.99e$+$01}&\textrm{ $\ \,$4.72e$-$02}&\textrm{$-$9.88e$+$01}&\textrm{ $\ \,$2.89e$-$03}&&\\
        &\textrm{$-$1.74e$+$01}&\textrm{ $\ \,$1.59e$-$01}&\textrm{$-$5.78e$+$01}&\textrm{ $\ \,$8.10e$-$02}&&\\
        &\textrm{$-$6.40e$+$01}&\textrm{ $\ \,$1.67e$-$01}&\textrm{$-$8.64e$+$01}&\textrm{ $\ \,$6.33e$-$02}&&\\\hline
\end{tabular}
\end{table}

\section{\label{sec:s10}The Device Calibration Data }
The noise model performed on QASM simulator is obtained from the calibration data of IBM Q device “\texttt{ibmq\_jakarta}”, as an example of the latest generation of noisy intermediate scale quantum device.

\begin{table}[H]
\centering
\small
\caption{\label{table:s10}The single-qubit calibration data of “\texttt{ibmq\_jakarta}” on Oct. $10^{th}$, 2021.}
\setlength\tabcolsep{8pt}
\begin{tabular}{|c|c|c|c|c|}
\hline
Qubit&Gate Error&Readout Error&$T_1 (\mu s)$& $T_2 (\mu s)$\\\hline
0&	0.0311\%&	4.65\%&	186.661	&37.528\\\hline
1&	0.0212\%&	1.90\%&	165.595	&22.414\\\hline
2&	0.0297\%&	1.71\%&	143.509	&24.961\\\hline
3&	0.0285\%&	1.65\%&	101.895	&26.525\\\hline
4&	0.0249\%&	3.77\%&	88.442	&52.854\\\hline
5&	0.0248\%&	2.93\%&	69.293	&48.840\\\hline
6&	0.0280\%&	2.72\%&	132.839	&22.288\\\hline
\end{tabular}
\end{table}

\begin{table}[H]
\centering
\small
\caption{\label{table:s11}The CNOT gate calibration data of “\texttt{ibmq\_jakarta}” on Oct. $10^{th}$, 2021.}
\setlength\tabcolsep{8pt}
\begin{tabular}{|c|c|c|}
\hline
Coupling pair&	Gate Error&	Gate time ($ns$)\\ \hline
[0,1]&	1.106\%&	235\\ \hline
[1,0]&	1.106\%&	270\\ \hline
[1,2]&	1.281\%&	284\\ \hline
[2,1]&	1.281\%&	249\\ \hline
[1,3]&	0.884\%&	384\\ \hline
[3,1]&	0.884\%&	420\\ \hline
[3,5]&	0.782\%&	341\\ \hline
[5,3]&	0.782\%&	377\\ \hline
[4,5]&	0.719\%&	405\\ \hline
[5,4]&	0.719\%&	370\\ \hline
[5,6]&	0.750\%&	313\\ \hline
[6,5]&	0.750\%&	277\\ \hline
\end{tabular}
\end{table}